\documentclass[12pt,oneside]{book}   

\usepackage{blindtext} 
\usepackage{amsmath} 
\usepackage{amssymb} 
\usepackage{cite} 
\usepackage[nottoc]{tocbibind} 
\usepackage{graphicx} 
\usepackage{placeins} 
\usepackage{longtable} 
\usepackage{bigstrut} 
\usepackage{enumerate} 
\usepackage{todonotes} 
\usepackage{makeidx} 
\makeindex

\usepackage[top=2cm, bottom=1.8cm,left=2.5cm,right=2.5cm]{geometry} 
\usepackage{fancyhdr} 

\makeatletter
\newcommand{\vast}{\bBigg@{4}}
\newcommand{\Vast}{\bBigg@{5}}
\makeatother

\begin{document}


\pagestyle{empty}
\title{Investigating Chaos by the Generalized Alignment Index (GALI) Method}
\author{Henok Tenaw Moges}
\date{} 
\maketitle
%
%
%
%
%
%


%

\begin{titlepage}

\newcommand{\HRule}{\rule{\linewidth}{0.5mm}} 
\setcounter{tocdepth}{1}
\setcounter{secnumdepth}{3}
\center 
 

\textsc{\LARGE University of Cape Town}\\[1.5cm] 
\textsc{\Large Faculty of Science}\\[0.5cm] 
\textsc{\large Department of Mathematics and Applied Mathematics}\\[0.5cm] 


\HRule \\[0.4cm]
{ \huge \bfseries Investigating Chaos by the Generalized Alignment Index (GALI) Method}\\[0.4cm] 
\HRule \\[1.5cm]
 

\begin{minipage}{0.4\textwidth}
\begin{flushleft} \large
\emph{Author:}\\
Henok Tenaw Moges 
\end{flushleft}
\end{minipage}
~
\begin{minipage}{0.4\textwidth}
\begin{flushright} \large
\emph{Supervisor:} \\
A/Prof Haris Skokos 
\end{flushright}
\end{minipage}\\[4cm]



{\large \date{March, 2020}}\\[3cm] 


\includegraphics[width=0.25\textwidth]{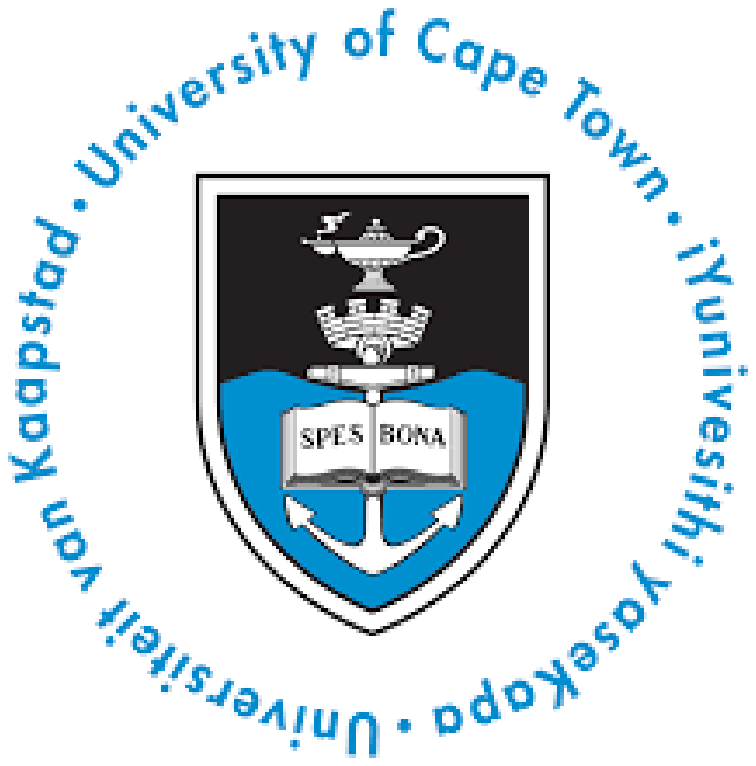}
 

\vfill 

\end{titlepage} 

\chapter*{Abstract}
\addcontentsline{toc}{chapter}{Abstract}  
One of the fundamental tasks in the study of dynamical systems is the discrimination between regular and chaotic behavior. Over the years several methods of chaos detection have been developed. Some of them, such as the construction of the system's Poincar{\'e} Surface of Section, are appropriate for low-dimensional systems. However, an enormous number of real-world problems are described by high-dimensional systems. Thus, modern numerical methods like the Smaller (SALI) and the Generalized (GALI) Alignment Index, which can also be used for lower-dimensional systems, are appropriate for investigating regular and chaotic motion in high-dimensional systems. In this work, we numerically investigate the behavior of the GALIs in the neighborhood of simple stable periodic orbits of the well-known Fermi-Pasta-Ulam-Tsingou lattice model. In particular, we study how the values of the GALIs depend on the width of the stability island and the system's energy. We find that the asymptotic GALI values increase when the studied regular orbits move closer to the edge of the stability island for fixed energy, while these indices decrease as the system's energy increases. We also investigate the dependence of the GALIs on the initial distribution of the coordinates of the deviation vectors used for their computation and the corresponding angles between these vectors. In this case, we show that the final constant values of the GALIs are independent of the choice of the initial deviation vectors needed for their computation.   
\vspace{4cm}
\section*{Plagiarism declaration}

\vspace{0.25cm}

I know the meaning of plagiarism and certify that this dissertation is my work, based on my study and I have acknowledged all the materials and resources used in the preparation.

\vspace{0.7cm}

\qquad \includegraphics[trim=6.0cm 5.cm 6.0cm 5.cm,width=0.1\textwidth,keepaspectratio]{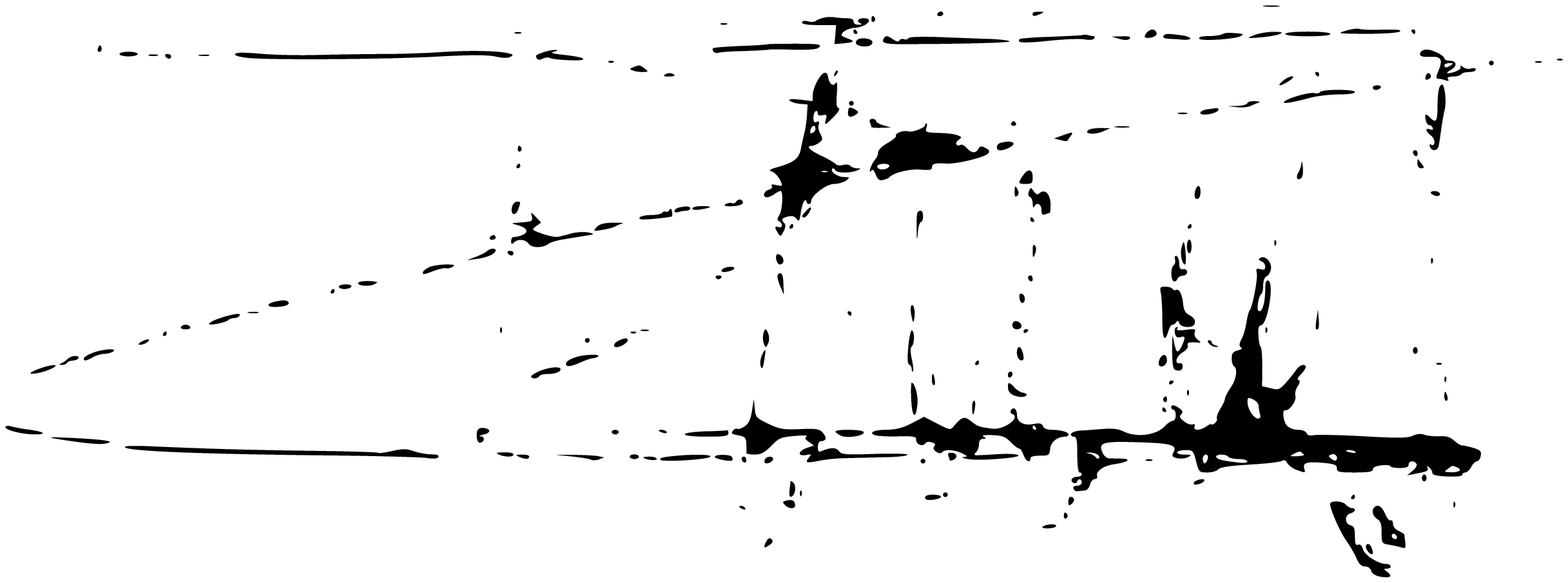}

\rule{3.2cm}{0.25mm}

\vspace{0.2cm}
\textsf{Henok Tenaw Moges} \
\chapter*{Acknowledgments}
\addcontentsline{toc}{chapter}{Acknowledgments}  
First and foremost, I would like to express my sincere gratitude to my supervisor, A/Prof Haris Skokos, for his continuous support and guidance. I am so grateful to have a supervisor who cares about my work. I would also like to thank A/Prof Thanos Manos for several useful discussions. I will forever be thankful to my colleagues in the `Nonlinear Dynamics and Chaos' group”; Bob Senyange, Many Manda, Malcolm Hillebrand, and the former group member Chinenye Ani. Thank you, guys. I would like to pay my special regards to the Department of Mathematics and Applied Mathematics of the University of Cape Town (UCT) as well. This work was fully funded by the Woldia University and the Ministry of Science and Higher Education of Ethiopia and partially supported by the UCT International and Refugee Grant 2019. Finally, I would like to thank the High-Performance Computing facility of UCT and the Center for High-Performance Computing for providing the computational resources needed for this work. 
\tableofcontents

\chapter*{List of abbreviations}
\addcontentsline{toc}{chapter}{Abbreviations}  
\begin{itemize}
	\item dof : degrees of freedom 
	\item FPUT : Fermi-Pasta-Ulam-Tsingou
	\item GALI : Generalized Alignment Index
	\item IC : Initial Condition
	\item LE : Lyapunov Exponent
	\item mLE : maximum Lyapunov Exponent
	\item $N$D : $N$-Dimensional
	\item PSS : Poincar{\'e} Surface of Section
	\item SALI : Smaller Alignment Index Method
	\item SI : Symplectic Integrator 
	\item SVD : Singular Value Decomposition 
	
\end{itemize}
\chapter*{Introduction} \label{Chapter-Zero}
\addcontentsline{toc}{chapter}{Introduction}

\pagestyle{fancy}
\fancyhf{}
\fancyhead[OC]{\leftmark}
\fancyhead[EC]{\rightmark}
\cfoot{\thepage}

Dynamical Systems theory attempts to understand, or at least describe, changes over time that occur in physical and artificial systems. It consists of a set of subaltern theories that can be applied to solve complex problems in numerous fields, such as weather predictability \cite{L93}, stochastic processes \cite{AAV02} and evolutions of astronomical systems \cite{C01}. Because of the remarkable applicability of this theory, it has become one of the top topics in modern scientific research. In particular, Hamiltonian systems of many degrees of freedom (dof) have been broadly used for a general study and a better understanding of energy transport and equipartition phenomena (see for example \cite{LL_92,MM1987,Wigg1988,Simo1999,BS_12} and references therein). Equipartition phenomena are associated with the dynamical nature of motion fostered in the respective phase space namely, the regular or chaotic evolution of the system's orbits \cite{LL_92}. A chaotic dynamical system is one in which nearby initial conditions (ICs) lead to very different states as time evolves. This means that small variations in ICs can lead to large, and frequently unpredictable, variations in final states. 

A fundamental aspect in studies of dynamical systems is the identification of chaotic behavior, both locally, i.e.~in the neighborhood of individual orbits and globally, meaning for large samples of ICs. There are several ways to identify the chaotic behavior of a given nonlinear system. The rapid and efficient detection of the regular or chaotic nature of motion in many dof systems has been a very active research topic over the years. The most commonly used method to characterize chaos is the computation of the Lyapunov Exponents (LEs) \cite{Lyapunov_1892,O_68,BGGS_80a,BGGS_80b,S10}. In general, LEs are asymptotic measures characterizing the average rate of growth or shrink of small perturbations to the orbits of dynamical systems. The development of efficient algorithms for the calculation of LEs \cite{BGGS_80a,BGGS_80b} (see also \cite{S10} for a recent review on this topic) led to a gallery of chaos detection methods (see e.g.~\cite{MDCG2011,DMCG2012,CMD2014,SkoGotLas2016}). The most commonly employed chaos indicators were recently reviewed in \cite{SkoGotLas2016}. 

This study focuses on the Generalized Alignment Index (GALI) method \cite{SBA07}, an efficient and fast chaos detection technique that can be used effectively for the description of the behavior of multidimensional systems. As its name would suggest, the GALI is a generalization of the Smaller Alignment Index (SALI) \cite{S01}. The main advantages of the GALI method are its ability to distinguish between regular and chaotic motion more quickly than other techniques and to determine the dimensionality of the torus on which the quasi-periodic motion occurs. In addition, it can predict slow diffusion in multidimensional Hamiltonian systems \cite{SBA08}. A concise overview of the theory and numerical applications of the SALI and the GALI methods can be found in \cite{SM16}.

Here we aim to understand the GALI's prediction efficiency and performance when the IC of an orbit moves gradually from a regular to a chaotic motion region in Hamiltonian systems, by studying in detail the variation of the asymptotic values of the indices. This transition can take place either by changing the total energy of the system or by moving the regular orbit's IC towards the edge of a stability island. In addition, we follow the time evolution of the deviation vectors, which are small perturbations of the IC and are needed for the computation of the GALIs, and examine if different initial distributions of the coordinates of these vectors affect the final GALI value. Our investigation was performed in phase space regions around some simple, stable, periodic orbits of the Fermi-Pasta-Ulam-Tsingou (FPUT) model \cite{FPUT55,F92}, which describes a chain of harmonic oscillators coupled through nonlinear interactions. The dynamics of this system has been studied extensively in the last decades (see e.g.~\cite{CamRosZas2005,Gal2008} and references therein) and this model is considered nowadays as a prototypical, multidimensional, nonlinear system. 

Moreover, an investigation of the behavior of the GALIs for regular orbits of coupled area-preserving standard mappings \cite{KG88} was carried out. More details on this mapping can be found in \cite{BMC2009}. In some sense, our study completes some previous works on the GALI method \cite{SBA07,SBA08} where other aspects of the index have been investigated, for example, its behavior for periodic orbits \cite{MSA12}. 

The thesis is organized as follows:

\begin{enumerate}
	\item Chapter~\ref{Chapter-One} provides a general overview of the general theory of Hamiltonian mechanics and chaotic dynamics. The Hamiltonian formulation of finite-dimensional systems is also presented.
	
	\item In Chapter~\ref{Chapter-Two}, we present the symplectic integration methods used for our computations along with the introduction of several chaos detection techniques, namely the Poincar{\'e} Surface of Section (PSS), the LEs, as well as the SALI and the GALI methods. Particular emphasis is given to the properties and the numerical computation of the GALI method. In order to illustrate the behavior of the LEs, as well as the SALI and the GALI methods for regular and chaotic motion, simple Hamiltonian models and symplectic mappings are used: two-dimensional (2D) and four-dimensional (4D) mappings, as well as the 2D H{\'e}non-Heiles system \cite{HH64}. Moreover, the behavior of these indicators for dissipative dynamical systems is briefly discussed.
	
	\item In Chapter~\ref{Chapter-Three}, aspects of the behavior of the GALI method in the case of multidimensional conservative Hamiltonian systems are investigated in detail. Most of the presented numerical simulations are performed using one of the most classical systems of statistical physics: the FPUT model \cite{FPUT55}. The equations of motion and variational equations of this model are presented. The asymptotic behavior of the GALI for regular orbits in the neighborhood of two simple periodic orbits of the $\beta-$FPUT model is analyzed. An investigation of the dynamical changes induced by the increase of the coupling between the 2D mappings and their influence on the behavior of the GALI using a coupled standard mapping \cite{BMC2009,SM16} is also undertaken.  
	
	\item In Chapter~\ref{Chapter-Four}, we finally summarize and discuss the results and findings of our study.
\end{enumerate}

\chapter{Hamiltonian mechanics} \label{Chapter-One}

\section{Overview of dynamical systems}
The phase space of a dynamical system is the collection of all possible states of the system in question. Each state represents a complete snapshot 
of the system at some moment in time. The dynamical evolution of the system is governed by rules which transform one point in the phase space, representing the state of 
the system ``now", into another point representing the state of the system one time step ``later".

There are two different classes of dynamical systems: 

\begin{enumerate}
	\item Discrete dynamical systems. They are described by recurrence relations iterated mappings or sets of difference equations
	       \begin{equation}
	          x_{n+1} = f(x_n), 
	         \label{eq:Ham-Map}
	       \end{equation} 
		where $f$ is a set of $n$ functions and $x_n$ is the state vector $x$ at the discrete time $t = n$, $n \in \mathbb{Z} $.  
		
 	 \item Continuous dynamical systems. They are described by differential equations 
 	       \begin{equation}
 	       	\dot{x} = \frac{dx}{dt} = f(x(t)),
 	       	\label{eq:dynamics_cont}
 	       \end{equation}
 where $x(t)\in$ $ \mathbb{R}^m$ is a vector of state variables, $f:\mathbb{R}^m \longrightarrow \mathbb{R}^m$, $m \in \mathbb{N}$ is a vector field, and $\dot{x}$ is the time-derivative, which we can write as $\dfrac{dx}{dt}$. We may regard \eqref{eq:dynamics_cont} as describing the evolution in continuous time $t$ of a dynamical system with finite-dimensional state $x(t)$ of dimension $m$. In component form, we write $x=(x_1,\dots, x_m)$, $f(x)=(f_1(x_1,\dots, x_m), \dots, f_m(x_1,\dots, x_m))$ \footnote{Notice that, here we do not use notations such as $\vec{x}$ to explicitly denote vectors. In addition, we also write vectors as row or column matrices.} and the system \eqref{eq:dynamics_cont} is given as
 \begin{eqnarray}
 \dot{x}_1 &=& f_1(x_1,\dots, x_m), \\ \nonumber
 \dot{x}_2 &=& f_2(x_1,\dots, x_m), \\ \nonumber
         &\vdots&                     \\ \nonumber
 \dot{x}_m &=& f_m(x_1,\dots, x_m). \nonumber
 \end{eqnarray} 
\end{enumerate}
 
\section{Chaos} 
Lets briefly discuss the definition of chaos. \textbf{Chaos theory} in mechanics and mathematics, studies the apparently random or unpredictable 
behavior in systems governed by deterministic laws. A more accurate term, \textit{deterministic chaos}, suggests a paradox because it 
connects two notions that are familiar and commonly regarded as incompatible \cite{EG_11}. The first is that of randomness or unpredictability, for example
in the trajectory of a molecule in a gas or the voting choice of a particular individual in a population. Usually, randomness is considered more apparent than real, arising from the ignorance of several causes in an experiment. In other words, it is
commonly believed that the world is unpredictable because it is too complicated. The second notion is that of deterministic behaviors/laws, like for example the ones appearing in the motion
of a pendulum or a planet, which have been used since the time of Isaac Newton, exemplifying the success of science in 
predicting the evolution of cases which are initially considered complex.

\textbf{Devaney's definition of chaos}: we present here a formal definition of chaos following the presentation of \cite{DV_12}. Let $(\chi,d)$ be a metric space, where $\chi$ is a set which could consist of vectors in $\mathbb{R^N}$ and $d: \chi \times \chi \rightarrow \mathbb{R}$ is the distance or metric function. A function $f: \chi \rightarrow \chi$ is called \textbf{chaotic} if and only if it satisfies the following three conditions: 

  \begin{enumerate}
  	\item $f$ has sensitive dependence on ICs. This means that $\exists$ $\delta > 0$ such that for any open set $U$ and for any point $x\in U$, there exists a point $y \in U$ such that $d(f^{k}(x),f^{k}(y))$ $>$ $\delta$ for some $k \in \mathbb{N}$, where $f^{k}$ denotes $k$ successive application of $f$. The positive real number $\delta$ is called a sensitivity constant and it only depends on the space $\chi$ and function $f$. 
  	\item $f$ is topologically transitive. Thus, for any two open sets $U$ and $V$ there exists $k$ such that $f^{k}(U) \cap V \ne \varnothing$.
  	\item The set of periodic points of $f$ is dense. A point $x$ is called periodic if $f^{k}(x) = x$ for some $k\ge1$.  	
  \end{enumerate}

This definition of chaos is widely used and accepted.  To classify a dynamical system as chaotic the above three properties, i.e.~the system being sensitive to ICs and topologically transitive, as well as having dense periodic orbits must be fulfilled. Sensitivity to ICs captures the idea that in chaotic systems minor errors or inaccuracies in the initial states can lead to large divergences in the system's evolution. In other words, small changes in the IC of the system can lead to very different long-term trajectories. Usually, the first condition is mainly considered as the central idea of chaos in physical systems although the last two properties are also relevant from a mathematical point of view.


\section{General theory of Hamiltonian systems} 
Newton's second law gives rise to systems of second-order differential equations in $\mathbb{R}^{n}$ and so to a system of first-order equations in $\mathbb{R}^{2n}$, i.e.~in an even-dimensional space. 
If the forces are derived from a potential function, the equations of motion of the mechanical system have many special properties, most of which follow from the fact that 
the equations of motion can be derived from a Hamiltonian formulation. The Hamiltonian formalism is the natural mathematical structure in which to develop the theory of 
conservative dynamical systems. 

In a Hamiltonian system with $N$ dof, the dynamics is derived from a Hamiltonian function 
\begin{equation}
H(p,q,t),
\label{eq:Ham-Eqn-t}
\end{equation}
where $q=(q_1,\dots,q_N)$ and $p=(p_1, \dots, p_N)$ 
are respectively the system's canonical coordinates and momenta. An orbit in the $2N$-dimensional ($2N$D) phase space $S$ of this system is defined by the state vector  
\begin{equation}
x(t)=(q_1(t),q_2(t),\dots,q_N(t),p_1(t),p_2(t),\dots,p_N(t),t), 
\label{eq:2N-Phase-Space}
\end{equation}
where $x_i=q_i, x_{i+N}=p_i$, and $i=1,2,\dots,N$. 

\subsection{Time independent Hamiltonian systems}
In many cases, the Hamiltonian does not explicitly depend on time and the energy
\begin{equation}
E = H(q,p),
\label{eq:Gen-Ham-Eqn}
\end{equation}
is conserved along trajectories, i.e.~it is a constant of motion \cite{KJ_89}. Examples of such systems are the pendulum, the harmonic oscillator, dynamical billiards or the motion of a particle with mass $m$ in a potential $V(q)$ described by the Hamiltonian
\begin{equation*}
H(p,q) = \frac{p^2}{2m} + V(q).
\end{equation*}

One might ask, when is the energy conserved? The answer to this question is given by the Noether`s theorem \cite{BR16}. This theorem applies to any action in classical mechanics weather it is described in the Lagrangian or Hamiltonian formulation. Focusing on Hamiltonian systems, let us consider a general action $I$ described by 
\begin{equation}
I[p_i,q_i] = \int dt(p_i \dot{q}_i-H(p,q)),
\label{Noether's_thm_int}
\end{equation}
and its Poisson bracket 
\begin{equation}
[F,G] = \frac{\partial F}{\partial q_i}\frac{\partial G}{\partial p_i} - \frac{\partial F}{\partial p_i}\frac{\partial G}{\partial q_i},
\end{equation}
where $F$ and $G$ are general functions of the generalized coordinates $q_i$ and $p_i$, for $i=1,2,\ldots, N$.  
The system's equations of motion are 
\begin{eqnarray*}
\dot{q}_i &=& \frac{\partial H}{\partial p_i} = [q_i,H], \\
\dot{p}_i &=& -\frac{\partial H}{\partial q_i} = [p_i,H],
\end{eqnarray*}

The time derivative of a function $G(p,q,t)$ of the canonical variables, which could explicitly depend on time $t$, is expressed in terms of Poisson brackets as 

\begin{eqnarray}
\frac{dG(p,q,t)}{dt} &=& \frac{\partial G}{\partial q_i}\dot{q}_i + \frac{\partial G}{\partial p_i}\dot{p}_i +\frac{\partial G}{\partial t}, \nonumber\\
  &=& \frac{\partial G}{\partial q_i}\frac{\partial H}{\partial p_i} - \frac{\partial G}{\partial p_i}\frac{\partial H}{\partial q_i} + \frac{\partial G}{\partial t} \nonumber, \\
  &=& [G,H] + \frac{\partial G}{\partial t}.
 \label{eq:Noethers_thm_canon}
\end{eqnarray}
Let us assume that we are interested in a quantity $Q(q,p,t)$ which is conserved under the dynamics. Then Equation \eqref{eq:Noethers_thm_canon} indicates that 
\begin{equation}
\frac{dQ(p,q,t)}{dt} = 0, \quad \Rightarrow \quad [Q,H] + \frac{\partial Q}{\partial t} = 0.
\label{eq:Noether_thrm_conser}
\end{equation}
In many cases of practical interest, the quantity $Q$ has no explicit time dependence and then Eq.~\eqref{eq:Noether_thrm_conser} reduces to having the Poisson bracket of $Q$ with the Hamiltonian $H$ equal to zero, i.e.~$\dfrac{dQ}{dt} = 0$ $\Rightarrow [Q, H] = 0$. 
For example in the case of the free motion of a unit mass particle on the plane, which is described by the Hamiltonian $H=\dfrac{1}{2}(p_x^2 + p_y^2)$, the angular momentum $Q=xp_y-yp_x$ is a constant of motion as $[H,Q] = -p_xp_y + p_yp_x = 0$.

If the Hamiltonian function \eqref{eq:Ham-Eqn-t} of a system does not explicitly depend on time then it is an integral of motion, i.e.~its value is conserved 

\begin{equation*}
H = \mbox{const},  
\end{equation*}
because from Eq.~\eqref{eq:Noether_thrm_conser} we see that $\dfrac{dH}{dt} = 0$. This behavior can be interpreted as follows: under stable conditions, performing an experiment today or tomorrow one expects to get the same results. In other words, for autonomous dynamical systems (i.e.~systems which do not explicitly depend on time $t$) the system's evolution is independent of the choice of the initial time value $t_0$.  
\subsection{Equations of motion and variational equations}
Consider a continuous autonomous Hamiltonian system \eqref{eq:Gen-Ham-Eqn} with $N$ dof described by $2N$ variables $p_i, q_i$, $i=1,\ldots, N$. The time evolution of an orbit of this system is governed by Hamilton equations of motion
\begin{equation}
\dot{x} = f(x) = \bigg[\frac{\partial H}{\partial p} - \frac{\partial H}{\partial q}\bigg]^{T} = J_{2N} \cdot D_H,  
\label{eq:Ham_Eqn_Motion}
\end{equation}
where $$J_{2N}=
 \begin{bmatrix} 
0_N & I_N\\
-I_N & 0_N 
\end{bmatrix}, $$ with $I_N$ and $0_N$ being the identity and zero $N\times N$ matrices, respectively, and $$D_H = \bigg[\dfrac{\partial H}{\partial q_1} \  \dfrac{\partial H}{\partial q_2} \ \ldots \ \dfrac{\partial H}{\partial q_N} \ \dfrac{\partial H}{\partial p_1} \ \dfrac{\partial H}{\partial p_2} \ \ldots \ \dfrac{\partial H}{\partial p_N}\bigg].$$

The time evolution of an initial deviation vector at time $t_0$,  $v(t_0)=\delta x(t_0)=(\delta q(t_0),\delta p(t_0))=(\delta q_1(t_0),
\dots,\delta q_N(t_0), \delta p_1(t_0), \dots, \delta p_N(t_0))$, from a given orbit with ICs $x(t_0)=(q(t_0),p(t_0))$ is 
defined by the so-called variational equations \cite{S10}:

\begin{equation}
\dot{v}(t) = \left[J_{2N} D_{H}^{2}(x(t))\right] \cdot { v}(t_0),
\label{eq:Ham_Eqn_variational} 
\end{equation}
where $D_{H}^{2}(z(t))$ is the $2N\times 2N$ Hessian matrix with elements 
\begin{equation}
D_{H}^{2}(x(t))_{i,j}=\frac{\partial^2 H}{\partial x_i \partial x_j}\Big|_{x(t)},
\label{eq:Ham_D2H}
\end{equation}
with $i,j=1,2,\dots, 2N$. Eq.~\eqref{eq:Ham_Eqn_variational} can be considered as the Hamilton equations of motion of the so-called tangent dynamics Hamiltonian $H_v$ of \eqref{eq:Gen-Ham-Eqn}   

\begin{equation}
H_v(\delta q, \delta p) = \frac{1}{2} \sum_{i=1}^{N} \delta p_i^2 + \frac{1}{2} \sum_{i,j=1}^{N} D^2_H(q(t))_{ij}\delta q_i\delta q_j.
\label{eq:2ND-tangen-dynamics}
\end{equation}

Let us now consider the $2N$D area-preserving \footnote{Any mapping $f \in (\mathbb{R^N}, \mathbb{R^N})$ is area-preserving if $\forall M \in \mathbb{R^N}$, $d(f(M))=d(M)$, where $d(M)$ is the $N$D measure of $M$.} symplectic mapping \eqref{eq:Ham-Map} of discrete time. The evolution of $x$ deviation vector $v_n$ at time $t=n$, $n \in \mathbb{Z}$ related to a reference orbit $x_n$, is the so-called the tangent map 
\begin{equation}
v_{n+1} = M_n \cdot w_n, 
\label{eq:2ND-tangen-map}
\end{equation}
where 
\begin{equation}
M_n=\frac{\partial f(x_{n+1})}{\partial x_{n+1}}=
\begin{pmatrix}
\frac{\partial f_1}{\partial x_1} & \frac{\partial f_1}{\partial x_2} &  \dots & \frac{\partial f_1}{\partial x_{2N}}\\[0.15cm]
\frac{\partial f_2}{\partial x_1} & \frac{\partial f_2}{\partial x_2} &  \dots & \frac{\partial f_2}{\partial x_{2N}}\\ 
\vdots & \vdots & \quad & \vdots \\ 
\frac{\partial f_{2N}}{\partial x_1} & \frac{\partial f_{2N}}{\partial x_2} &  \dots & \frac{\partial f_{2N}}{\partial x_{2N}}\\
\end{pmatrix},  
\end{equation} 
is the system's Jacobian matrix.

The evolution of the deviation vectors is needed for the computation of the chaos indicators. For a continuous Hamiltonian system, this evolution is done along the simultaneous integration of the equations of motion. In the case of symplectic mappings, the evolution of the deviation vectors is computed by simultaneously iterating the mapping \eqref{eq:Ham-Map} and the tangent map \eqref{eq:2ND-tangen-map}. 


\chapter{Numerical techniques and models} \label{Chapter-Two}

\section{Numerical integration: Symplectic Integrators} \label{sec:SIs}
Symplectic integrators (SIs) are numerical schemes aiming to determine the solution of Hamilton equations of motion, preserving at the same time the Hamiltonian system's underlying symplectic structure \cite{YD90,BCFLMM13,SG10,SS18,DMMS19}. An advantage of SIs is that their application transforms the numerical integration of the Hamilton equations of motion into the application of a symplectic mapping. The use of SIs is facilitated by the presence of a separable Hamiltonian function, i.e.~when the whole Hamiltonian system can be written as a sum of Hamiltonian terms, whose solution is known explicitly. 

Let us present a general way of constructing explicit SIs for the separable Hamiltonian 
\begin{equation}
H(q,p) = T(p) + V(q),
\label{eq:SIs_Separable}
\end{equation}
where $T(p)$ is the system's kinetic energy and $V(q)$ is the potential energy, following an approach based on Lie algebraic notion, Eq.~\eqref{eq:Ham_Eqn_Motion} can be simply expressed as (see e.g. \cite{SG10}) 

\begin{equation}
\frac{dx}{dt} = L_{H}x,
\label{eq:CH1_Ham_pois}
\end{equation}
where $L_{H}$ is a differential operator defined by the Poisson bracket $L_{H}f=[f,H]$. The solution of this set of equations, for ICs $x(0)$=$x_0$ is formally given as 
\begin{equation}
x = e^{tL_H} x_0.
\end{equation} 

In the common case of Eq.~\eqref{eq:SIs_Separable} the Hamiltonian function can be split into two integrable parts as $H(q,p)= A(p) + B(q)$, with $A(p)$ being the kinetic energy $T(p)$, which is  a function of only the momenta $p_i$, and $B(q)$ being the potential energy $V(q)$ depending only on the coordinates $q_i$. A symplectic scheme for integrating \eqref{eq:CH1_Ham_pois} 
from time $t$ to $t+\tau$, with $\tau$ being the integration step, consists of approximating the operator $e^{\tau L_H}$ by an integrator of $j$ steps involving products of operators
$e^{c_i\tau L_A }$ and $e^{d_i\tau L_B }$, $i = 1, 2, . . ., j$, which are exact integrations over times $c_i\tau$ and $d_i\tau$ of 
the integrable Hamiltonian functions $A$ and $B$. $c_i$ and $d_i$ being carefully chosen constants in order to improve the accuracy of the integration scheme. Thus, a SI approximates the action of the operator of $e^{\tau L_H}$ by a product of the form  
\begin{equation}
e^{\tau L_H} = \prod_{i=1}^{j} e^{(c_i\tau) L_A} e^{(d_i\tau) L_B} + O(\tau^{j+1}),
\label{eqn:Gen-SI-oper-prod}
\end{equation}
where $c_i$ and $d_i$ are constants such that $\sum_{i=1}^{j} c_i = \sum_{i=1}^{j} d_i = 1$ and $j \in \mathbb{N}$ is the so-called \textit{order of the integrator}. Each operator $e^{c_i\tau L_A }$ and $e^{d_i\tau L_B }$ corresponds to a symplectic mapping, and consequently the product appearing on the right-hand side of Eq.~\eqref{eqn:Gen-SI-oper-prod} is also a symplectic mapping. Various approaches have been developed over the years in order to determine the values of the coefficients $c_i$ and $d_i$ resulting to schemes of different orders (see e.g. \cite{FR89,MS90,YD90,LR20,EF06,BCFLMM13,SS18,DMMS19} and references therein).  

\subsection{Second order symplectic integrators} 
A basic second order SI can be written in the form
\begin{equation}
S_{2nd}(\tau) = e^{(c_1 \tau) L_A} e^{(d_1 \tau) L_B} e^{(c_2 \tau) L_A},
\label{SI_order2}
\end{equation}
like for example: the so-called \textit{leapfrog method} having three steps (i.e.~number of applications of the simple operator $e^{c_i \tau L_A}$ and $e^{d_i \tau L_A}$) with constants $c_1=c_2=0.5$ and $d=1$. Thus
	\begin{equation}
	e^{\tau L_H} = e^{(\tau/2) L_A} e^{\tau L_B} e^{(\tau/2) L_A}.
	\label{SI_lepfog}
	\end{equation} 
Let us now present some other 2nd order SIs like the 5 step SABA2 SI with composition constants $c_1=\frac{1}{2} - \frac{1}{2\sqrt{3}}, 
	c_2 = \frac{1}{\sqrt{3}}$ and $d_1=\frac{1}{2}$
	\begin{equation}
	e^{\tau L_H} = e^{(c_1 \tau) L_A} e^{(d_1 \tau) L_B} e^{(c_2 \tau) L_A} e^{(d_1 \tau) L_B} e^{(c_1 \tau) L_A}, 
	\label{SI_SABA2}
	\end{equation} 
and the SBAB2 scheme with constants $c_1=\frac{1}{2}, d_1=\frac{1}{6}$ and $d_2=\frac{1}{2}$ \cite{LR20}
	\begin{equation}
	e^{\tau L_H} = e^{(d_1 \tau) L_A} e^{(c_1 \tau) L_B} e^{(d_2 \tau) L_A} e^{(c_1 \tau) L_B} e^{(d_1 \tau) L_A}. 
	\label{SI_SBAB2}
	\end{equation} 
Another SI of this kind is the 9 step ABA82 scheme \cite{BCFLMM13}  
	\begin{equation}
	e^{\tau L_H} = e^{(c_1 \tau) L_A} e^{(d_1 \tau) L_B} e^{(c_2 \tau) L_A} e^{(d_2 \tau) L_B} e^{(c_3 \tau) L_A} e^{(d_2 \tau) L_B} e^{(c_2 \tau) L_A} 
	e^{(d_1 \tau) L_B} e^{(c_1 \tau) L_A},
	\label{SI_ABA82}
	\end{equation} 
where 
	\begin{equation*}
	\begin{matrix}
	c_1=\frac{1}{2} - \frac{\sqrt{525+70\sqrt{30}}}{70}, & c_2=\frac{\sqrt{525+70\sqrt{30}} - \sqrt{525-70\sqrt{30}}}{70}, & c_3 = \frac{\sqrt{525-70\sqrt{30}}}{35},\\[0.2cm]
	d_1=\frac{1}{4}-\frac{\sqrt{30}}{72}, & d_2=\frac{1}{4} + \frac{\sqrt{30}}{72}.   
	\end{matrix}
	\end{equation*}

\subsection{Fourth order symplectic integrators}
A 4th order SI can be obtained by a symmetric repetition (product) of 2nd order SIs \eqref{SI_order2} in the form 
\begin{equation}
S_{4th}(\tau) = S_{2nd}(r_1\tau)S_{2nd}(r_0\tau)S_{2nd}(r_1\tau),
\label{SI_order4}
\end{equation} 
where $r_0$ and $r_1$ are two real adequately determined constants. The construction can lead to the integrator developed by Forest and Ruth \cite{FR89} involving 7 step
	\begin{equation}
	e^{\tau L_H} = e^{(c_1\tau) L_A} e^{(d_1\tau) L_B} e^{(c_2\tau) L_A}e^{(d_2\tau) L_B} e^{(c_3\tau) L_A} e^{(d_3\tau) L_B} e^{(c_4\tau) L_A},
	\label{SI_Ruth_Forest}
	\end{equation}
with composition constants 
	\begin{equation*}
	\begin{matrix}
	c_1=c_4=\frac{1}{2(2-2^{1/3})}, & c_2=c_3=\frac{1-2^{1/3}}{2(2-2^{1/3})},\\[0.2cm]
	d_1=d_3=\frac{1}{2-2^{1/3}}, & d_4=-\frac{2^{1/3}}{2-2^{1/3}}.
	\end{matrix}
	\end{equation*}
The second order schemes \eqref{SI_SABA2} and \eqref{SI_SBAB2} can be used to derive the 9 step fourth order SI SABA4 and SBAB4 \cite{LR20} with  constants $c_i,d_i,i=1, 2, 3$. Higher order coefficients for this family of SIs is found in \cite{LR20}. 

Quite efficient fourth order schemes named ABA864 and ABAH864 having respectively 15 and 17 steps were developed in \cite{BCFLMM13}. The corresponding coefficients $c_i,d_i, i=1,2,3,4$ can be found in Table 3 and 4 of \cite{BCFLMM13}.

\subsection{Sixth order symplectic integrators} 
Once a 4th order SI is found, it is easy to obtain a 6th order scheme using the 4th order and implementing the same composition process, i.e. 
\begin{equation}
S_{6th}(\tau) = S_{4nd}(e_1 \tau)S_{4nd}(e_0 \tau)S_{4nd}(e_1 \tau).
\label{SI_order6}
\end{equation} 
An example of this construction is the SI developed in \cite{YD90} which has 19 steps. 

More generally, if a SI of order $2n$, $S_{2n}$, is already known, a SI of order $(2n+2)$ can be obtained through the composition   
\begin{equation}
S_{2n+2}(\tau) = S_{2n}(z_1\tau)S_{2n}(z_0\tau)S_{2n}(z_1\tau),
\label{SI_ordern}
\end{equation} 
with 
\begin{equation*}
z_0 = - \frac{2^{1/(2n+1)}}{2-2^{1/2(2n+1)}}, \quad \mbox{and} \quad z_1 =  \frac{1}{2-2^{1/2(2n+1)}}.  
\end{equation*}
\section{Chaos detection methods} \label{sec:chaos_indicators}

\subsection{The Poincar{\'e} Surface of Section (PSS)}
\begin{figure}[!h]
	\centering
	\includegraphics[width=0.5\textwidth,keepaspectratio]{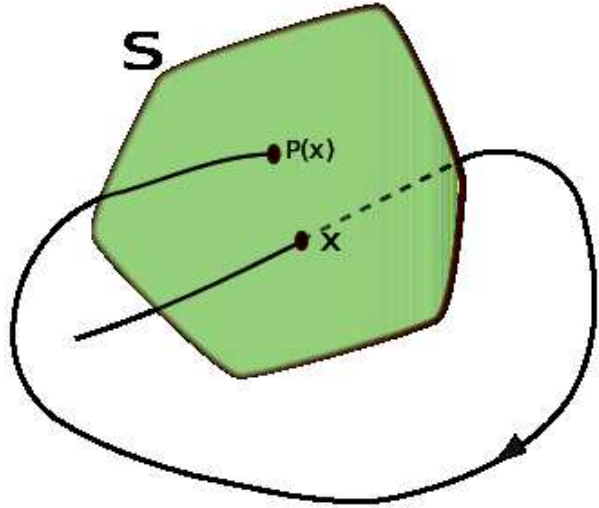}
	\caption{The Poincar{\'e} mapping $P$ evolves point $x$ onto point $P(x)$ (taken from \cite{Wikip2019}).}
	\label{fig:Poincare_map}
\end{figure}

An efficient numerical technique for visualizing the behavior of a dynamical system is the so-called PSS (see for e.g. \cite{HH64,LL_92}) which is named after Henri Poincar{\'e}. According to this method the dynamics in the phase space of a high-dimensional system is understood by observing the behavior induced by the flow on a particular section of the phase space. In particular, the dynamics is represented by the successive intersections of orbits with the PSS when this section is crossed in the same direction. In this way, a mapping is defined (see Fig.~\ref{fig:Poincare_map}). The created Poincar{\'e} mapping is a discrete dynamical system which represents the continuous flow of the original dynamical system. 
\FloatBarrier
Let us discuss this in more detail by considering an $N$D dynamical system defined by the set of ordinary differential equations  
\begin{equation}
\dot{x} = f(x).
\label{eq:ODE-1D}
\end{equation}
Let $S_1$ be an $(N-1)$D PSS. This surface must be transverse to the flow, meaning that all trajectories starting on $S_1$ should cross it and should not evolve parallel to it. The Poincar{\'e} mapping is a mapping $P: S_1 \rightarrow S_1$, which is obtained by following trajectories from one intersection after the other. Let us denote by  $x_k$ the $k$th intersection of an orbit with the PSS, $k \in \mathbb{N}$, and define the Poincar{\'e} mapping as 
\begin{equation}
x_{k+1} = P(x_k).
\label{eq:PSS-map}
\end{equation} 

If $x^{*}$ is a fixed point of the mapping, the trajectory starting at this point comes back after some iterations $m$ (equivalently after some time $T$ of the original dynamical system). Indicating that this is a $T$ periodic orbit of the original system \eqref{eq:ODE-1D}. This fixed point corresponds to an $m$-periodic orbit of the Poincar{\'e} mapping \eqref{eq:PSS-map}.    

An efficient numerical approach to determine the Poincar{\'e} mapping of a dynamical system was proposed by H{\'e}non \cite{HM82}. Let us discuss this approach in more detail. Given an $N$D autonomous dynamical system of the form of Eq.~\eqref{eq:ODE-1D},    
\begin{eqnarray} \nonumber
\dfrac{dx_1}{dt} &=& f_1(x_1,\ldots,{x_{N}}),\\ \nonumber
\dfrac{dx_2}{dt} &=& f_2(x_1,\ldots,{x_{N}}),\\ \nonumber
                 &\vdots& \\
\dfrac{dx_{N}}{dt} &=& f_{N}(x_1,\ldots,{x_{N}}),\\ \nonumber
\label{eq:PSS_matrix}
\end{eqnarray}
the PSS $S_1$ can be defined by a function of the form 
\begin{equation}
S(x_1,x_2,\ldots,x_{N}) = 0.
\label{eq:PSS_S} 
\end{equation}
Then \eqref{eq:PSS_matrix} implicitly defines the Poincar{\'e} mapping of dimension $(N-1)$. Usually we define the PSS by a function of the form $x_N - a = 0$ where $a$ is a constant number. In order to find the crossing points of a trajectory with the PSS defined by $x_N = a$, H{\'e}non introduced in \cite{HM82} a simple idea using a system of differential equation which is equivalent to the system \eqref{eq:PSS_matrix} by using $x_N$ as the independent integration variable instead of time $t$   
\begin{eqnarray}
\dfrac{dt}{dx_1} &=& \dfrac{1}{f_1}, \nonumber\\ 
\dfrac{dx_2}{dx_1} &=& \dfrac{f_2}{f_1}, \nonumber\\  
&\vdots& \nonumber \\
\dfrac{dx_{2N}}{dx_1} &=& \dfrac{f_{2N}}{f_1}. \\ \nonumber
\label{eq:PSS_matrix2}
\end{eqnarray}


\subsection{Lyapunov Exponents}
The LEs were introduced by Lyapunov \cite{Lyapunov_1892} and they measure the exponential divergence of nearby orbits in phase space. LEs are quantities which measure the system's sensitive dependence on ICs, i.e.~the rate at which the information is lost as the system evolves in time.

The maximum Lyapunov Exponent (mLE), $\sigma_1$ (see \cite{S10} and references therein), is the average rate of divergence (or convergence) of two neighboring trajectories in the phase space of a dynamical system and it is given by 
\begin{equation}
\sigma_{1} = \lim\limits_{t\to \infty} \frac{1}{t}\ln \lVert d_x\Phi^t v\Vert,
\end{equation}
where the operator $d_x\Phi^t$ maps the deviation vector $v$ from the tangent space at one point of the trajectory to the tangent space of the next point along the orbit. Practically the mLE can be computed as the limit for $t\longrightarrow \infty$ of the quantity 

\begin{equation}
\chi_{1}(t) = \frac{1}{t}\ln \frac{\lVert{v(t)}\Vert}{\lVert{v(0)}\Vert},
\label{eq:mLEs_chi}
\end{equation}
which is usually called \textit{finite-time mLE} \cite{S10}. In \eqref{eq:mLEs_chi} $v(0)$ and $v(t)$ are deviation vectors from a given orbit, at times $t=0$ and $t>0$, 
respectively, and $\lVert{.}\Vert$ denotes any norm of a vector. Taking the limit as time goes to infinity, we have 

\begin{equation}
\sigma_{1} = \lim\limits_{t\to \infty} \chi_{1}(t). 
\label{eq:mLEs_sigma}
\end{equation}

We note that the direct application of Eq.~\eqref{eq:mLEs_chi} faces some computational problems \cite{BGGS_80a} as in general, it results in an exponential increase of $\lVert v(t) \lVert$ in the case of chaotic orbit. More specifically, with increasing $t$ this norm rapidly exceeds the possibilities of ordinary numerical computations. Luckily, this can be avoided by taking into account the linearity of the evolution operator $d_x\Phi^t$ and the composition law of these mappings \cite{CF70}. So the quantity $\sigma_{1}$ can be computed as 

\begin{equation}
\sigma_{1} = \lim\limits_{t\to \infty} \frac{1}{k\tau} \sum_{i=1}^{k} \ln (\alpha_i).  
\end{equation}          
where $\alpha_i=\dfrac{\lVert v(i\tau) \Vert}{\lVert{v(0)}\Vert}$, $\tau$ is a small time interval of the integration where time $t=k\tau$, $k=1,2,\ldots$. Then after time $\tau$ the deviation vector is renormalized in order to avoid overflow problems. The mLE, in the case of autonomous Hamiltonian systems, is positive for chaotic orbits whereas for regular orbits it goes to zero by following a power law $\sigma_{1} \propto t^{-1}$ \cite{BGGS_80a}. 

\subsection{The spectrum of LEs}
The spectrum of LEs is another useful tool for estimating the stability and chaos of $2N$D dynamical systems. While the information from the mLE $\sigma_1$ can be used to determine the regular $(\sigma_1 = 0)$ or chaotic $(\sigma_1 > 0)$ nature of orbits, the knowledge of part, 
or of the whole set of LEs, $\sigma_{1}, \sigma_{2}, \dots, \sigma_{2N}$, provides additional information on the underlying dynamics and the statistical properties of the system as they can be used for example to measure the fractal dimension of strange attractors in dissipative systems \cite{S10}. Following \cite{S10} the ordering of the spectrum of LEs is given by  
\begin{equation}
\sigma_1(x) \ge \sigma_2(x) \ge \ldots\ge\sigma_N(x) \ge -\sigma_N(x)  \ge \ldots  \ge -\sigma_2(x) \ge -\sigma_1(x).
\end{equation}
The sum $s$ of all LEs (i.e.~$\sum \limits_{i=1}^{2N} \sigma_{i}$) measures the contraction rate of volumes in the phase space. In the so-called dissipative systems, $s < 0$, meaning that volumes visited by generic trajectories shrink exponentially. In autonomous Hamiltonian systems, $s=0$, i.e.~volumes are preserved. Similarly the sum of LEs for area-preserving symplectic mappings is zero   
\begin{equation}
\sum_{i=1}^{2N} \sigma_{i}(x) = 0.
\end{equation}
The LEs come in pairs of values having opposite signs  
\begin{equation}
\sigma_i(x) = -\sigma_{2N-i+1}(x), i=1,\ldots, N. 
\end{equation}
In addition, at least one pair of LEs is by default zero 
\begin{equation}
\sigma_N(x) = -\sigma_{N+1}(x) = 0.  
\label{eq:LEs-Symmetry}
\end{equation}
Other vanishing exponents may signal the existence of additional constants of motion apart from the Hamiltonian itself. 

For $2N$D Hamiltonian systems, the computation of the first $p$ LEs, with $1 < p \le 2N$ of an orbit is performed using the so-called standard method in \cite{BC79}. This method involves the time evolution of $p$ initial linearly independent and orthonormal deviation vectors, with a new set of orthonormal vectors, obtained by the Gram-Schmidt orthonormalization process, replacing the evolved deviation vectors, along with the simultaneous integration of the equations of motions.  

\newpage 
\subsection{The Smaller Alignment Index (SALI) method}

The idea behind the introduction of a simple, fast and efficient chaos indicator, such as the SALI \cite{S01,SM16} was the need to overcome the slow limit convergence of $\sigma_{1}(t)$ \eqref{eq:mLEs_sigma}. Instead of estimating the average rate of exponential growth, i.e.~$\sigma_{1}(t)$, SALI uses the possible alignment of any two normalized deviation vectors to identify the chaotic nature of orbits. In order to compute the SALI, we follow the evolution of the orbit and two deviation vectors, $v_1(t)$ and $v_2(t)$. The SALI is computed at any unit time by \cite{S01} 
\begin{equation}
\mbox{SALI}(t) = min \Big \{ \lVert \hat{v}_1(t) + \hat{v}_2(t) \Vert, \lVert \hat{v}_1(t) - \hat{v}_2(t) \Vert \Big \},
\label{eq:SALI}
\end{equation}
where $t$ is either continuous or discrete time and $ \hat{v}_1$ and $\hat{v}_2$ are unit vectors given by the relation
\begin{equation}
\hat{v}_i = \frac{v_i}{\lVert v_i \Vert}, \qquad i =1,2.
\label{eq:Unit_vectors}
\end{equation}
Note that $0 \le$ SALI$(t) \le  \sqrt{2}$. SALI $=0$ means the two deviation vectors, $v_1(t)$ and $v_2(t)$ align (i.e.~both being either parallel or antiparallel) and SALI $=\sqrt{2}$ when the two vectors are perpendicular. 
  
\subsubsection{Properties of the SALI}
The asymptotic behavior of the SALI for regular and chaotic motion is given as follows:
\begin{enumerate}
	\item In the case of regular motion SALI does not become zero \cite{SABV03}. The two deviation vectors, $v_1$ and $v_2$, tend to fall on the tangent space of the torus, following a $t^{-1}$ time evolution and having in general two different directions. Thus, in this case, SALI attains a constant positive value, i.e. 
	\begin{equation}
	\mbox{SALI} \propto \mbox{const}.
	\label{SALI_Poperty_R}
	\end{equation}
	\item On other hand, for  chaotic orbits, the deviation vectors align in the direction defined by the mLE and the SALI tends exponentially fast to zero at a rate which is related to the difference of the two largest LEs, $\sigma_1$ and $\sigma_2$ as discussed in \cite{SABV04}, i.e. 
	\begin{equation}
	\mbox{SALI}(t) \propto e^{-(\sigma_1 - \sigma_2)t}.
	\label{SALI_Poperty_C}
	\end{equation}

\end{enumerate}

Rather than evaluating SALI \eqref{eq:SALI}, in order to see if the two deviation vectors are aligned or not we can use the wedge product of these vectors \cite{SM16}
\begin{equation}
\lVert \hat{v}_1 \wedge \hat{v}_2 \Vert = \frac{\lVert \hat{v}_1 + \hat{v}_2 \Vert \cdot \lVert \hat{v}_1 - \hat{v}_2 \Vert}{2},
\end{equation}
which represents the area of the parallelogram formed by the two deviation vectors. Then $\lVert \hat{v}_1 \wedge \hat{v}_2 \Vert \rightarrow 0$ indicates chaos and $\lVert \hat{v}_1 \wedge \hat{v}_2 \Vert \neq 0$ corresponds to regular motion. If we take more than two deviation vectors, say $v_1, v_2, \dots, v_k$, $2 \le k \le 2N$, then the wedge product of the corresponding unit vectors evaluates the volume of the parallelepiped formed by these deviation vectors, which leads to the introduction of the GALI Method.

\subsection{The Generalized Alignment Index (GALI) Method}
Consider the $2N$D phase space of a conservative dynamical system represented by either a Hamiltonian flow of $N$ dof or a $2N$D symplectic mapping. In order to study whether an orbit is chaotic or not, we examine the asymptotic behavior of $k$ initially linearly independent deviations 
from this orbit, denoted by vectors $v_1, v_2, \dots, v_k$, $2 \le k \le 2N$. Thus, we follow the orbit, using the Hamilton equations of motion or the mapping equations and simultaneously we solve the corresponding variational equations or the related tangent map to study the behavior of deviation vectors from this orbit.

The Generalized Alignment Index of order $k$ (GALI$_k$) represents the volume of the generalized parallelogram defined by the evolved $k$ unit deviation vectors at any given time and it is determined as the norm of the wedge (or exterior) product of these vectors \cite{SBA07}
\begin{equation}
\mbox{GALI}_k(t) = \lVert \hat{v}_1 \wedge \hat{v}_2 \wedge \hat{v}_3 \wedge \dots \wedge \hat{v}_k \Vert.
\label{Def:GALI}
\end{equation}

In the case that the number of deviation vectors $k$ exceeds the dimension $2N$ of the system's phase space, by definition the vectors will be linearly dependent and the corresponding volume (the value of GALI$_k$, for $k>2N$) will be zero. 

\subsubsection{Properties of the GALI}
Let us consider $N$ dof Hamiltonian systems. For regular orbits, if we start with $k \le N$ linearly independent initial deviation vectors, then the deviation vectors eventually fall on the $N$D tangent space of the torus \cite{SBA07}. In this case, the asymptotic GALI value will be practically constant. Whereas, if we start with $N < k \leq 2N$ linearly independent initial deviation vectors, then the asymptotic GALI value will be zero since some deviation vectors will eventually become linearly dependent as again all of them will fall on the $N$D tangent space of the torus. The general behavior of the GALI$_k$ for regular orbits lying on an $N$D torus is given by \cite{SBA07}
\begin{equation}
\label{eq:GALI_reg}
\mbox{GALI}_k (t) \propto \left\{ \begin{array}{lll} \mbox{constant} &
\mbox{if} &  2 \leq k \leq N,  \\
& & \\
t^{-2(k-N)} & \mbox{if} & N< k \leq
2N . \\
\end{array} \right.
\end{equation}
We note that, the behavior of GALI$_k$ for regular orbits on lower-dimensional tori i.e. $k$D torus with $k<N$ is given by \cite{MSA12}. 

\begin{equation}
\label{eq:GALI_st}
\mbox{GALI}_k (t) \propto \left\{ \begin{array}{lll} t^{-(k-1)} &
\mbox{if} &  2 \leq k \leq 2N-1,  \\
& & \\
t^{-2N} & \mbox{if} &  k=2N. \\
\end{array} \right.
\end{equation}

On the other hand, for chaotic and unstable periodic orbits all deviation vectors align in the direction defined by the mLE and the  value of GALI$_k$ decays exponentially fast to zero following a rate which depends on the values of several LEs \cite{SBA07,MSA12}
\begin{equation}
\label{eq:GALI_chaos}
\mbox{GALI}_k(t) \propto e^{-[(\sigma_1-\sigma_2)+(\sigma_1-\sigma_3)+\dots+(\sigma_1-\sigma_k)]t},
\end{equation}
where $\sigma_1 \ge \sigma_2 \ge \dots \ge \sigma_k$ are approximations of the $k$ largest LEs of the orbit. In particular when $k=2$, GALI$_2$ tends to zero exponentially fast as follows: 
\begin{equation}
\mbox{GALI}_2(t) \propto e^{-(\sigma_1 - \sigma_2)t},
\label{Prop:GALI_C_N2}
\end{equation}
which is similar to the behavior of the SALI shown in \eqref{SALI_Poperty_C}. This is due to the fact that the SALI is equivalent to the GALI$_2$ \cite{SBA07}. The behavior of the GALI$_2$ for regular orbits in 2D mappings requires special attention. Since in this case the motion exists on a 1D torus, the two deviation vectors tend to fall on the tangent space of this torus, which is again 1D. Thus the two vectors will eventually become linearly dependent as in the case of chaotic orbits, but this happens with a different time rate given by \cite{SM16}
\begin{equation}
\mbox{GALI}_2(n) \propto \frac{1}{n^2}.
\label{Prop:GALI_2-STM}
\end{equation}
where $n$ is the number of iterations.

An efficient way of computing the value of GALI$_k$ is through the Singular Value Decomposition (SVD) of the matrix
\begin{equation}
\label{eq:mat_dev}
\mathbf{A}=
\begin{pmatrix}
\hat{v}_{1} & \hat{v}_{2} & \ldots & \hat{v}_{k}\\
\end{pmatrix}
=
\begin{pmatrix}
\hat{v}_{1,1} & \hat{v}_{2,1} &  \ldots & \hat{v}_{k,1}\\
\hat{v}_{1,2} & \hat{v}_{2,2} &  \ldots & \hat{v}_{k,2}\\
\vdots & \vdots & \quad & \vdots \\
\hat{v}_{1,2N} & \hat{v}_{2,2N} &  \ldots & \hat{v}_{k,2N}\\
\end{pmatrix},
\end{equation}
having as columns the coordinates of the $k$ unitary vectors $\hat{v}_i(t) = \frac{v_i(t)}{\Vert v_i(t) \Vert}=\left( \hat{v}_{i,1}, \hat{v}_{i,2},\dots,\hat{v}_{i,2N} \right)$ \cite{SBA08}. Following \cite{SBA08} we see that the value of GALI$_k$ can be given by  
\begin{equation}
\mbox{GALI}_k (t) = \sqrt{det[A(t) \cdot A^T(t)]}.
\label{Def:GALI_2}
\end{equation}
where `$det(A)$' denotes the determinant of a matrix $A$. The matrix product in \eqref{Def:GALI_2} is a $k \times k$ symmetric positive definite matrix
\begin{equation}
AA^{T} =
\begin{bmatrix}
<\hat{v}_1,\hat{v}_1> & <\hat{v}_1,\hat{v}_2> & <\hat{v}_1,\hat{v}_3> &\dots & <\hat{v}_1,\hat{v}_k> \\
<\hat{v}_1,\hat{v}_2> & <\hat{v}_2,\hat{v}_2> & <\hat{v}_2,\hat{v}_3> &\dots & <\hat{v}_2,\hat{v}_k> \\
\vdots & \vdots & \vdots  & \quad & \vdots \\
<\hat{v}_1,\hat{v}_k> & <\hat{v}_2,\hat{v}_k> & <\hat{v}_3,\hat{v}_k> &\dots & <\hat{v}_k,\hat{v}_k> \\
\end{bmatrix},
\label{eqn:matrix_S}
\end{equation}
where each element of $i$th row and $j$th column is the inner product of the unit deviation vectors $\hat{v}_i$ and $\hat{v_j}$, so that 
\begin{equation}
<\hat{v}_i,\hat{v}_j> = \cos \theta_{ij}, \qquad i,j=1,2,\dots,k,
\end{equation}
where $\theta_{ij}$ is the angle between vectors $\hat{v}_i$ and $\hat{v_j}$. Thus, the matrix of Eq.~\eqref{eqn:matrix_S} can be written as 
\begin{equation}
AA^T=
\begin{bmatrix}
1 & \cos \theta_{12} & \cos \theta_{13} &\dots & \cos \theta_{1k} \\
\cos \theta_{12} & 1 & \cos \theta_{23} &\dots & \cos \theta_{2k} \\
\vdots & \vdots & \vdots  & \quad & \vdots \\
\cos \theta_{1k} & \cos \theta_{2k} & \cos \theta_{3k} &\dots & 1 \\
\end{bmatrix}.
\label{eqn:matrix_cos}
\end{equation}

The value of the GALI$_k$ can be computed through the norm of the wedge product of the $k$ deviation vectors defined in \eqref{Def:GALI} as 
\begin{equation}
\mbox{GALI}_k = \Vast\{ \sum_{1 \le i_1<i_2<\dots<i_k \le 2N}\Vast(det\Vast[
\begin{pmatrix}
v_{1i_1} & v_{1i_2} &  \dots & v_{1i_k}\\
v_{2i_1} & v_{2i_2} &  \dots & v_{2i_k}\\
\vdots & \vdots & \quad & \dots \\
v_{ki_1} & v_{ki_2} &  \dots & v_{ki_k}\\
\end{pmatrix}  
\Vast]\Vast)^2\Vast\}^{1/2},
\label{eqn:GALI-1}
\end{equation}
where the sum is performed over all possible combinations of $k$ indices out of $2N$ (more details can be found in \cite{SBA07}). This means that in our computation we have to consider all the possible $k\times k$ determinants of $A$. From a practical point of view this technique is not numerically efficient for systems with many dof because of the large number of determinants in Eq.~\eqref{eqn:GALI-1}. Nevertheless, Eq.~\eqref{eqn:GALI-1} is ideal for the theoretical treatment of the GALI's asymptotic behavior for chaotic and regular orbits as has shown in \cite{SBA07}.

According to the SVD method \cite{WikiSVD} the $2N\times k$ matrix $A^T$ can be written as the product of a $2N\times k$ column-orthogonal matrix $U$, a $k\times k$ diagonal matrix $Z$ with non-negative real numbers $z_i$, $i=1,\dots,k$ on its diagonal and the transpose of a $k\times k$ orthogonal matrix $V$:
\begin{equation}
A^T=U Z V^T.
\label{Def:SVD}
\end{equation}  
Then the GALI$_k$ is computed using Eqs.~\eqref{Def:GALI_2} and \eqref{Def:SVD} and by keeping in mind that $U^T\cdot U = V^T\cdot V = I_k$, where $I_k$ is the $k\times k$ unit matrix, since $U$ and $V$ are orthogonal. Thus, 
\begin{eqnarray}
\mbox{GALI}_k &=& \sqrt{det(AA^T)}, \nonumber\\  
&=& \sqrt{det\big[(VZ^TU^T)\cdot(UZV^T)\big]},\nonumber\\ 
&=& \sqrt{det\big[Vdiag(z_i^2)V^T\big]}, \nonumber\\
&=&  \sqrt{det\big[diag(z_i^2)\big]}, \nonumber\\ 
&=&\prod_{i=1}^{k} z_i \label{eq:SVD}, 
\end{eqnarray}
where $z_i$, $i=1,\dots,k$, are the so-called singular values of $A$, obtained through
the SVD procedure.

In our study, in order to compute the value of the GALI$_k$ we follow the time evolution of $k$ initially linearly independent random unit deviation vectors $\hat{v}_{1}(0), \hat{v}_{2}(0),\dots,\hat{v}_{k}(0)$ and implement the approach of Eq.~\eqref{eq:SVD}. Furthermore, in order to statistically analyze the behavior of the GALIs we average the values of the indices over several different choices of the set of initial deviation vectors. The random choice of the initial vectors leads to different GALI$_k(0)$ values. Thus, in order to fairly and adequately compare the behavior of the indices for different initial sets of vectors we normalize the GALI's evolution by keeping the ratio $\mbox{GALI}_k(t)/\mbox{GALI}_k(0)$, i.e.~we measure the change of the volume defined by the $k$ deviation vectors with respect to the initially defined volume. Another option is to start the evolution of the dynamics by considering a set of $k$ \emph{orthonormal} vectors so that GALI$_k(0)=1$. As we will see later on, both approaches lead to similar results, so in our study, we will follow the former procedure unless otherwise stated.

An algorithm to compute the SALI according to Eq.~\eqref{eq:SALI} and the GALI using the SVD procedure \eqref{eq:SVD} can be found in \cite{SM16}.

\section{Illustrative applications to simple dynamical systems} \label{sec:Simple_DS}
Here, we consider some simple dynamical systems and use them to illustrate the behavior of the mLE, the SALI and the GALI for regular and chaotic orbits. 
\subsection{Conservative systems}
\subsubsection{Some Symplectic Mapping cases} \label{sec:some SM cases}
Let us start our presentation by considering a simple area-preserving 2D symplectic mapping the so-called standard mapping \cite{BC79}: 
 
\begin{equation} 
\begin{matrix} 
x'_1 & = & x_1 + x_2,           \\ 
x'_2 & = & x_2 - K\sin (x_1 + x_2),
\label{eq:2DSM} 
\end{matrix}
\end{equation}
where $K$ is a real parameter and $x'_1$ and $x'_2$ denote the values of coordinates $x_1x_2$ after one iteration of the motion. We note that all coordinates have (mod $1$), i.e.~$0 \le x_i < 1 $ for $i = 1, 2$.   

\begin{figure}[!h]
	\centering
	\includegraphics[width=0.45\textwidth,keepaspectratio]{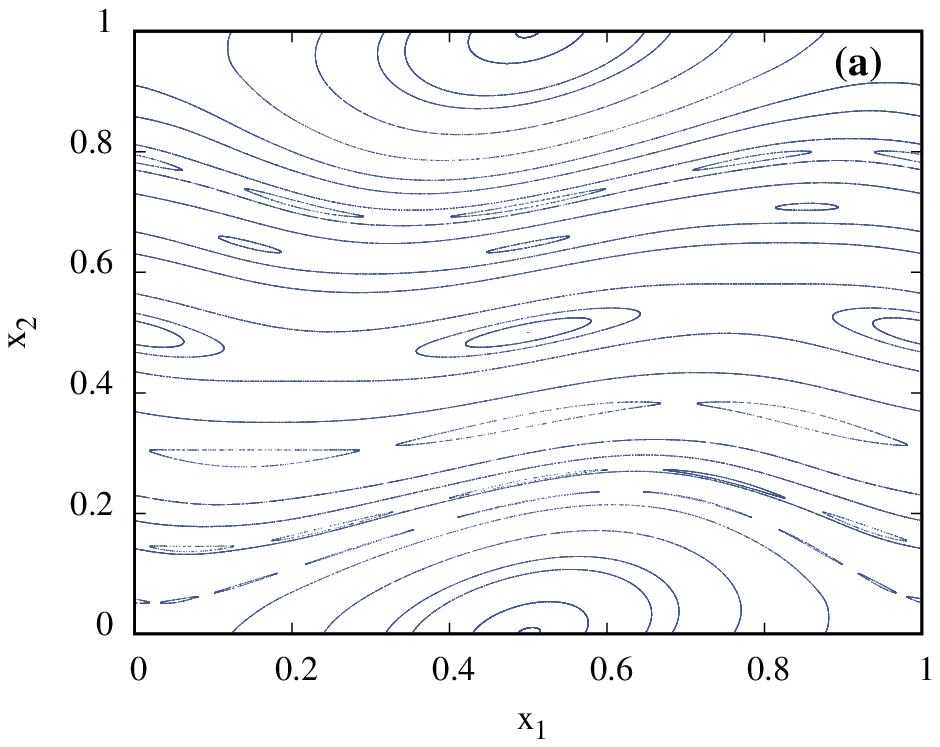}
	\includegraphics[width=0.45\textwidth,keepaspectratio]{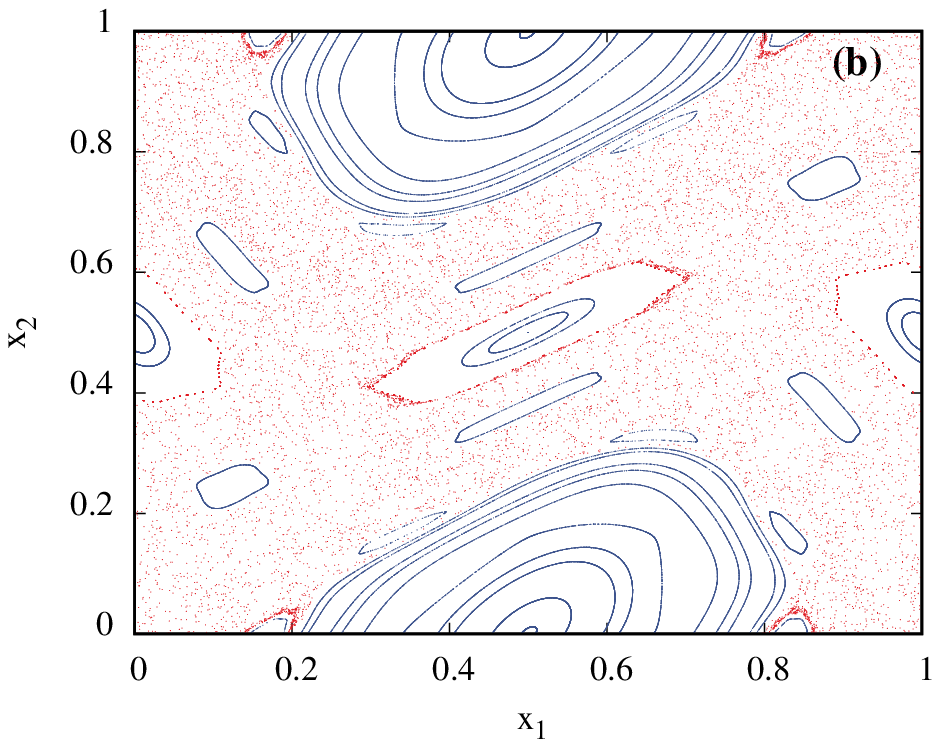}
	\includegraphics[width=0.45\textwidth,keepaspectratio]{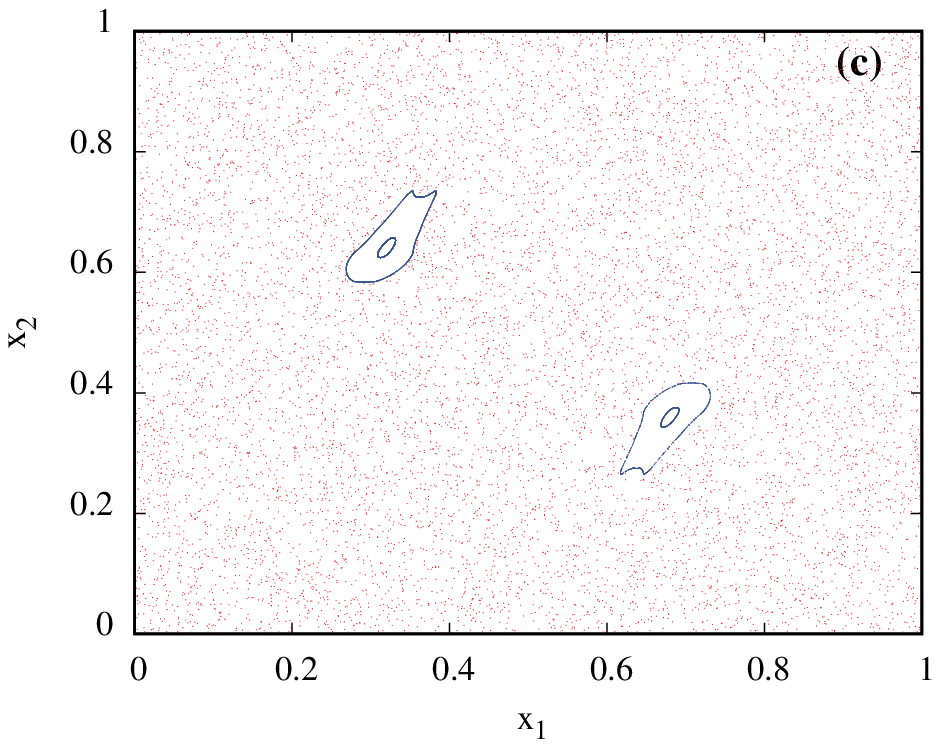}
	\caption{Phase space portraits of the standard mapping \eqref{eq:2DSM} for (a) $K=0.5$,  (b) $K=1.2$, and (c) $K=5$. The portraits are produced by $10^6$ iterations of several ICs. Different colors are used to help distinguishing between regular (blue points) and chaotic (red points) orbits.}
	\label{fig:2DSTM_Pss}
\end{figure}

Phase space portraits of the standard mapping for various values of the parameter $K$ are shown in Figure \ref{fig:2DSTM_Pss}. More specifically, Fig.~\ref{fig:2DSTM_Pss}a shows that the phase space is practically filled with stability islands for a small value of the parameter $K=0.5$. When we increase the value $K$ to $K=1.2$ (Fig.~\ref{fig:2DSTM_Pss}b), the stability islands become embedded within a chaotic sea represented by the scattered point zone. For $K=5$, the chaotic sea dominates as we see in Fig.~\ref{fig:2DSTM_Pss}c.     

\FloatBarrier
The tangent map \eqref{eq:2ND-tangen-map} of the 2D mapping \eqref{eq:2DSM} which is needed in order to compute chaos indicators is given by
\begin{eqnarray}
\delta x'_1 &=& \delta x_1 + \delta x_2, \nonumber   \\
\delta x'_2 &=& \delta x_2 - K(\delta x_1 + \delta x_2 )\cos (x_1 + x_2).
\label{eq:2DTM}         
\end{eqnarray}

\begin{figure}[!h]
	\centering
		\includegraphics[width=0.45\textwidth,keepaspectratio]{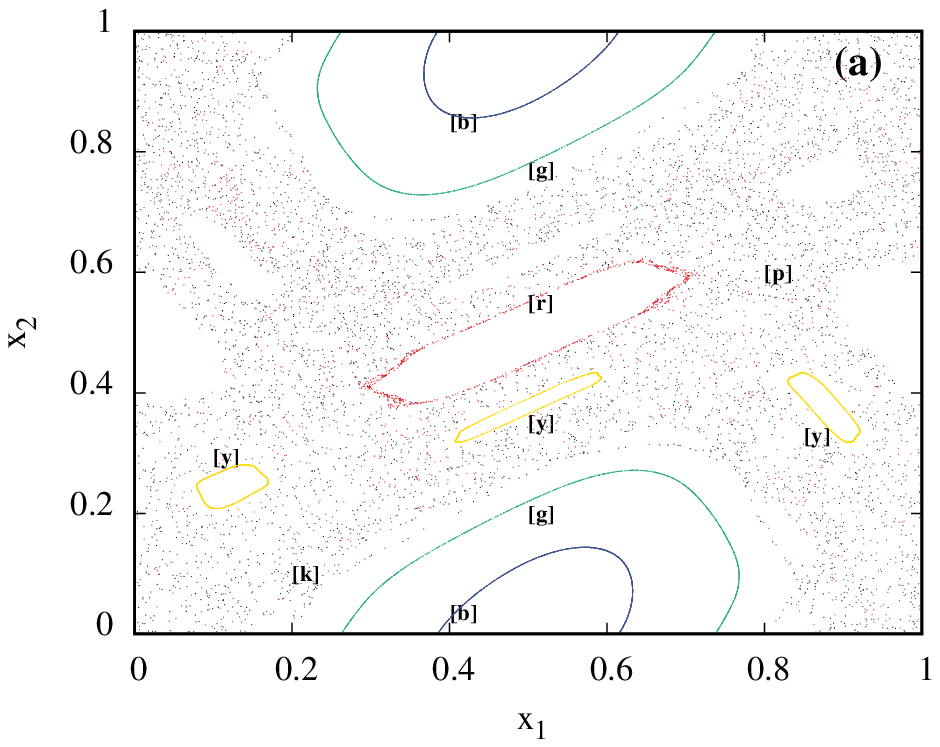}
	\includegraphics[width=0.45\textwidth,keepaspectratio]{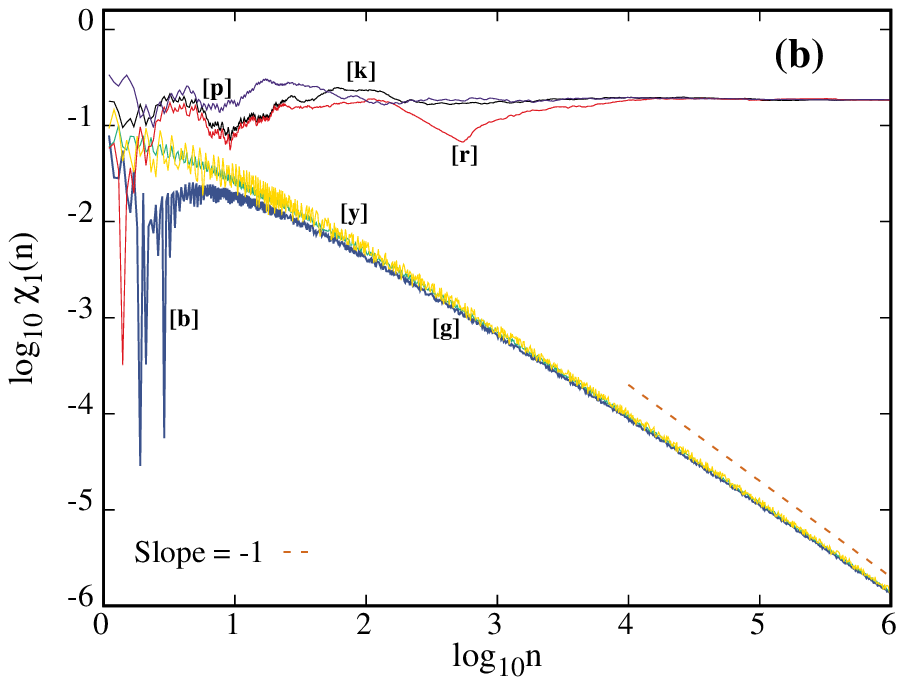}
	\includegraphics[width=0.45\textwidth,keepaspectratio]{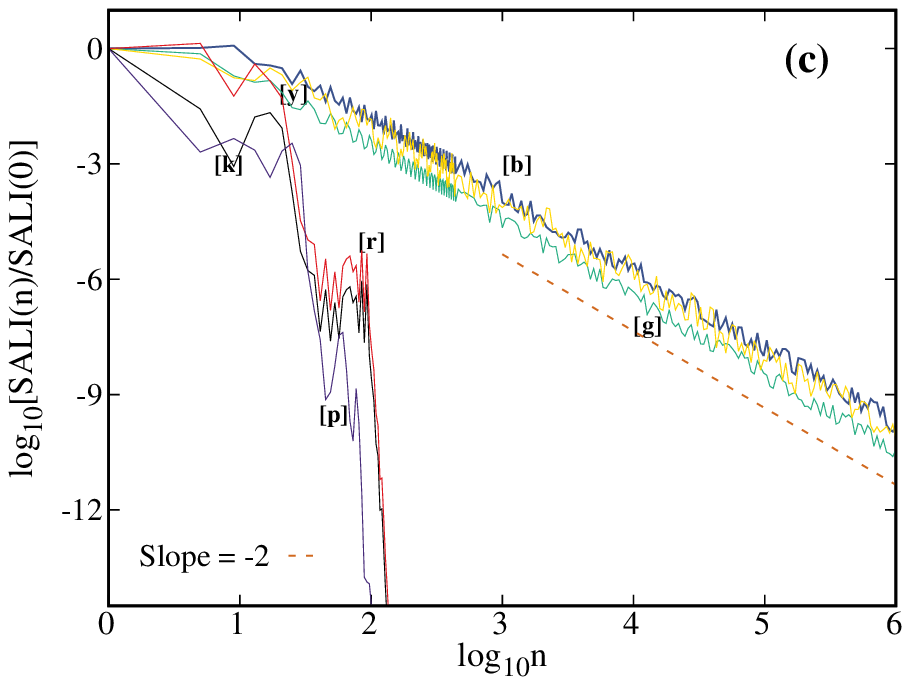}
	\caption{(a) Phase space portraits of the standard mapping \eqref{eq:2DSM} for three different regular orbits with ICs $(0.5, 0.12)$ (blue `[b]'), $(0.5, 0.22)$ (green `[g]') and $(0.5, 0.365)$ (orange `[o]') and three different chaotic orbits with ICs $(0.5, 0.4)$ (black `[k]'), $(0.5, 0.6)$ (red `[r]') and $(0.5, 0.7)$ (purple `[p]') for $K=1.2$. We see the evolution of the mLEs and the SALIs of these orbits, with respect to the number of iteration $n$, respectively in panels (b) and (c). In (b) $\sigma_1(n)$ tends to zero following a power law $n^{-1}$ for regular orbits and becomes constant for chaotic ones. In (c) SALI$(n)$ tends to zero following a power law $n^{-2}$ for regular orbits and it goes to zero exponentially fast for chaotic orbits. The dotted straight lines in (b) and (c) correspond respectively to functions proportional to $n^{-1}$ and $n^{-2}$.}
	\label{fig:2DSTM_mLCE-SALI}
\end{figure}

We consider several orbits in the 2D mapping \eqref{eq:2DSM} for $K=1.2$ (Fig.~\ref{fig:2DSTM_Pss}b) for which it has both well defined chaotic and regular regions and we determine the nature of each orbit by using the following chaos indicators: the mLE and the SALI/GALI$_2$. The obtained results are shown in Fig.~\ref{fig:2DSTM_mLCE-SALI}. In order to to illustrate the behavior of the chaos indicators, we consider three different regular and chaotic orbits from Fig.~\ref{fig:2DSTM_Pss}b. The phase space portraits of the standard mapping \eqref{eq:2DSM} for these orbits is shown in Fig.~\ref{fig:2DSTM_mLCE-SALI}a. 

Fig.~\ref{fig:2DSTM_mLCE-SALI}b shows that the mLE goes to zero for the regular orbits following an evolution which is proportional to the power law $n^{-1}$ (which is indicated by the dashed straight line) whereas mLE eventually saturates to a constant value for chaotic orbits. Fig.~\ref{fig:2DSTM_mLCE-SALI}c illustrates the time evolution of the SALI for the same regular and chaotic orbits. In the case of regular orbits, the evolution of the SALI is proportional to the theoretical prediction $n^{-2}$ \eqref{Prop:GALI_2-STM}, while for chaotic orbits the SALI goes to zero exponentially fast following the exponential decay in \eqref{SALI_Poperty_C}. If we compare the time evolution of the mLE and SALI for chaotic orbits of Figs.~\ref{fig:2DSTM_mLCE-SALI}b and \ref{fig:2DSTM_mLCE-SALI}c we can notice that SALI discriminates the chaotic orbits very fast.  

\FloatBarrier

Similar behaviors are also observed in the case of the 4D mapping \cite{CF70, CG88,SCP96}
\begin{equation} 
\begin{matrix}
x'_1 &=& x_1 + x_2,   \\

x'_2 &=& x_2 - K_1\sin(x_1 + x_2)-\mu[1-\cos(x_1+x_2+x_3+x_4)],  \\         
x'_3 &=& x_3 + x_4,   \\ 

x'_4 &=& x_4 - K_2 \sin(x_3 + x_4)-\mu[1-\cos(x_1+x_2+x_3+x_4)],\\   
\end {matrix}
\label{eq:4DSM}
\end{equation}
Consisting of two coupled 2D standard mappings with parameters $K_1$ and $K_2$, which are coupled through a term whose strength is defined by the parameter $\mu$. In this case all coordinates $x_i$, $i=1,\dots, 4$ are given (mod $2\pi$), i.e.~$-\pi \le x_i < \pi$.

\begin{figure}[!h]
	\centering
	\includegraphics[width=0.45\textwidth,keepaspectratio]{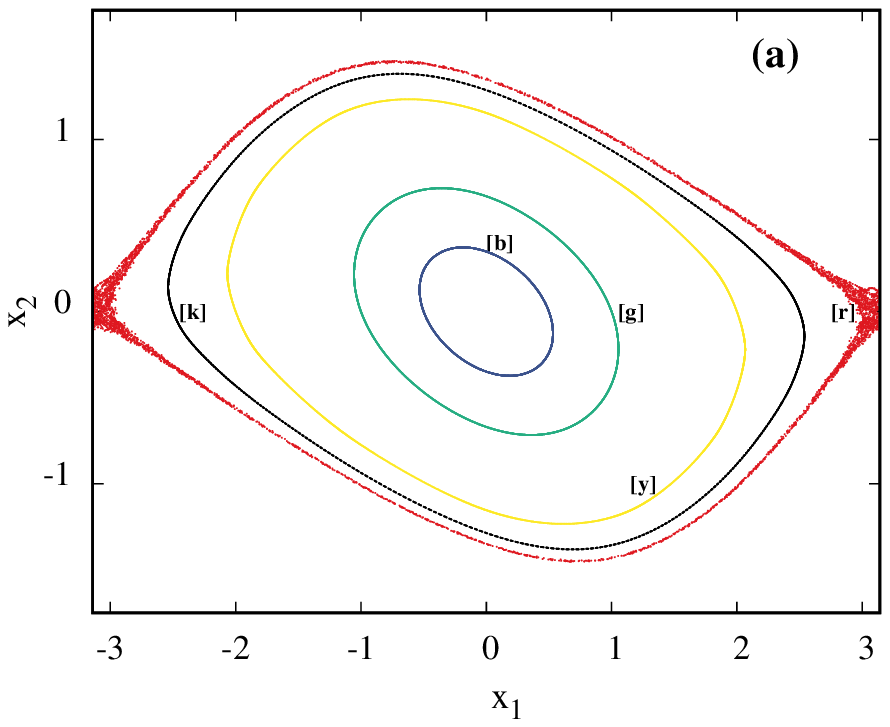}
	\includegraphics[width=0.45\textwidth,keepaspectratio]{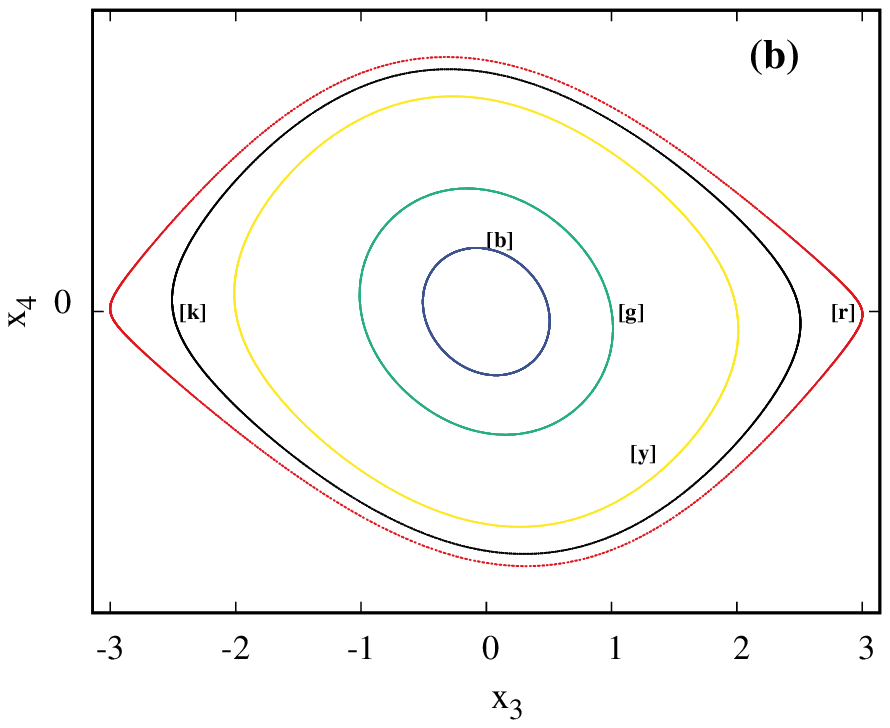}
	\caption{Two-dimensional projection of orbits of the 4D mapping \eqref{eq:4DSM} for $\mu=0$ with ICs $(0.5, 0, 0.5, 0)$ (blue `[b]'), $(1, 0, 1, 0)$ (green `[g]'), $(2, 0, 2, 0)$ (yellow `[y]'), $(2.5,0, 2.5, 0)$ (black `[k]') and $(3, 0,2, 0.)$ (red `[r]') on (a) the $x_1x_2$ plane and (b) the $x_3x_4$ plane.}
	\label{fig:4DSTM_PSS1}
\end{figure}

\begin{figure}[!h]
	\centering
	\includegraphics[width=0.45\textwidth,keepaspectratio]{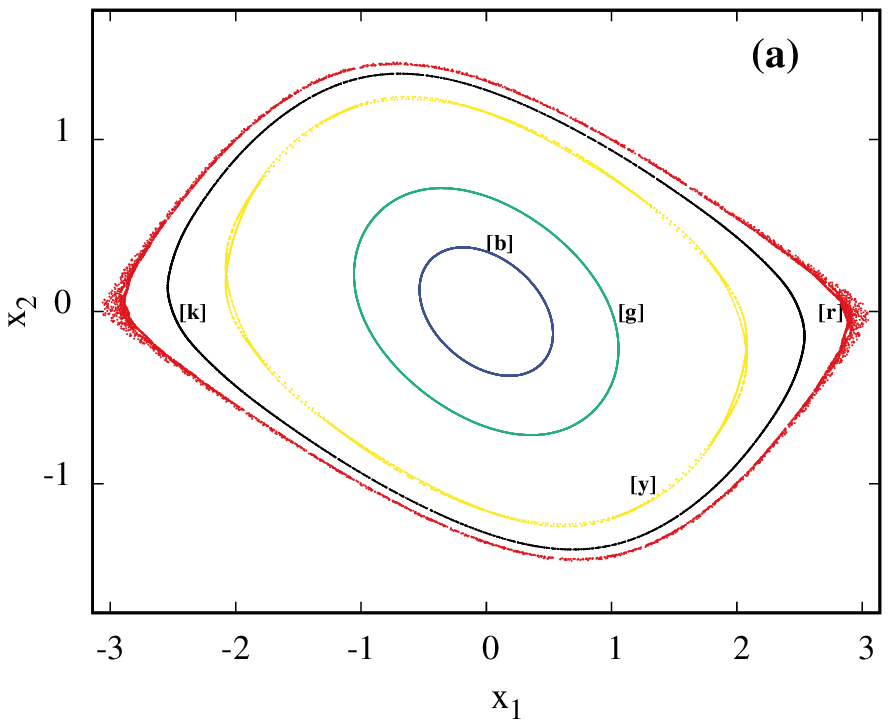}
	\includegraphics[width=0.45\textwidth,keepaspectratio]{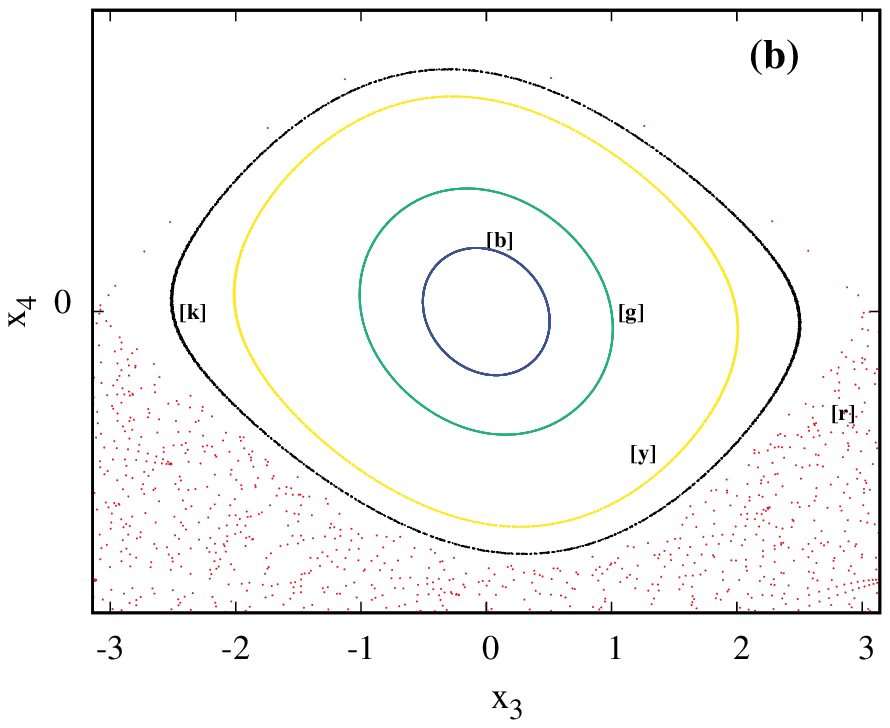}
	\caption{Similar to Fig.~\ref{fig:4DSTM_PSS1} but for $\mu=10^{-4}$.}
	\label{fig:4DSTM_PSS2}
\end{figure}

\begin{figure}[!h]
	\centering
	\includegraphics[width=0.45\textwidth,keepaspectratio]{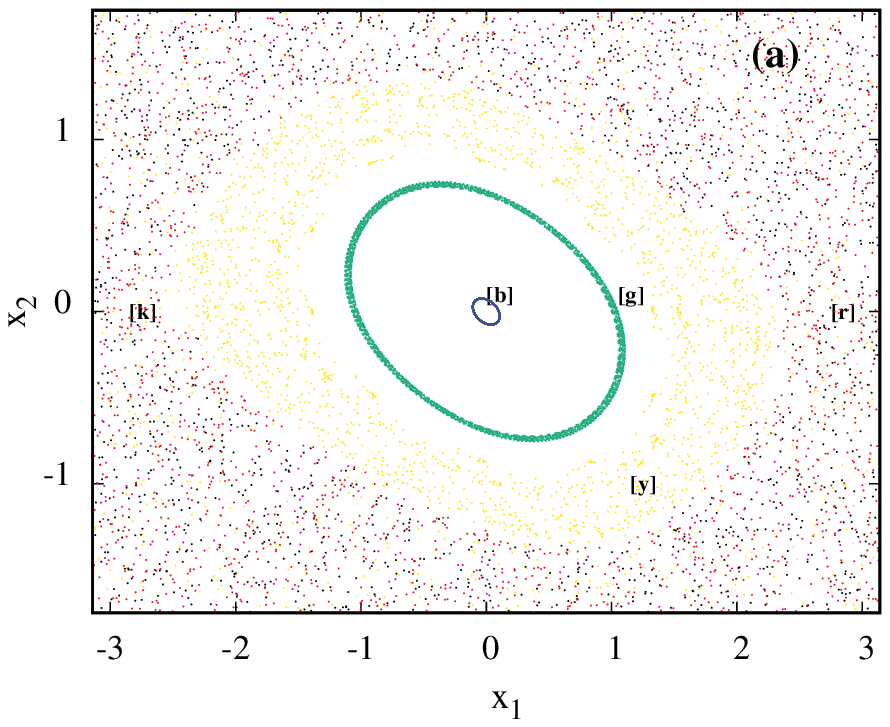}
	\includegraphics[width=0.45\textwidth,keepaspectratio]{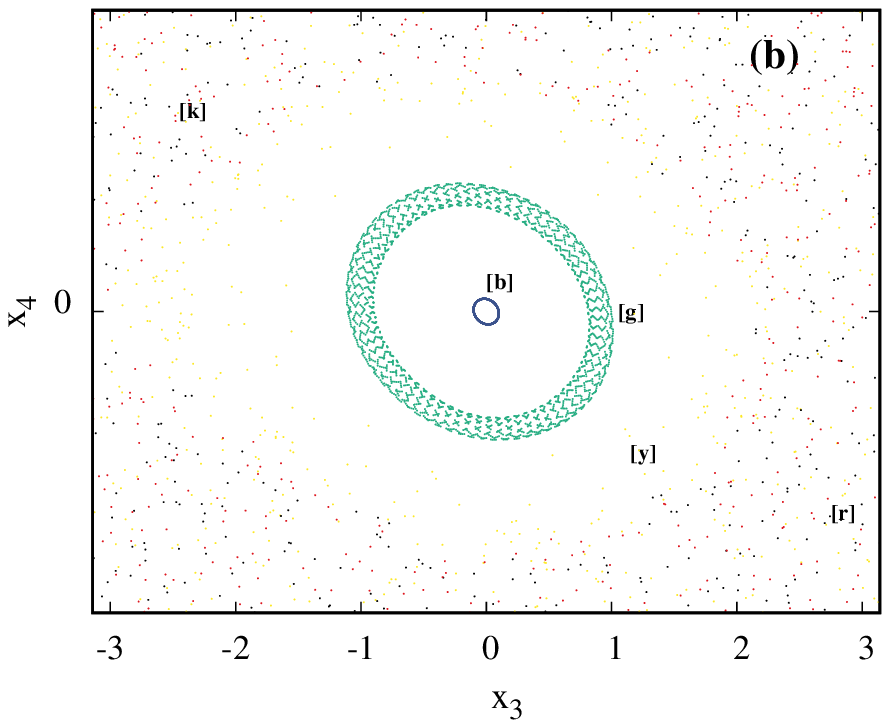}
	\caption{Similar to Fig.~\ref{fig:4DSTM_PSS1} but for $\mu=10^{-2}$.}
	\label{fig:4DSTM_PSS3}
\end{figure}

Figures \ref{fig:4DSTM_PSS1} to \ref{fig:4DSTM_PSS3} show 2D projection of the 4D mapping for different values of $\mu$ by fixing the parameters $K_1=0.5$ and $K_2=0.1$. In each case we iterated the mapping \eqref{eq:4DSM} by using five different ICs of the form $(x_1,x_2,x_3,x_4)=(c_i,0,c_i,0)$, for $i=1,\ldots,5$, where $c_1=0.5$, $c_2=2$, $c_3=2$, $c_4=2.5$, $c_5=3$. From the results of Fig.~\ref{fig:4DSTM_PSS1} we see that the considered orbits are mainly regular for $\mu = 0$ except for the orbit with $x_1=x_4=3$ which shows a weak chaotic behavior. As the value of the coupling parameter increases to $\mu=10^{-4}$ (Fig.~\ref{fig:4DSTM_PSS2}) and $\mu = 10^{-2}$ (Fig.~\ref{fig:4DSTM_PSS3}) more orbits become chaotic.    

\FloatBarrier

\FloatBarrier

\FloatBarrier
The tangent map of system \eqref{eq:4DSM} is given by  
\begin{equation} \label{eq: 4DTM}
\begin{matrix}
\delta x'_1 &=& \delta x_1 + \delta x_2,   \\
\delta x'_2 &=& \delta x_2 - K_1(\delta x_1 + \delta x_2 )\cos (x_1 + x_2)-\mu(\delta x_1 + \delta x_2 + \delta x_3 + \delta x_4) \sin(x_1+x_2+x_3+x_4),  \\         
\delta x'_3 &=& \delta x_3 + \delta x_4,   \\

\delta x'_4 &=& \delta x_4 - K_2(\delta x_3 + \delta x_4 )\cos (x_3 + x_4)-\mu(\delta x_1 + \delta x_2 + \delta x_3 + \delta x_4) \sin(x_1+x_2+x_3+x_4),       
\end {matrix}
\end{equation}

\begin{figure}[!h]
	\centering
	\includegraphics[width=0.45\textwidth,keepaspectratio]{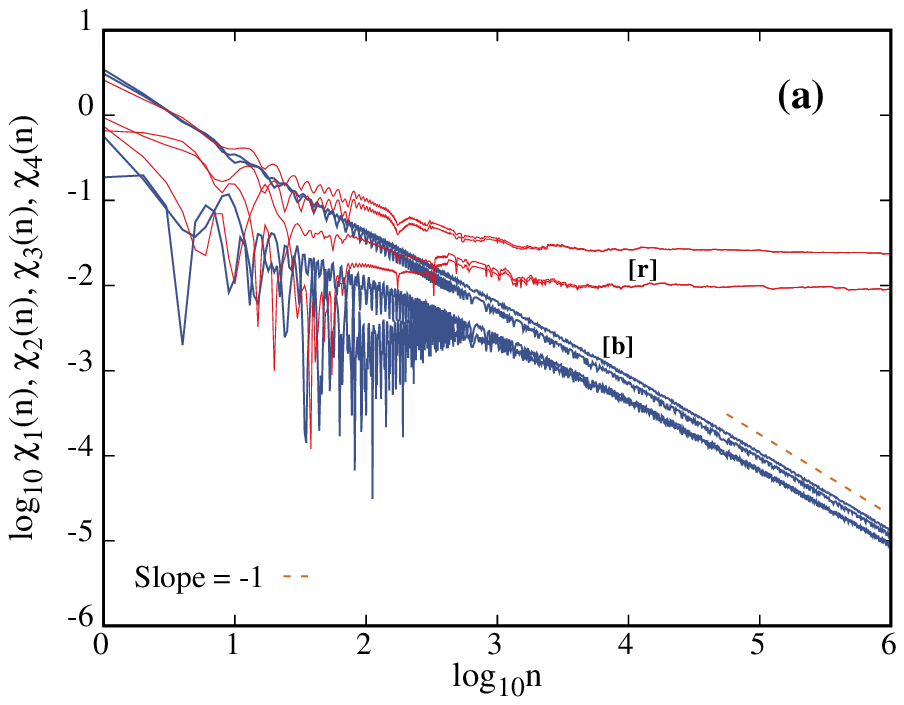}
	\includegraphics[width=0.45\textwidth,keepaspectratio]{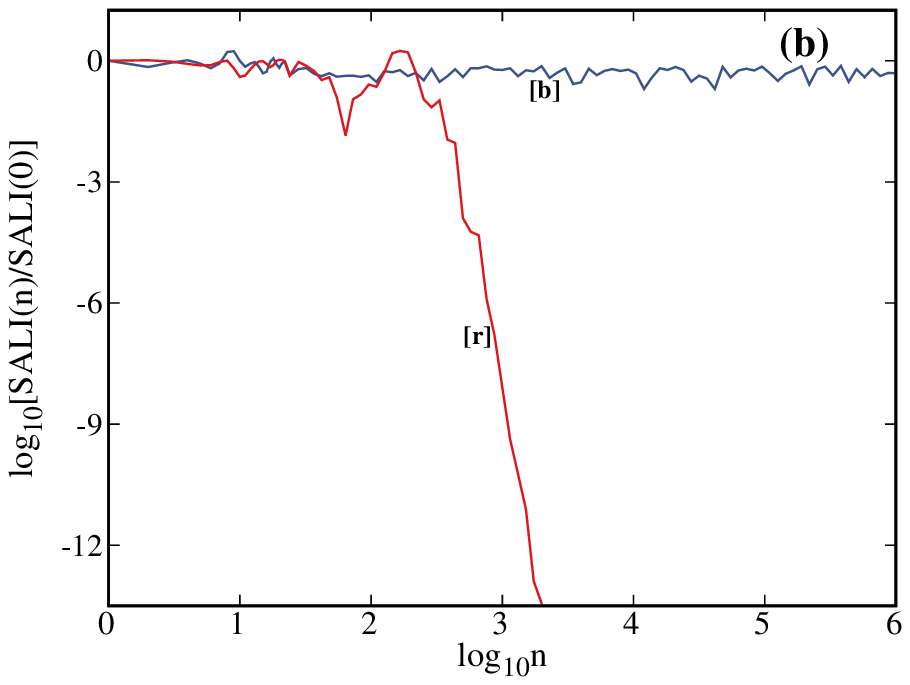}
	\caption{Evolution of (a) the four LEs and (b) the SALIs with respect to the number of iteration $n$ of the 4D mapping \eqref{eq:4DSM} with $\mu = 10^{-2}$, $K_1 = 0.5$, $K_2 = 0.1$ (Fig.~\ref{fig:4DSTM_PSS3}) for the regular orbit with IC $x_1=0.5, x_2=0, x_3=0.5$ and $x_4=0$ (blue `[b]' curves) and the chaotic orbit with IC, $x_1=3, x_2=0, x_3=3$ and $x_4=0$ (red `[r]' curves). In (a) the LEs of regular orbit tend to zero following a power law $n^{-1}$ while they attain a constant positive value for chaotic orbits. We note that for the chaotic orbit we plot the quantities $\chi_{1}$, $\chi_{2}$, $|\chi_{3}|$, $|\chi_{4}|$ in order to avoid the logarithm of negative values. In (b) SALI$(n)$ reach a non-negative constant value for the regular orbit and it goes to zero exponentially fast for the chaotic one. The dashed straight line in (a) corresponds to a function proportional to $n^{-1}$. Axes in both panels are in logarithmic scale.}
	\label{fig:4DSTM_LEs-SALI}
\end{figure}

Let us consider the orbits of the 4D mapping \eqref{eq:4DSM} with ICs $x_1=0.5, x_2=0, x_3=0.5$ and $x_4=0$ as a representative of a regular (blue points in Fig.~\ref{fig:4DSTM_PSS3}) and $x_1=3, x_2=0, x_3=3$ and $x_4=0$ for a chaotic orbit (red points in Fig.~\ref{fig:4DSTM_PSS3}). In Fig.~\ref{fig:4DSTM_LEs-SALI}a we plot the four LEs for the regular and the chaotic orbit. For regular orbit, the LEs go to zero following an evolution which is proportional to the power law $n^{-1}$. For the chaotic orbit, the LEs come in pairs of opposite values, i.e.~$\chi_{1} = -\chi_{4}$ and $\chi_{2} = -\chi_{3}$. For this reason, we plot $\chi_{1}$, $\chi_{2}$, $|\chi_{3}|$, $|\chi_{4}|$. The four finite time LEs eventually saturate to a constant value for the chaotic orbit. In Fig.~\ref{fig:4DSTM_LEs-SALI}b we see that the time evolution of the SALI for the regular orbit leads to a constant value whereas the SALI goes to zero exponentially fast for the chaotic orbit. 

\begin{figure}[h!]
	\centering
	\includegraphics[width=0.45\textwidth,keepaspectratio]{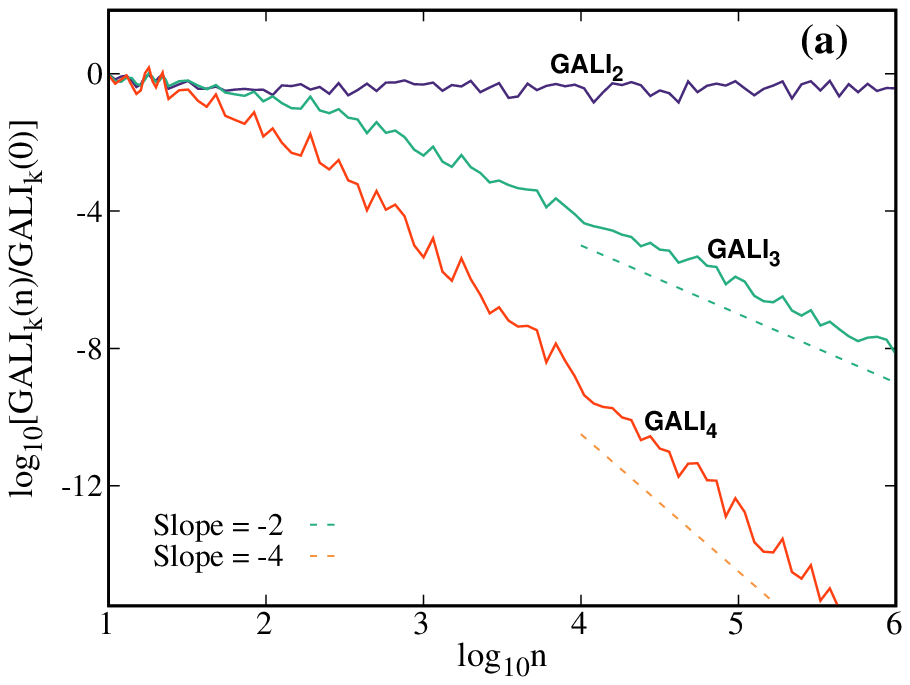}
	\includegraphics[width=0.45\textwidth,keepaspectratio]{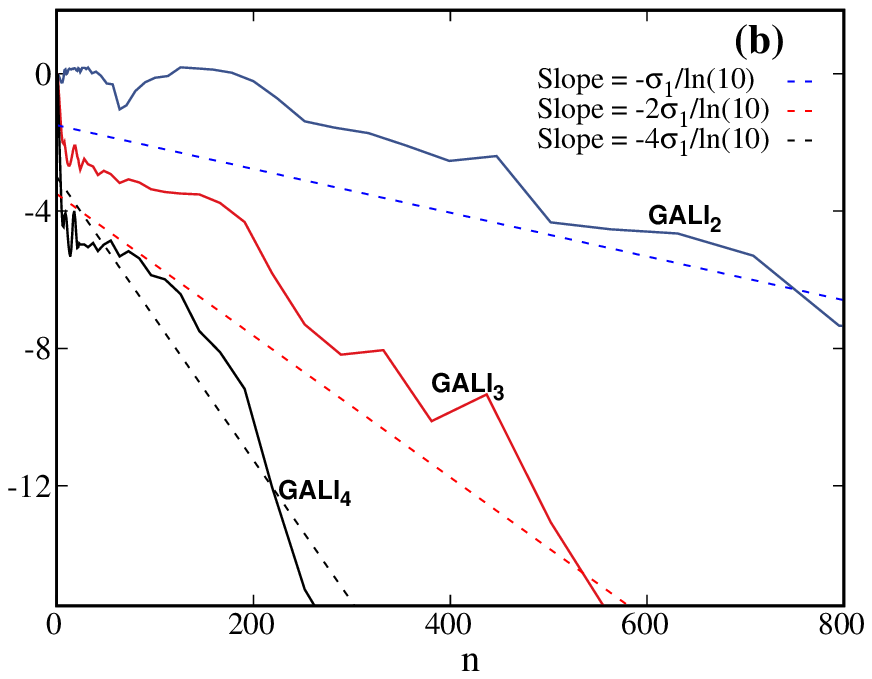}
	\caption{Evolution of the GALI$_k(n)$, $k=2,3,4$ with respect to the number of iteration $n$ of the 4D mapping \eqref{eq:4DSM} with $\mu = 10^{-2} $, $K_1 = 0.5$ and $K_2 = 0.1$ for (a) the regular orbit with IC $x_1=0.5, x_2=0, x_3=0.5$ and $x_4=0$ and (b) the chaotic orbit with IC $x_1=3, x_2=0, x_3=3$ and $x_4=0$. The dashed straight lines in (a) correspond to functions proportional to $n^{-1}$ and $n^{-2}$ while in (b) they represented functions proportional to $\exp [-\sigma_1n]$, $\exp [-2\sigma_1n]$ and $\exp [-4\sigma_1n]$ for $\sigma_1 = 0.023765$.}
	\label{fig:4DSTM_GALI}
\end{figure}

\FloatBarrier
Furthermore, in Fig.~\ref{fig:4DSTM_GALI} we computed the evolution of the GALIs for these two orbits of Fig.~\ref{fig:4DSTM_LEs-SALI}. In Fig.~\ref{fig:4DSTM_GALI}a we plot the evolution of GALI$_2$, GALI$_3$ and GALI$_4$ for the regular orbit. We see that GALI$_2$ eventually saturates to a non-zero constant value while GALI$_3$ and GALI$_4$ go to zero following respectively the asymptotic power law decays $t^{-1}$ and $t^{-2}$. In Fig.~\ref{fig:4DSTM_GALI}b we see the evolution of GALI$_2$, GALI$_3$ and GALI$_4$ for the chaotic orbit. In this case, the GALIs go to zero following an exponential decay which is proportional to $\exp[-\sigma_1n]$, $\exp[-2 \sigma_1 n]$ and $\exp[-4\sigma_1n]$ for $\sigma_1 = 0.023765$ which is the estimation of the orbit's mLE (see Fig.~\ref{fig:4DSTM_LEs-SALI}).
\FloatBarrier
\subsubsection{The H{\'e}non-Heiles system}

The H{\'e}non-Heiles system is a prototypical 2D Hamiltonian dynamical system initially used in \cite{HH64} to investigate the motion of a star in a simplified galactic potential. The star is assumed to move on the galactic plane with coordinates $x$ and $y$. The Hamiltonian function of this system is 

\begin{equation}
H_2 = \frac{1}{2}(p_x^2+p_y^2) + \frac{1}{2}(x^2+y^2) + x^2y - \frac{1}{3} y^3 ,
\label{eq:2DHH}
\end{equation}
where $p_x$ and $p_y$ are the conjugate momentum. The system's equations of motion are 
\begin{eqnarray}
\dot{x} &=& p_x, \nonumber \\ 
\dot{y} &=& p_y,  \nonumber \\ 
\dot{p}_x &=& -x(1+2y), \nonumber \\
\dot{p}_y &=& y^2 - x^2 - y. 
\label{eq:Henon-Heiles}
\end{eqnarray}		

\begin{figure}[h!]
\centering
\includegraphics[width=0.7\textwidth,keepaspectratio]{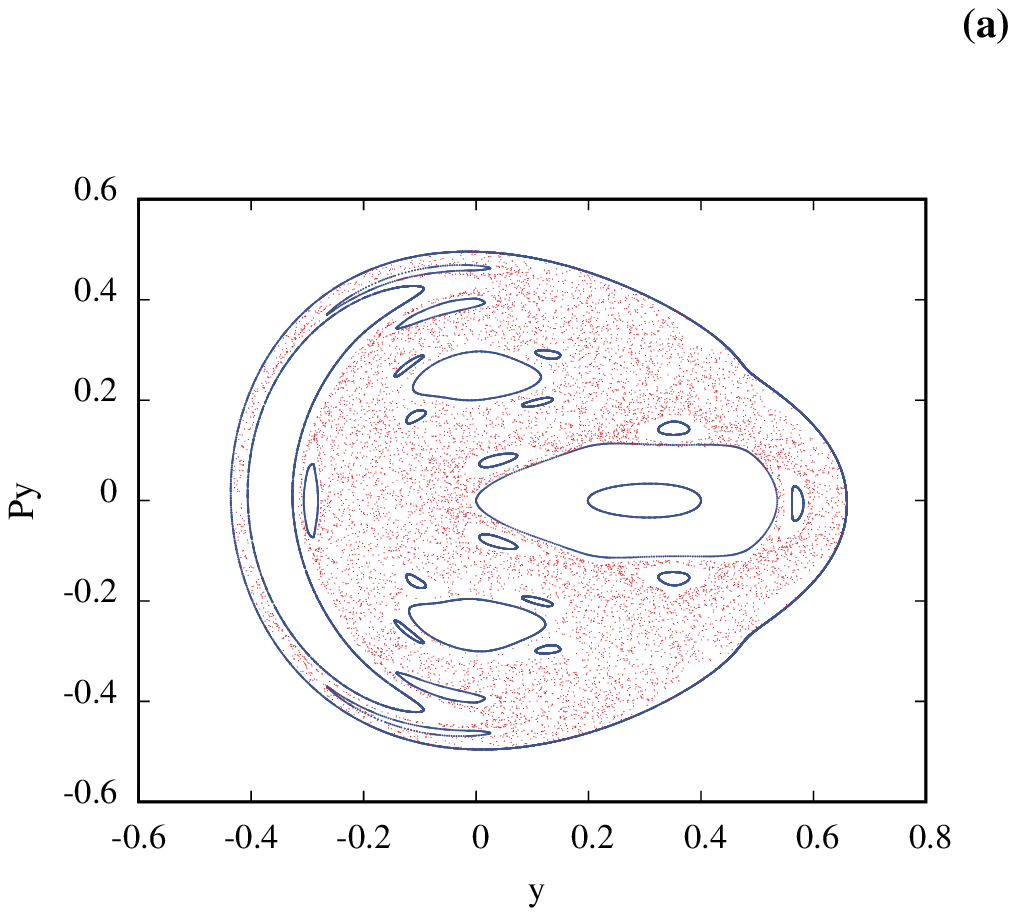}
\caption{The PSS of the H{\'e}non-Heiles system \eqref{eq:2DHH} with energy $H_2 = 0.125$ for $x=0$, $p_x\geq0$. Orbits were integrated up to a final time $t = 10,000$ time units using the SABA2 SI scheme with an integration time step of $\tau = 0.01$.}
\label{fig:H-H_PSS}
\end{figure}

In Fig.~\ref{fig:H-H_PSS} we present the PSS of the H{\'e}non-Heiles system \eqref{eq:2DHH} with  $H_2 = 0.125$, defined by the condition $x=0$ and $p_x\geq0$. There ordered orbits correspond to closed smooth curves, while the chaotic orbits are represented by scattered points.  
\FloatBarrier 

In order to compute chaos indicators like the mLE, the SALI and the GALI we also need to integrate the system's variational equations  
\begin{eqnarray}
\delta \dot{x} &=& \delta p_x, \nonumber \\ 
\delta \dot{y} &=& \delta p_y,  \nonumber \\
\delta \dot{p}_x &=& -\big[(1+2y)\delta x + 2x\delta y \big], \nonumber \\
\delta \dot{p}_y &=& -\big[2x\delta x +(1+2y)\delta y\big]. 
\label{eq:HH-Variational}
\end{eqnarray}		
In our simulations we simultaneously integrated Eqs.~\eqref{eq:Henon-Heiles} and \eqref{eq:HH-Variational} using the SABA2 SI scheme with an appropriate integration time step which keep the relative energy error below $10^{-5}$.

\begin{figure}[h!]
	\centering
	\includegraphics[width=0.45\textwidth,keepaspectratio]{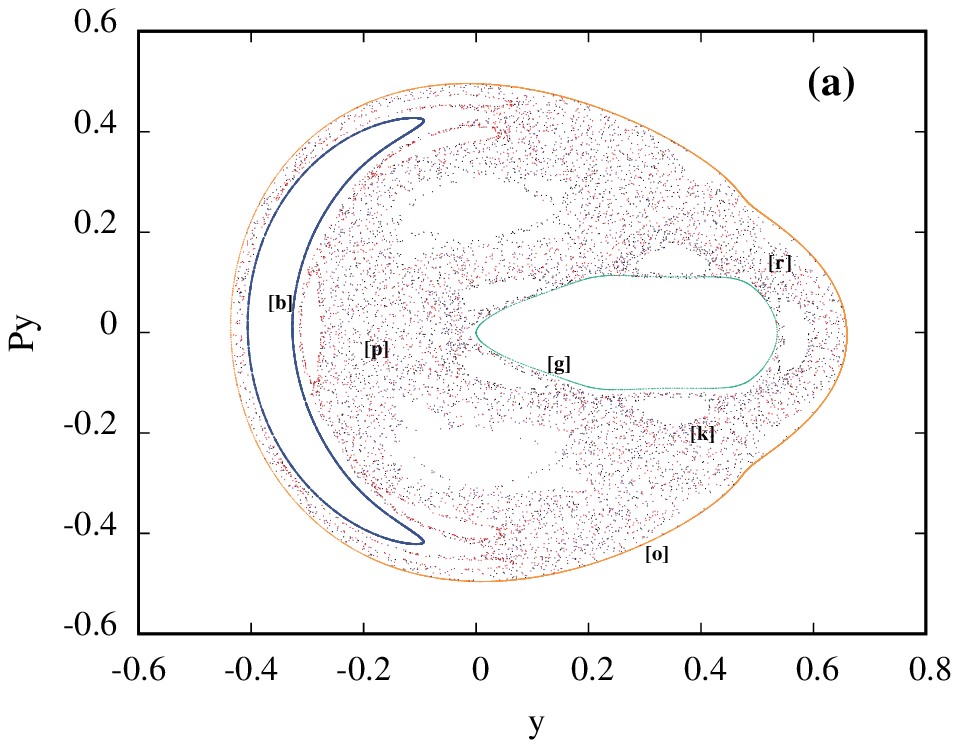}
	\includegraphics[width=0.45\textwidth,keepaspectratio]{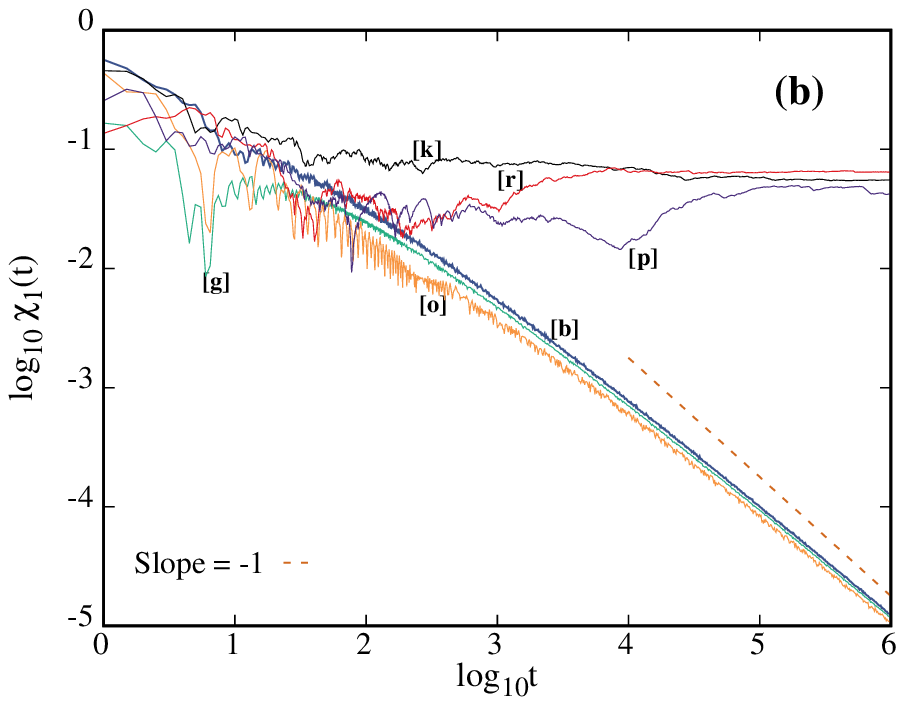}
	\includegraphics[width=0.45\textwidth,keepaspectratio]{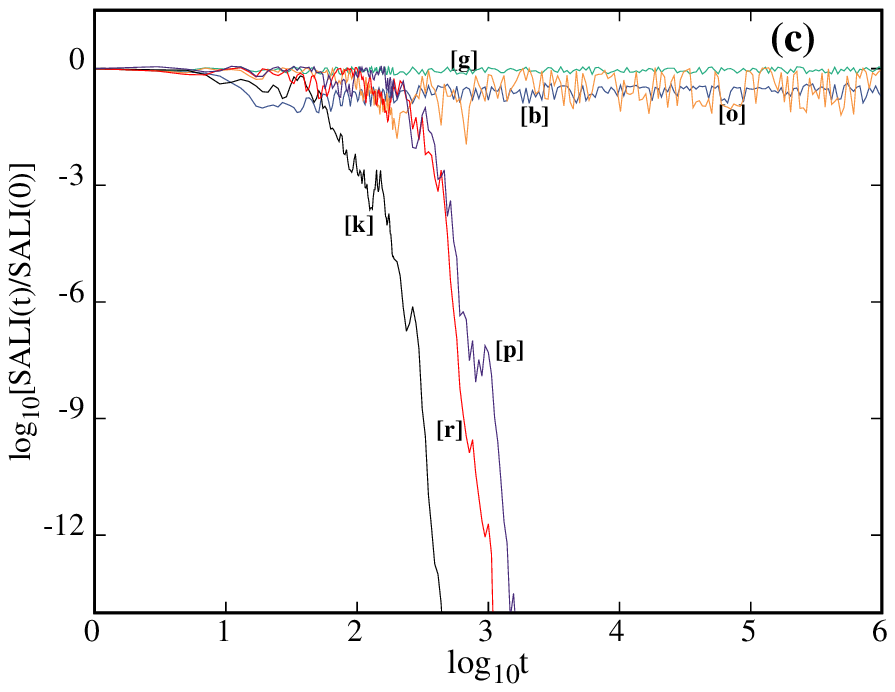}
	\caption{(a) The PSS ($x=0, p_x \ge 0$) of the 2D system \eqref{eq:2DHH} with energy $H_2 = 0.125$ for regular orbits with ICs $(y,p_x,p_y)$ $(-0.1,0.267, -0.41)$ `blue [b]', $(0.4, -0.3, 0)$ `green [g]', and $(0.5, 0.144, -0.25)$ `orange [o]' and chaotic orbits with ICs $(0.52, 0.232, 0.14)$ `black [k]', $(-0.38, 0.17, -0.2)$ `red [r]', and $(-0.1, 0.488, -0.03)$ `purple [p]'. The time evolution of the mLEs and the SALIs are respectively shown in (b) and (c). In (b) the plotted line corresponds to a function proportional to $t^{-1}$.}
\label{fig:H-H_LEs}
\end{figure}

We compute the mLEs and the SALIs for three different regular orbits which are taken from the islands of stability around periodic orbits of different periods \footnote{An orbit $x(t)$, in the system's phase space, is called periodic if $x(t+T)=x(T)$, for a non-zero constant number $T$} (Fig.~\ref{fig:H-H_LEs}a). Their ICs are $(0, -0.1,0.267, -0.41)$, $(0, 0.4, -0.3, 0)$, and $(0, 0.5, 0.144, -0.25)$ for the regular orbits and $(0, 0.52, 0.232, 0.14)$, $(0, -0.38, 0.17, -0.2)$, and $(0, -0.1, 0.488, -0.03)$ for the chaotic ones. Fig.~\ref{fig:H-H_LEs}b shows that the mLEs go to zero for regular orbits following an evolution which is proportional to the power law $t^{-1}$, while they eventually attain constant positive values for chaotic orbits. In Fig.~\ref{fig:H-H_LEs}c we present the time evolution of the SALIs for the same regular and chaotic orbits. In the case of regular orbits, the SALIs obtain non-zero constant values. On the other hand, the SALIs go to zero exponentially fast for chaotic orbits. Similar to the discrete system of Section \ref{sec:some SM cases} (Figs.~\ref{fig:2DSTM_mLCE-SALI} and \ref{fig:4DSTM_LEs-SALI}). The SALIs discriminate chaotic orbits very fast. 
 
\FloatBarrier

\begin{figure}[h!]
	\centering
	\includegraphics[width=0.45\textwidth,keepaspectratio]{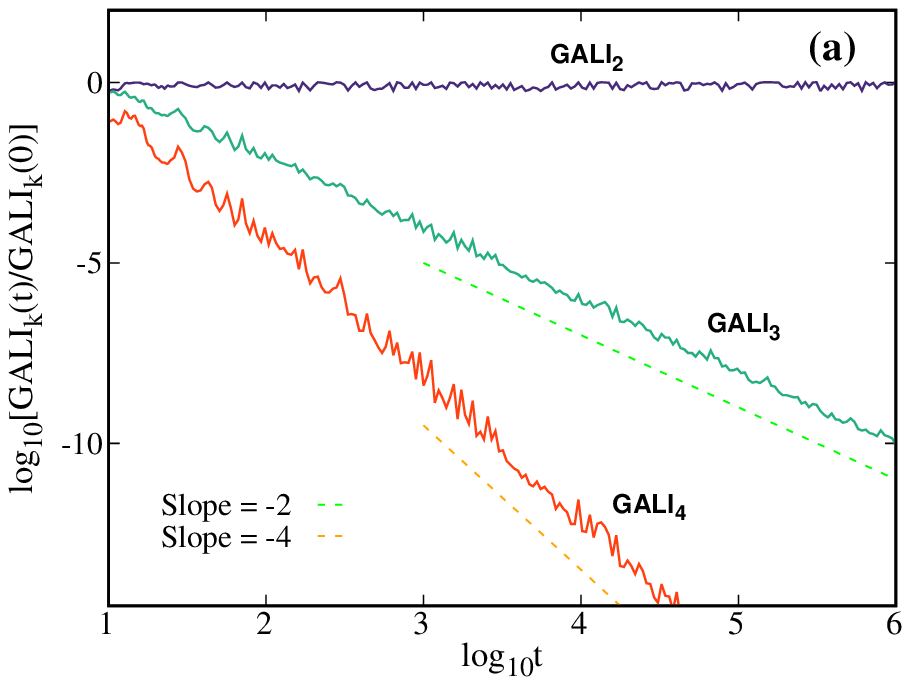}
	\includegraphics[width=0.45\textwidth,keepaspectratio]{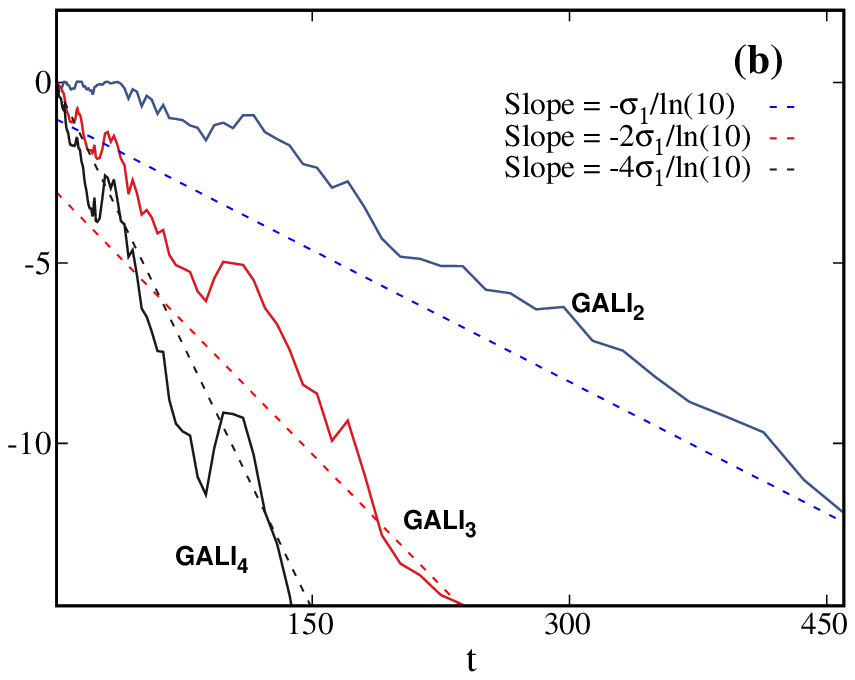}
	\caption{The time evolution of GALI$_k(t)$, $k=2,3,4$ of (a) the regular orbit with IC $x=0, y=0.1, p_x=0.495$ and $p_y=0$ and (b) the chaotic orbit with IC $x=0, p_x=-0.25, x_3=0.421$ and $x_4=0$ for the system \eqref{eq:2DHH} with energy $H_2 = 0.125$. The plotted straight lines in (a) correspond to functions proportional to $t^{-1}$ and $t^{-2}$ while in (b) they correspond to functions proportional to $\exp [(-\sigma_1)t]$, $\exp [(-2\sigma_1)t]$ and $\exp [(-4\sigma_1)t]$ for $\sigma_1=0.055984$ which is the orbits mLE. Both axes in (a) are in logarithmic scale whereas in (b) the $x$-axis is linear while the $y$-axis is in logarithmic scale.}
	\label{fig:H-H_GALIs}
\end{figure}

In Fig.~\ref{fig:H-H_GALIs} we also study the behavior of the GALIs for the regular orbit with IC $x=0., y=0.1, p_x=0.495$ and $p_y=0$, and the chaotic orbit with IC $x=0, p_x=-0.25, x_3=0.421$ and $x_4=0$. The regular orbit lies on a 2D torus and therefore its GALI$_2$ eventually saturates to a non-zero constant value, whereas the GALI$_3$ and the GALI$_4$ go to zero respectively following the asymptotic power law decays $t^{-1}$ and $t^{-2}$ (Fig.~\ref{fig:H-H_GALIs}a). The GALIs for the chaotic orbit are shown in Fig.~\ref{fig:H-H_GALIs}b. All of them go to zero exponentially fast following the exponential decays which are described in Eq.~\eqref{eq:GALI_chaos}. 

\FloatBarrier

Overall all the methods we implemented can correctly capture the dynamical nature of the studied orbits. GALI$_k$, with higher-order $k$, is the fastest in discriminating between regular and chaotic motion whereas the mLE is the least efficient technique. For example, consider the evolution of the GALI$_4$ and the mLE for the 4D mapping \eqref{eq:4DSM} and 2D Hamiltonian \eqref{eq:Henon-Heiles}. The exponential decay of the GALI$_4$ for the chaotic orbit with IC $x_1=3, x_2=0, x_3=3$ and $x_4=0$ (black curve in \ref{fig:4DSTM_GALI}b) of \eqref{eq:4DSM} reaches the value $10^{-16}$ around time $t=10^{2.4}$, while the mLE (red curve in \ref{fig:4DSTM_LEs-SALI}a) clearly indicates that the orbit is chaotic around time $t=10^{3.8}$. Also, the exponential decay of the GALI$_4$ for the chaotic orbit with IC $x=0, p_x=-0.25, x_3=0.421$ and $x_4=0$ (black curve in \ref{fig:H-H_GALIs}b) of \eqref{eq:Henon-Heiles} reaches the value $10^{-16}$ around time $t=10^{1.3}$, while the mLE (purple curve in \ref{fig:H-H_LEs}b) shows that the orbit is chaotic around time $10^{5}$. Based on our results and discussion, we can conclude that the GALI method is the most effective among the methods we have considered. Even though the convergence of the LEs is theoretically guaranteed, its rate of convergence can become too slow, while the PSS is impractical for systems with many dof.    
\subsection{Dissipative systems}

A dynamical system is called non-conservative if its phase space volume changes over time. The determinant of the system's Jacobian matrix is an important tool to explore the properties of the dynamic, in particular to describe the change in volume. If this determinant is smaller than one, then the volume of phase space decreases over time \cite{HW14}.  

For dissipative systems, in the limit of $t\rightarrow \infty$ the dynamics converges to a set of values for various ICs, the so-called \textit{attractor}. A dynamical system can have several different attractors. We should note that the concept of an attractor is only relevant when discussing the asymptotic behavior of orbits. 

In this section, we compute various chaos indicator of two dissipative systems. The aim is to briefly investigate the behavior of the GALIs for these systems. Our results suggest that further work is certainly required. However, this study provides a good starting point for further investigations.
\subsubsection{The H{\'e}non Mapping}
\begin{figure}[h!]
 	\centering
 	\includegraphics[width=0.7\textwidth,keepaspectratio]{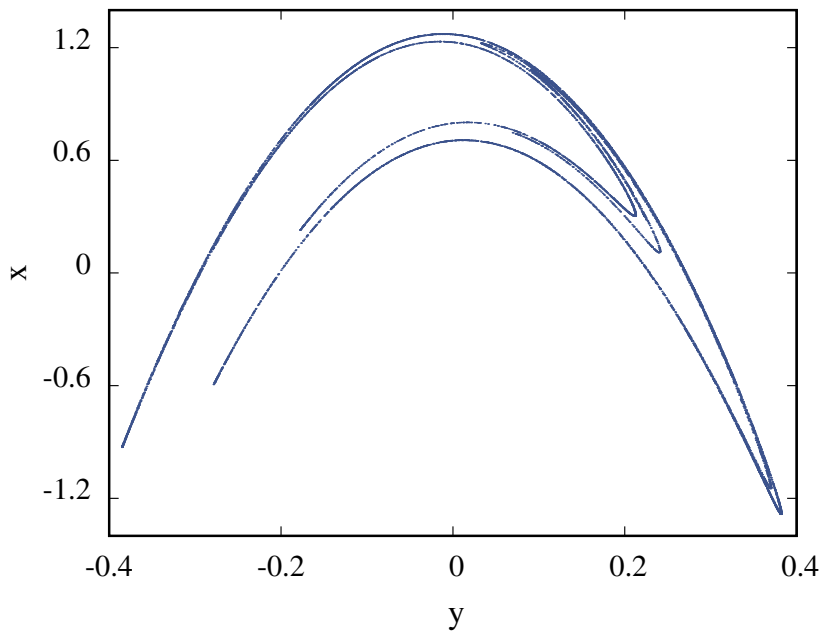}
 	\caption{Phase plot of the strange attractor of the H{\'e}non mapping \eqref{eq:Hen_Map} with $\alpha = 1.4$ and $\beta = 0.3$. An IC $(0.5, 0.5)$ is iterated $10^{5}$ times in order to produce this plot.}
 	\label{fig:Henon_Map_x=y=0}
 \end{figure}

First we consider a mapping which was introduced by H{\'e}non \cite{HM76} based on concepts of stretching and folding of areas in the phase space. The H{\'e}non mapping takes a point $(x, y)$ in the plane and maps it to a new point $(x', y')$ according to the rule
 \begin{eqnarray}
 x' &=& y + 1 - \alpha x^2, \nonumber \\
 y' &=& \beta x.
 \label{eq:Hen_Map} 
 \end{eqnarray}
According to \cite{HM76}, the dynamics converges to a strange attractor for parameter values $\alpha = 1.4$ and $\beta = 0.3$. This attractor is seen in Fig.~\ref{fig:Henon_Map_x=y=0} 
\FloatBarrier 
The rate of expansion of an area in the map's phase space is given by the determinant of the Jacobian matrix of the mapping \eqref{eq:Hen_Map}
\begin{equation}
J = \left\lvert det
\begin{bmatrix}
-2\alpha x_n & 1 \\
\beta & 0 \\
\end{bmatrix}
\right\lvert
= \left\lvert \beta \right\lvert .
\end{equation}
Thus for $\beta=1$ the mapping is area preserving but when $\beta<1$, as we have done for the case $\beta = 0.3$, the mapping is dissipative and phase space areas contract.  


There is an equally simple (as the H{\'e}non mapping \eqref{eq:Hen_Map}) 2D quadratic mapping admitting several attractors \cite{ZS09}. This mapping, has two parameters, $\alpha$ and $\beta$, and it is given by     
\begin{eqnarray}
x' &=& 1- \alpha y^{2} + \beta x, \nonumber \\
y' &=& x.
\label{eq:2NHen_Map} 
\end{eqnarray}
Note that this mapping is conservative for values of the parameter $\beta \in(-1.6,0.5)$. The associated tangent map of the 2D quadratic mapping \eqref{eq:2NHen_Map} is given by    
\begin{eqnarray}
\delta x' &=& - \beta \delta x -2 \alpha y^{2}\delta y,  \nonumber \\
\delta y' &=& \delta x.
\label{eq:NHen_Map-Var} 
\end{eqnarray}

\begin{figure}[h!]
		\centering
		\centering
		\includegraphics[width=0.45\textwidth,keepaspectratio]{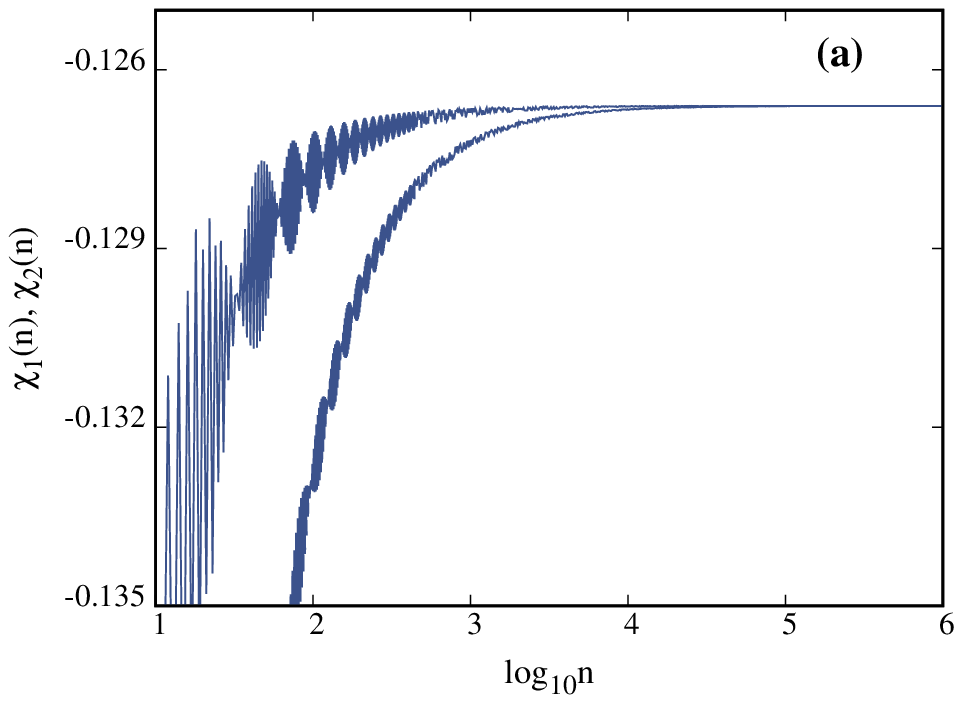}
		\includegraphics[width=0.45\textwidth,keepaspectratio]{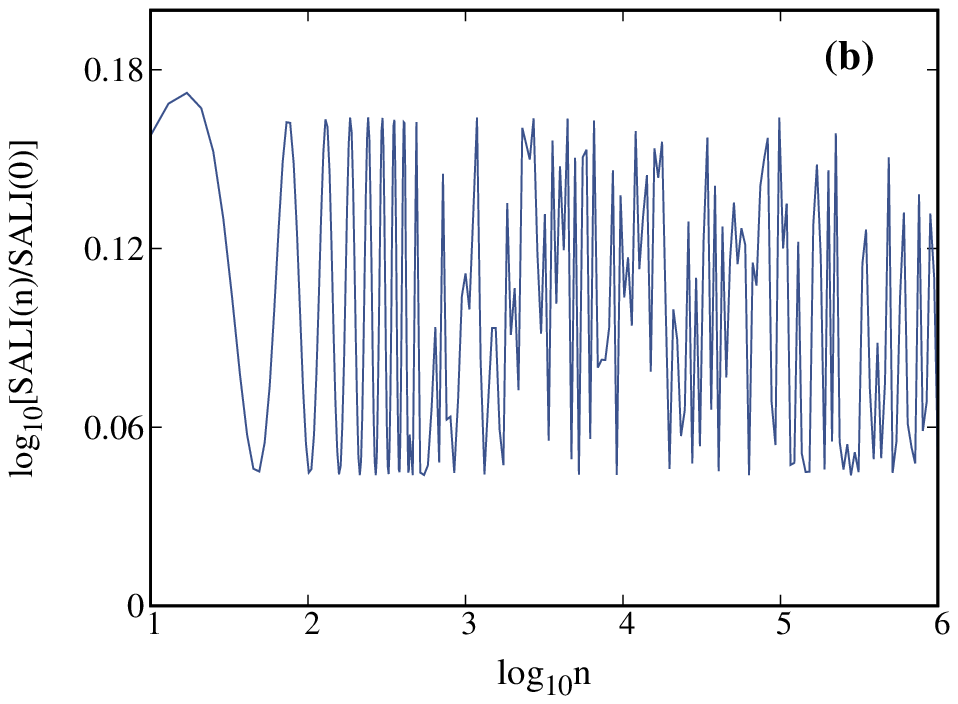}
    	\caption{The evolution of (a) the two LEs and (b) the SALI with respect to the number of iteration $n$ of the mapping \eqref{eq:2NHen_Map} for the regular orbit with IC $(x,y)=(0.5,0.5)$, for $\alpha = 0.5$ and $\beta = 0.1$.}
	\label{fig:NHMap-mLCE-SALI_1} 
\end{figure}

We compute the LEs as well as the SALI of the mapping \eqref{eq:2NHen_Map} of a representative regular and a chaotic orbit. For the regular orbit we use the parameters $(\alpha, \beta) = (0.5, 0.1)$ and the IC $x=0.5$, $y=0.5$. In Fig.~\ref{fig:NHMap-mLCE-SALI_1}a we plot the evolution of the two LEs of this orbit. We see that the LEs eventually attain negative values for this regular orbit, i.e.~$\chi_1 = -0.126605$ and $\chi_2 = -0.126604$. Fig.~\ref{fig:NHMap-mLCE-SALI_1}b displays the time evolution of the SALI for this orbit. We see that, the SALI remains practically constant. 

\FloatBarrier

\begin{figure}[h!]
	\centering
	\centering
	\includegraphics[width=0.45\textwidth,keepaspectratio]{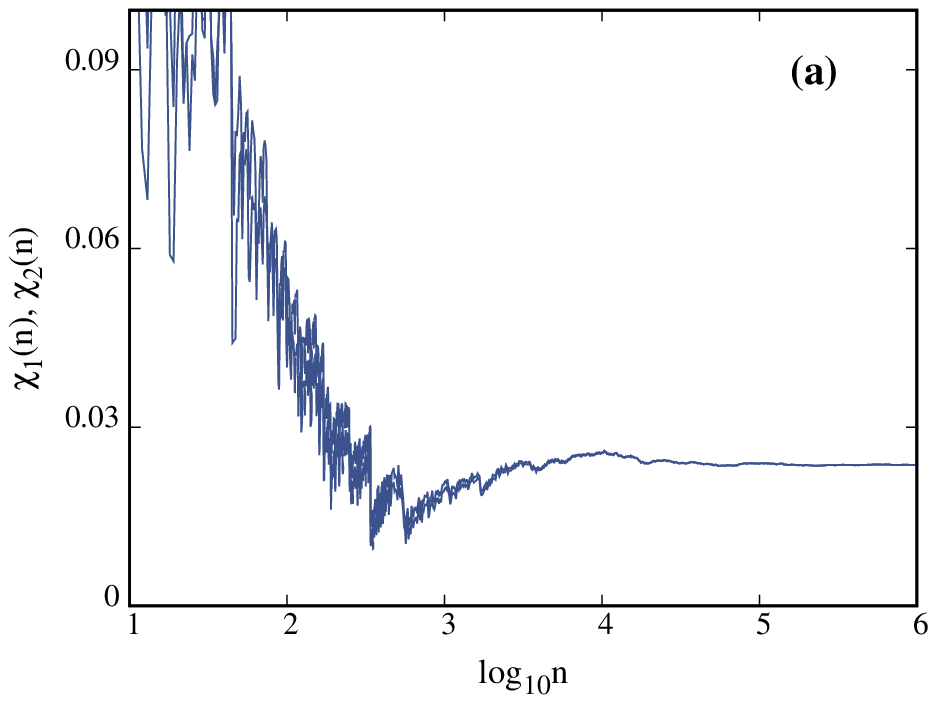}
	\includegraphics[width=0.45\textwidth,keepaspectratio]{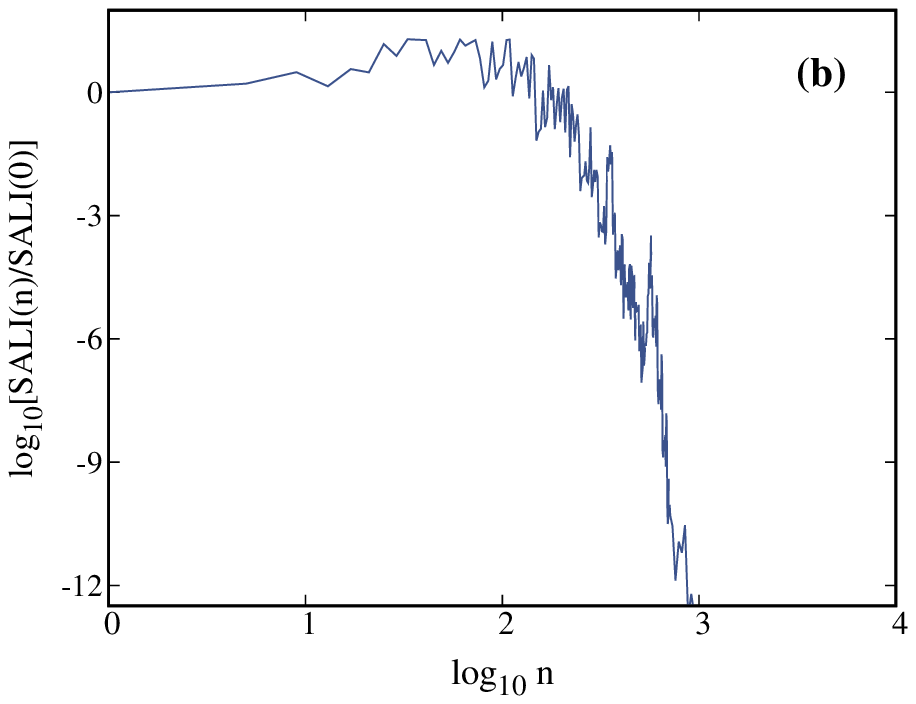}
	\caption{Similar to Fig.~\ref{fig:NHMap-mLCE-SALI_1} but for a chaotic orbit with IC $(x,y)=(0.5,0.5)$, and for $\alpha = 0.9$ and $\beta = 0.6$.}
	\label{fig:NHMap-mLCE-SALI_2} 
\end{figure}

In Fig.~\ref{fig:NHMap-mLCE-SALI_2} we use the parameters $(\alpha, \beta) = (0.9, 0.6)$ for the mapping \eqref{eq:2NHen_Map} and compute the chaos indicators for the chaotic orbit with IC $(x,y)=(0.5,0.5)$. Fig.~\ref{fig:NHMap-mLCE-SALI_2}a shows the evolution of two LEs for this orbit. We see that the two finite time LEs eventually saturate to positive constant values $\chi_1 = 0.0236519$ and $\chi_2 = 0.0236530$. In Fig.~\ref{fig:NHMap-mLCE-SALI_2}b we plot the evolution of the SALI for the same orbit and we can see that it goes to zero exponentially fast.  
\FloatBarrier
\subsubsection{The case of a 4D mapping}
Similar behaviors are observed for a more complicated mapping model given by \cite{TB86}
\begin{eqnarray}
x_{n+1} &=& -D x_{n-1} + 2x_{n} \left[ C_1+\frac{BS_1}{Q_1}\frac{1-exp{(\frac{-r_n^2}{2})}}{r_n^2}\right] + E\xi_i, \nonumber \\
y_{n+1} &=& -D y_{n-1} + 2y_{n} \left[ C_2+\frac{BS_2}{Q_2}\frac{1-exp{(\frac{-r_n^2}{2})}}{r_n^2}\right] + E\eta_i,  
\label{eq:4DDMap} 
\end{eqnarray}
where $r_n^2=x_n^2+y_n^2$, $n=0,1,2,\dots$ and $Q_1,Q_2, B \in \mathbb{R}$, $0 < D < 1$ is the dissipation parameter, $\xi_i, \eta_i$ are randomly generated noise variables, within the interval $(-0.0005, +0.0005)$, and $c_i=\cos 2\pi Q_i$, $S_i=\sin 2\pi Q_i$, $i=1,2$. The corresponding tangent map is given by 
\begin{eqnarray}
\delta x_{n+1} &=& -D\delta x_{n-1} + 2C_1\delta x_{n} + \frac{BS_1}{Q_1}\left[2(x_{n}+\delta x_{n}) 
\frac{1-exp(\frac{-r_n^2}{2}-p_n)}{r_n^2+2p_n} - 2x_n\frac{1-exp(\frac{-r_n^2}{2})}{r_n^2}\right], \nonumber \\
\delta y_{n+1} &=& -D\delta y_{n-1} + 2C_2\delta y_{n} + \frac{BS_2}{Q_2}\left[2(y_{n}+\delta y_{n}) 
\frac{1-exp(\frac{-r_n^2}{2}-p_n)}{r_n^2+2p_n} - 2y_n\frac{1-exp(\frac{-r_n^2}{2})}{r_n^2}\right],
\label{eq:4DDMap_TM} 
\end{eqnarray}
where $p_n^2=x_n\delta x_n+y_n\delta y_n$. Mapping \eqref{eq:4DDMap} is conservative when $D = 1$. 

\begin{figure}[h!]
		\centering
		\centering
	\includegraphics[width=0.45\textwidth,keepaspectratio]{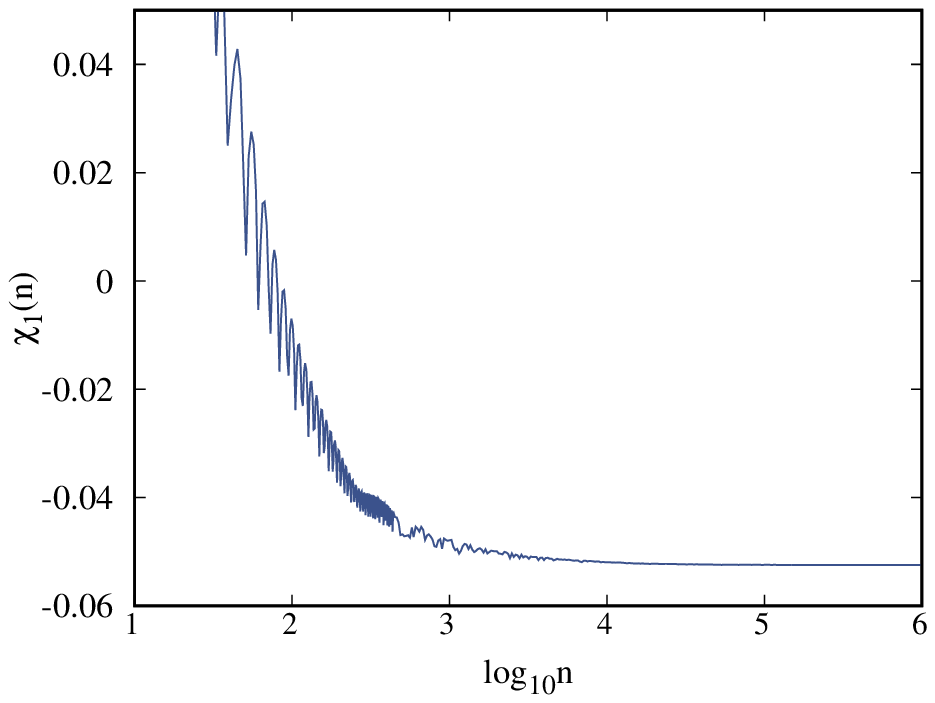}
	\includegraphics[width=0.45\textwidth,keepaspectratio]{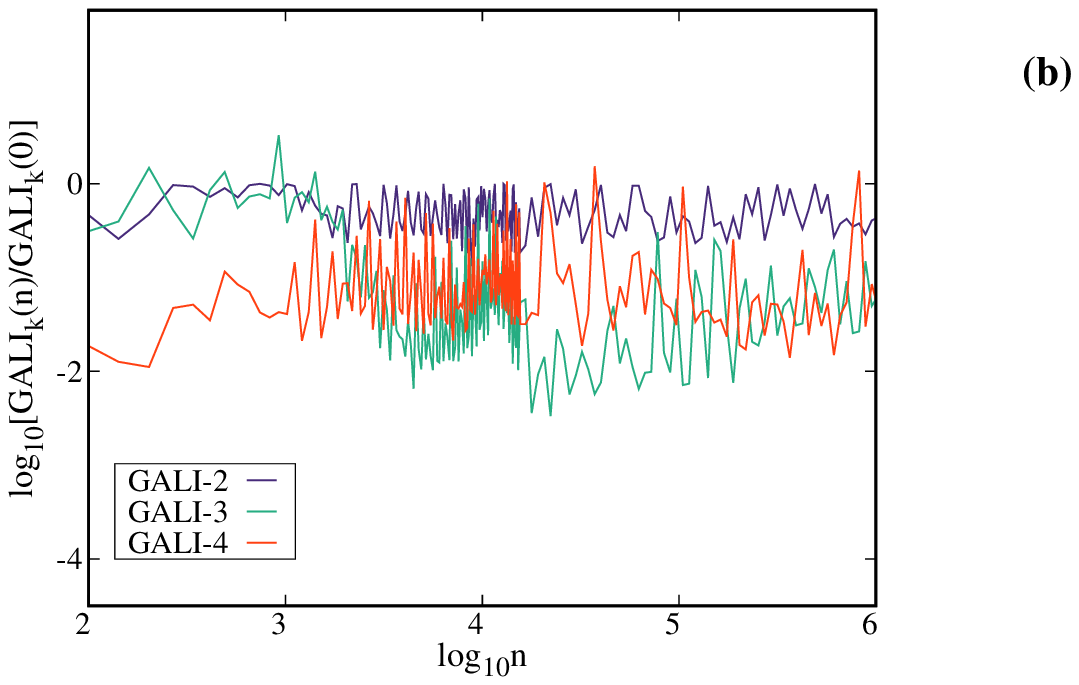}
\caption{Evolution of (a) the mLE and (b) the GALI$_k(n)$, $k=2,3,4$ with respect to the number of iterations $n$ of the mapping \eqref{eq:4DDMap} for a regular orbit with IC $(x_0,y_0,x_1,y_1)=(\sqrt{0.1},0,0,\sqrt{0.1})$ when the mapping parameters are set to $E=1.0$, $Q2=0.9$, $\sigma=0.98$, $D=0.9$, $B=0.33301$.}
	\label{fig:4DDMap-mLCE-GALI1} 
\end{figure}

\begin{figure}[h!]
		\centering
	\includegraphics[width=0.45\textwidth,keepaspectratio]{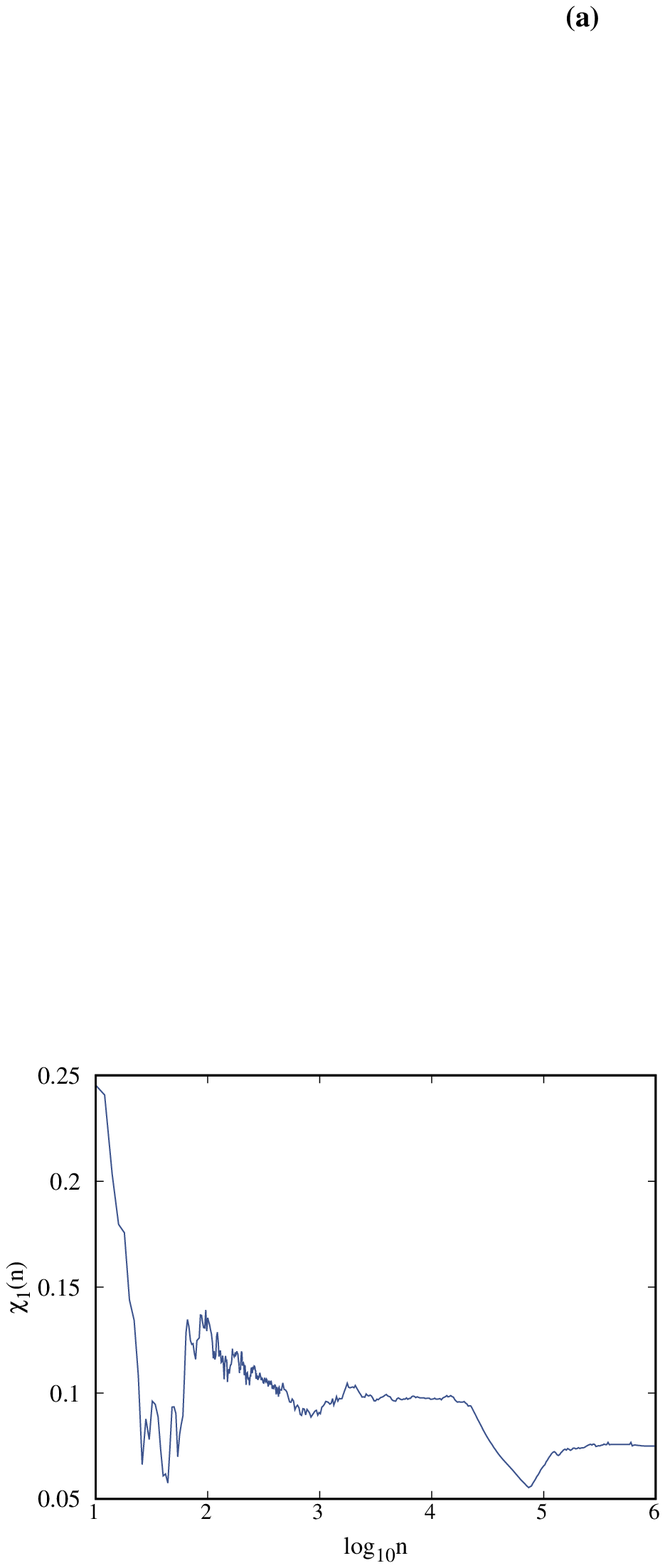}
\includegraphics[width=0.45\textwidth,keepaspectratio]{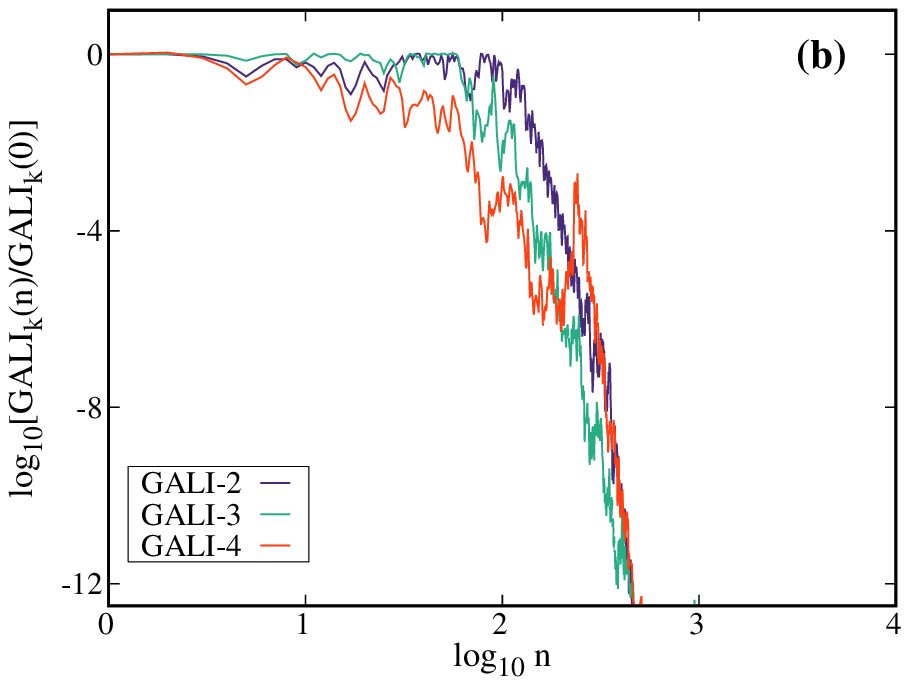}
\caption{Similar to Fig.~\ref{fig:4DDMap-mLCE-GALI1} but for a chaotic orbit with IC $(x_0,y_0,x_1,y_1)=(\sqrt{0.1},0,0,\sqrt{0.1})$ for $D=1.0$ (all other parameters are the same as in Fig.~\ref{fig:4DDMap-mLCE-GALI1}).}
\label{fig:4DDMap-mLCE-GALI2} 
\end{figure}

We also studied the time evolution of the mLE and the GALI$_k$, $k=2,3,4$ of mapping \eqref{eq:4DDMap} for a regular and chaotic orbit. From Fig.~\ref{fig:4DDMap-mLCE-GALI1}a we see that the mLE reaches a negative constant value for the regular orbit ($\chi_{1} = -0.052461$) with IC $(x_0,y_0,x_1,y_1)=(\sqrt{0.1},0,0,\sqrt{0.1})$ when the mapping parameters are set to $E=1.0$, $Q2=0.9$, $\sigma=0.98$, $D=0.9$, $B=0.33301$. Whereas it becomes positive value for the chaotic orbit with IC $(x_0,y_0,x_1,y_1)=(\sqrt{0.1},0,0,\sqrt{0.1})$ when the mapping parameters are set to $E=1.0$, $Q2=0.9$, $\sigma=0.98$, $D=1.0$, $B=0.33301$ ($ \chi_{1} = 0.065738$) (Fig.~\ref{fig:4DDMap-mLCE-GALI2}a). In Fig.~\ref{fig:4DDMap-mLCE-GALI1}b we can see that the GALIs remain practically constant for the regular orbit while they go to zero exponentially fast for the chaotic orbit (Fig.~\ref{fig:4DDMap-mLCE-GALI2}b). 

\FloatBarrier
    
\chapter{Behavior of the GALI for regular motion in multidimensional Hamiltonian systems} \label{Chapter-Three} 
\section[Fermi\text{-}Pasta\text{-}Ulam\text{-}Tsingou model]{The Fermi\text{-}Pasta\text{-}Ulam\text{-}Tsingou (FPUT) model}

\begin{figure}[h!]
\centering
\includegraphics[trim=.0cm 0.cm .0cm 0.525cm,width=0.8\textwidth,keepaspectratio]{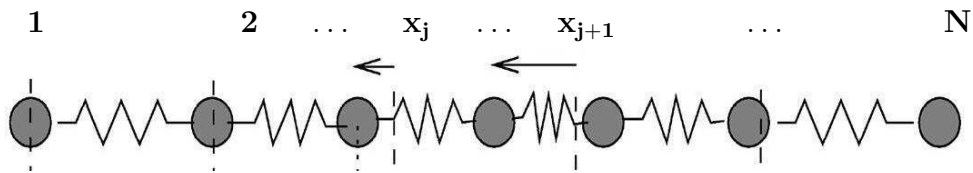}
\caption{Schematic model of the FPUT model: identical masses that can move only in one spatial dimension are coupled by non-linear springs. $x_j$ is the displacement from the equilibrium position of the $j$th mass. The boundary conditions are fixed.}
\label{fig:FPUchainedependules}
\end{figure}

The FPUT model \cite{FPUT55, F92} describes a multidimensional lattice system and it is related to the famous paradox of the appearance of a recurrent behavior in a nonlinear system. It is traditionally called the $(\alpha + \beta)-$FPUT model and represents a one-dimensional chain of $N$ identical particles with nearest-neighbor interactions (see Fig.~\ref{fig:FPUchainedependules}). 

\hfill

$  \quad\quad \quad	 \mathbf{1}	\quad\quad\quad \quad\quad \quad \;	\mathbf{2}	\; \;\quad  \dots \quad\;\;	 \mathbf{x_j} 	\quad \; \dots \quad \;	\mathbf{x_{j+1}} 	 \hfil \hfil \dots \hfil \hfil \quad	\mathbf{N} \hfil$
\FloatBarrier 
The $(\alpha + \beta)-$FPUT Hamiltonian is 
\begin{equation}
H_N = \frac{1}{2}\sum_{j=1}^{N} p_j^2 + \sum_{j=0}^{N} \left[ \frac{1}{2} (x_{j+1}-x_j)^2 +\frac{1}{3}\alpha (x_{j+1}-x_j)^3+ \frac{1}{4}\beta (x_{j+1}-x_j)^4 \right], 
\label{eq:FPUT_AB}
\end{equation}
where $x_j$ is the displacement of the $j$th particle from its equilibrium position and $p_j$ is the corresponding conjugate momentum. In our study we impose fixed boundary condition to system \eqref{eq:FPUT_AB}, i.e.~$x_0=x_{N+1}=p_0=p_{N+1}=0$. We also chose to work with the $\beta-$FPUT system (i.e.~$\alpha = 0$ in \eqref{eq:FPUT_AB})
\begin{equation}
H_N = \frac{1}{2}\sum_{j=1}^{N} p_j^2 + \sum_{j=0}^{N} \left[ \frac{1}{2} (x_{j+1}-x_j)^2 + \frac{1}{4}\beta (x_{j+1}-x_j)^4 \right], 
\label{eq:FPUT_B}
\end{equation}
because it does not allow particles to escape.

In order to use a SI scheme for the integration of the system's equations of motion and variational equations (using the so-called tangent map method \cite{SG10,GS11, GES12}) we split the Hamiltonian in two integrable parts $H_N= A + B$ in which $A$ is a function of the momenta $p_j$ and $B$ is a function of the coordinates $q_j$:  
\begin{equation}
A = \frac{1}{2}\sum_{j=1}^{N} p_j^2, \quad \text{and} \quad B =  \sum_{j=0}^{N} \left[ \frac{1}{2} (x_{j+1}-x_j)^2 + \frac{1}{4}\beta (x_{j+1}-x_j)^4 \right].
\end{equation} 

Then the corresponding equations of motion for the Hamiltonian functions $A$ and $B$, for $ 1 \le j \le N$, are 
\begin{equation}\label{prop:FPU-LAZ}
\dot{X} = L_{AZ}X:
\begin{cases}
\dot{x}_j &= \quad p_j  \\
\dot{p}_j &= \quad 0     \\
\dot{\delta x}_j &= \quad \delta p_j   \\
\dot{\delta p}_j &= \quad 0    \\
\end{cases},
\end{equation}
and 
\begin{equation} \label{prop:FPU-LAB}
\dot{X} = L_{BZ}X:
\begin{cases}
\dot{x}_j &= \quad 0  \\
\dot{p}_j &= \quad x_{j+1}-2x_j+x_{j-1} -\beta(x_{j+1}-x_j)^3 + \beta (x_{j}-x_{j-1})^3      \\
\dot{\delta x}_j &= \quad 0  \\
\dot{\delta p}_j &= \quad (1+3\beta (x_{i}-x_{i-1})^2)\delta x_{i-1} -(2+3\beta(x_{i}-x_{i-1})^2-3\beta(x_{i+1}-x_{i})^2)\delta x_{i} \\
& \qquad + (1+3\beta (x_{i+1}-x_{i})^2)\delta x_{i+1}    
\end{cases}.
\end{equation}
These two sets of differential equations can be easily integrated. The operators $e^{\tau L_{A}z}$ and $e^{\tau L_{B}z}$ which propagate the set of ICs $x_j, \  p_j, \ \delta x_j$ and $\delta p_j$ at time $t$ to $x'_j$, $y'_j$, $\delta x'_j$ and $\delta p'_j$ time $t + \tau$ ($\tau$ being the integration time step), for $1 \le j \le N$, are  

\begin{equation}\label{prop:FPU-eLAZ}
e^{\tau LAz}X =
\begin{cases}
x'_j &= \quad x_j + \tau p_j  \\
p'_j &= \quad p_i  \hspace{0.5in}    \\
\delta x'_j &= \quad \delta x_j + \tau \delta p_j  \\
\delta p'_j &= \quad \delta p_j    \\
\end{cases},
\end{equation}
and 
\begin{equation}\label{prop:FPU-eLAB}
e^{\tau LBz}X =
\begin{cases}
x'_j &= \quad x_j  \\
p'_j &= \quad p_j + \tau \big[x_{j+1}-2x_j+x_{j-1} -\beta(x_{j+1}-x_j)^3 + \beta (x_{j}-x_{j-1})^3\big]     \\
\delta x'_j &= \quad \delta x_j  \\
\delta p'_j &= \quad \delta p_j + \tau \big[(1+3\beta (x_{i}-x_{i-1})^2)\delta x_{i-1} -(2+3\beta(x_{i}-x_{i-1})^2-3\beta(x_{i+1}-x_{i})^2)\delta x_{i} \\
& \qquad + (1+3\beta (x_{i+1}-x_{i})^2)\delta x_{i+1}\big].
\end{cases}.
\end{equation}

\subsection*{Computational considerations}
Here we discuss some practical aspects of our numerical simulation. In order to compute the time evolution of GALI$_k$ of a $2N$ dof Hamiltonian system, we need $k$ initially linearly independent deviation vectors. In most considered cases the coordinates of these vectors are numbers taken from a uniform random distribution in the interval $[-0.5, 0.5]$. To statistically analyze the behavior of the GALIs their values are averaged over several different choices of sets of initial deviation vectors. In particular, we use $n_v=10$ sets of initial vectors. The random choice of the initial vectors leads to different GALI$_k (0)$ values, i.e.~the value of the GALI$_k$ at the beginning of the evolution. Thus, to fairly and adequately compare the behavior of the indices for different initial sets of vectors we normalize the GALIs' evolution by registering the ratio GALI$_k (t)$/GALI$_k (0)$, i.e.~we measure the change of the volume defined by the $k$ deviation vectors with respect to the initially defined volume. Another option is to start the evolution of the dynamics by considering a set of $k$ orthonormal vectors so that GALI$_k(0) = 1$. As we will see later on, both approaches lead to similar results, so in our study we will follow the first procedure, unless otherwise specifically stated. 
	
In our investigation, we implement an efficient fourth-order symplectic integration scheme the so-called ABA864 method, to integrate the equations of motion and the variational equations of the $\beta-$FPUT system \eqref{eq:FPUT_B}. In our numerical simulations, we typically integrate the $\beta-$FPUT system \eqref{eq:FPUT_B} up to a final time $t_f$ of $10^8$ time units. By adequately adjusting the used integration time step we always keep the absolute relative energy error \footnote{The relative energy error, $RE$, is defined as $RE = \left| \frac{H_c - H_0}{H_0} \right|$, where $H_c$ is the computed Hamiltonian value at the current time unit and $H_0$ is the initial Hamiltonian value.} below $10^{-8}$.

The scope of our work is to quantitatively analyze the chaotic and regular behavior of orbits of the multidimensional $\beta-$FPUT model \eqref{eq:FPUT_B}. In Chapter \ref{Chapter-Two}, we described in detail the behavior of the mLE, and the spectrum of LEs, as well as of the SALI and the GALI for 2D and 4D systems. These behaviors should hold for multidimensional systems. In order to illustrate the behavior of these indices let us take the $\beta$-FPUT model \eqref{eq:FPUT_B} with $N=8$ dof and consider some typical regular and chaotic orbits. More specifically we consider the regular orbit \textbf{$R_1$} with total energy $H_8 = 0.0633$ and ICs $-x_1=x_2=-x_4=x_5=-x_7=x_8=0.067$, $x_3=0$, $x_6=0.08$, $p_1=p_3=p_6=0$, $-p_2=p_4=-p_5=p_7=-p_8=0.08$, and $\beta = 1$ and the chaotic orbit \textbf{$C_1$} with total energy $H_8 = 10.495$ and ICs $x_1=x_4=2$, $x_2=x_5=1$, $x_3=x_6=0.5$, $x_7=x_8=0.1$, $p_i=0$, $i=1,2,\dots, 8$, and $\beta = 1.04$.    

\begin{figure}[!h]
	\centering
	\includegraphics[width=0.45\textwidth,keepaspectratio]{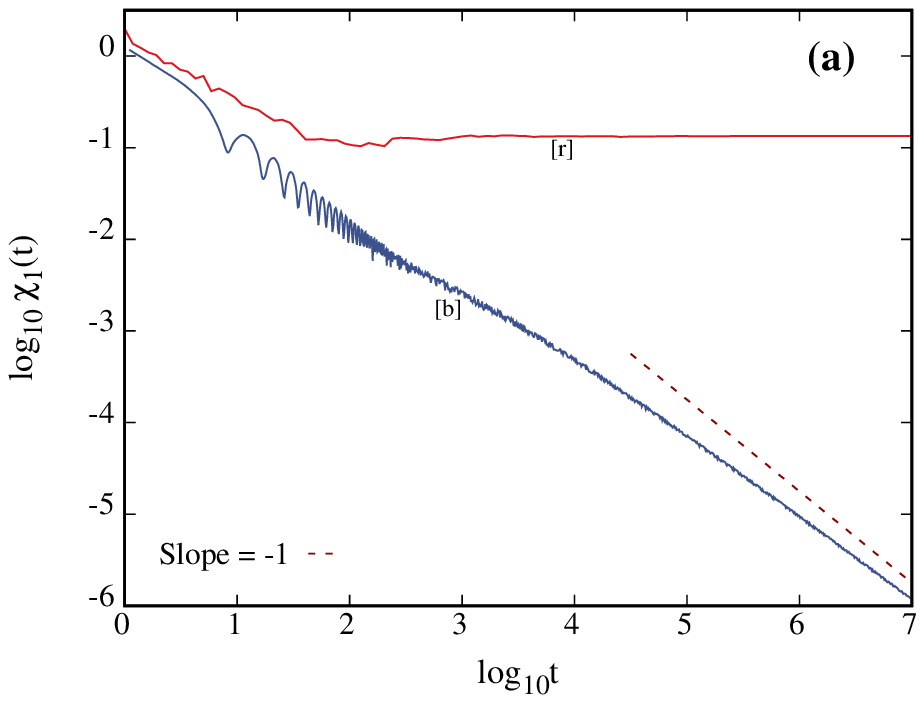}
	\includegraphics[width=0.45\textwidth,keepaspectratio]{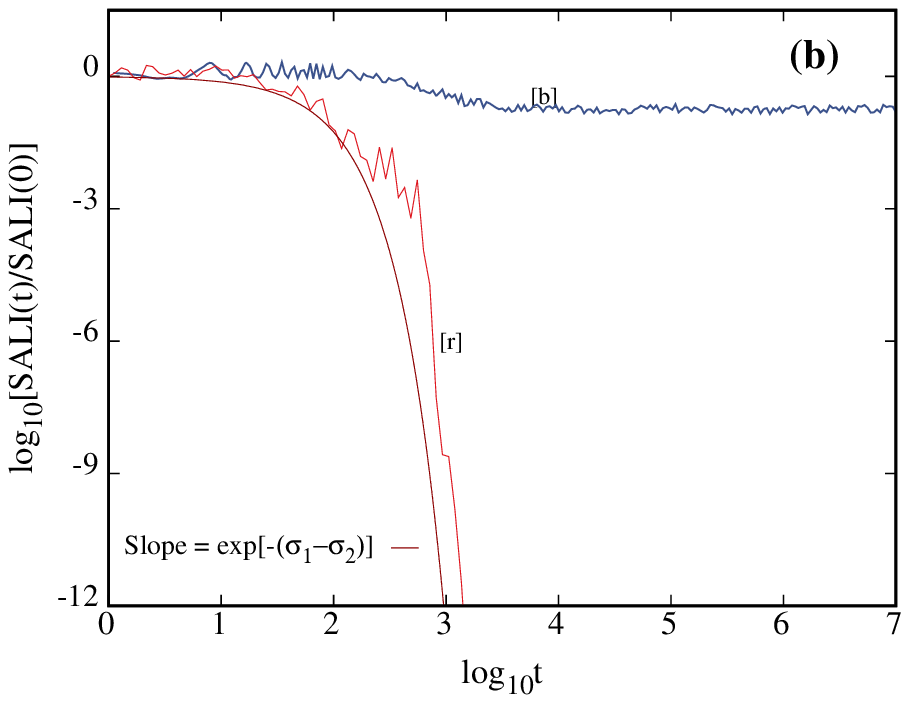}
	\caption{Time evolution of (a) the mLE and (b) the SALI of the regular orbit \textbf{$R_1$} (blue curves denoted by [b]) and the chaotic orbit \textbf{$C_1$} (red curves denoted by [r]) of the $\beta-$FPUT system \eqref{eq:FPUT_B} for $N=8$. The dotted curve indicates a function proportional to (a) $t^{-1}$ (and (b) $e^{-(\sigma_1 - \sigma_2)t}$ for $\sigma_1=0.170$ and $\sigma_2=0.141$). Both axes are in logarithmic scale.}
	\label{fig:FPU_mLCE-SALI}
\end{figure}

The evolution of the mLE for these orbits is shown in Fig.~\ref{fig:FPU_mLCE-SALI}a while their SALI is depicted in Fig.~\ref{fig:FPU_mLCE-SALI}b. In Fig.~\ref{fig:FPU_mLCE-SALI}a, we see that for the regular orbit (blue curve) the mLE goes to zero with a decrease proportional to the function $t^{-1}$ whereas the $\sigma_1(t)$ remains practically constant for the chaotic orbit (red curve), which is exactly what we expect from a chaotic trajectory. In Fig.~\ref{fig:FPU_mLCE-SALI}b we see that SALI goes to zero exponentially fast following the law of \eqref{SALI_Poperty_C}, i.e.~SALI$(t) \propto exp[-(\sigma_1 - \sigma_2)t]$ for $\sigma_1=0.170$ and $\sigma_2=0.141$, which are good estimations of the first and the second largest LEs. On the other hand, it remains practically constant and different from zero for the regular orbit (blue curve). 
     
\begin{figure}[!h]
	\centering
	\includegraphics[width=0.45\textwidth,keepaspectratio]{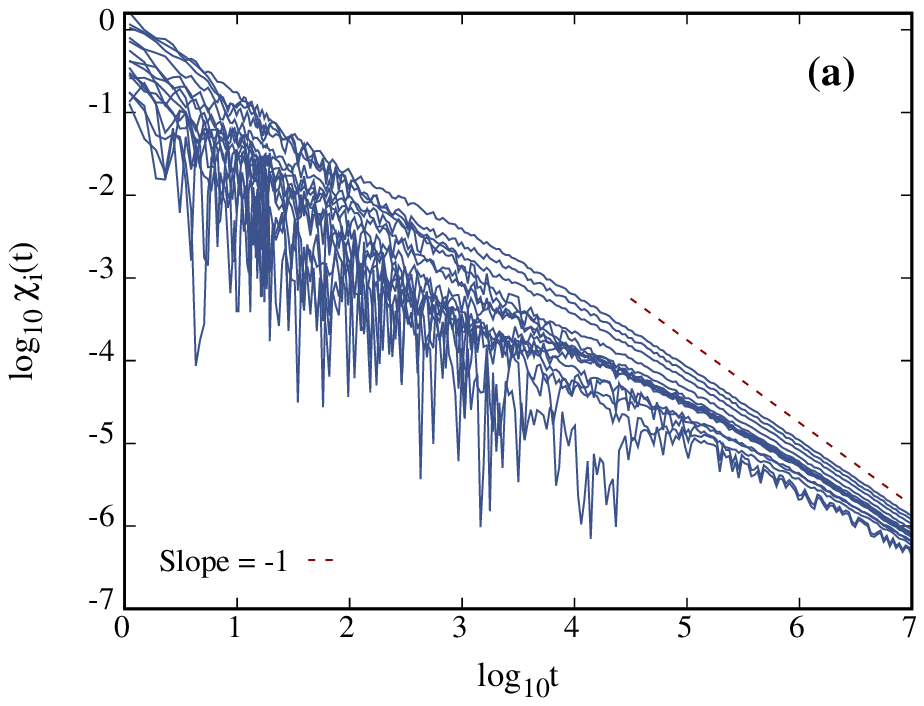}
	\includegraphics[width=0.45\textwidth,keepaspectratio]{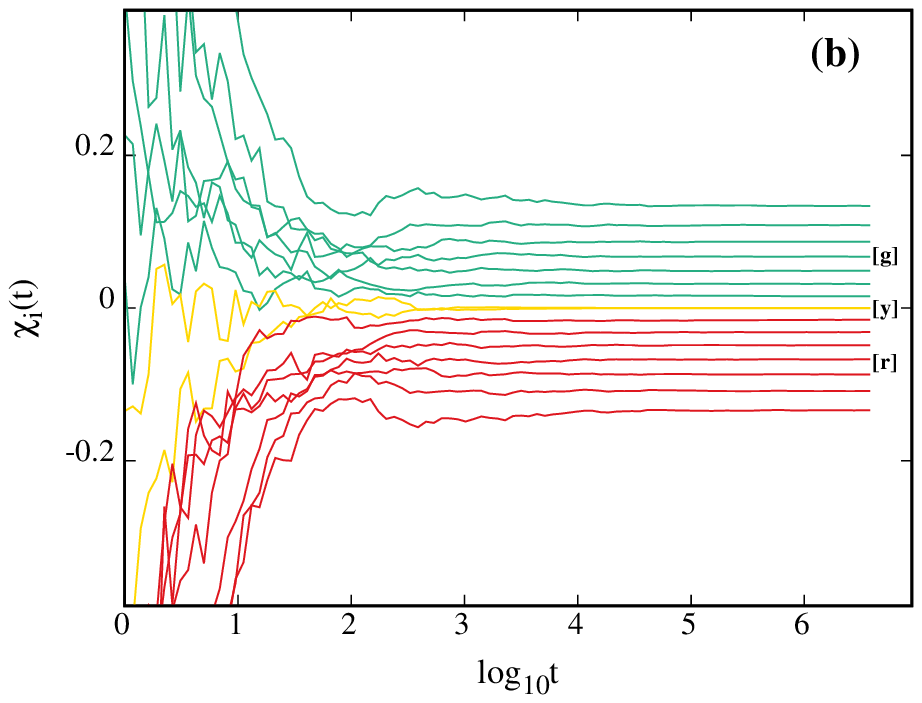}
	\caption{Time evolution of the 16 LEs, $\chi_{i}$, $i=1,\ldots,16$ of the $\beta-$FPUT system \eqref{eq:FPUT_B} with $N=8$ for (a) the regular orbit \textbf{$R_1$} and (b) the chaotic orbit \textbf{$C_1$}. The plotted straight line in (a) corresponds to a function proportional to $t^{-1}$. In (b) labels indicate green `[g]', yellow `[y]' and red `[r]' curves. Axes in (a) are in logarithmic scale, while in (b) the $x$- and $y$-axis is in logarithmic and linear scale, respectively.}
	\label{fig:FPU_LCEs}
\end{figure}

\begin{figure}[!h]
	\centering
	\includegraphics[trim=.0cm 0.25cm .0cm 2.cm,width=0.85\textwidth,keepaspectratio]{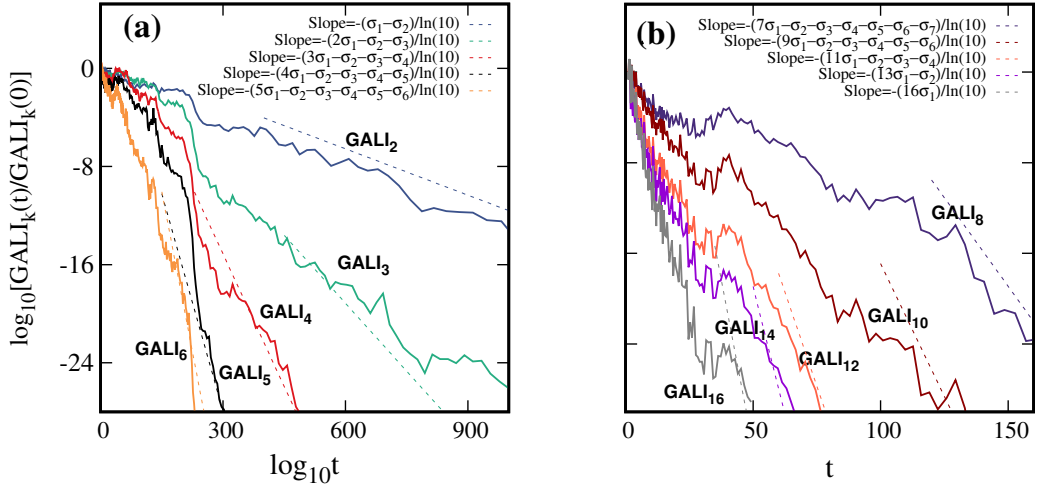}
\caption{Time evolution of the GALI$_k$ for (a) $k=2,\ldots,6$ and (b) $k=8,10,12,14,16$ of the chaotic orbit \textbf{$C_1$} of the $\beta-$FPUT system \eqref{eq:FPUT_B}. The plotted straight lines correspond to functions that follow the asymptotic decay laws in Eq.~\eqref{eq:GALI_chaos} for $\sigma_1=0.170$, $\sigma_2=0.141$, $\sigma_3=0.114$, $\sigma_4=0.089$, $\sigma_5=0.064$, $\sigma_6=0.042$ and $\sigma_7=0.020$.}
	\label{fig:FPU_GALI_C8}
\end{figure}

\FloatBarrier 
In Fig.~\ref{fig:FPU_LCEs} we plot the whole spectrum of LEs consisting in total of 16 LEs, for orbits \textbf{$R_1$} and \textbf{$C_1$}. In the case of the regular orbit (Fig.~\ref{fig:FPU_LCEs}a)  all 16 LEs approach zero following an evolution proportional to $t^{-1}$, while in the case of the chaotic orbit (Fig.~\ref{fig:FPU_LCEs}b) the set of LEs consist of pairs of values having opposite signs. Positive LEs are plotted in green while negative values in red. Moreover, two of the exponents, $\chi_8$ and $\chi_9$ (yellow curves in Fig.~\ref{fig:FPU_LCEs}b), eventually become zero. We will use the values of these 16 exponents later on when we will discuss the results of Fig.~\ref{fig:FPU_GALI_C8}.      

\begin{figure}[!h]
	\centering
   \includegraphics[trim=.0cm 0.25cm .0cm 2.cm,width=0.8\textwidth,keepaspectratio]{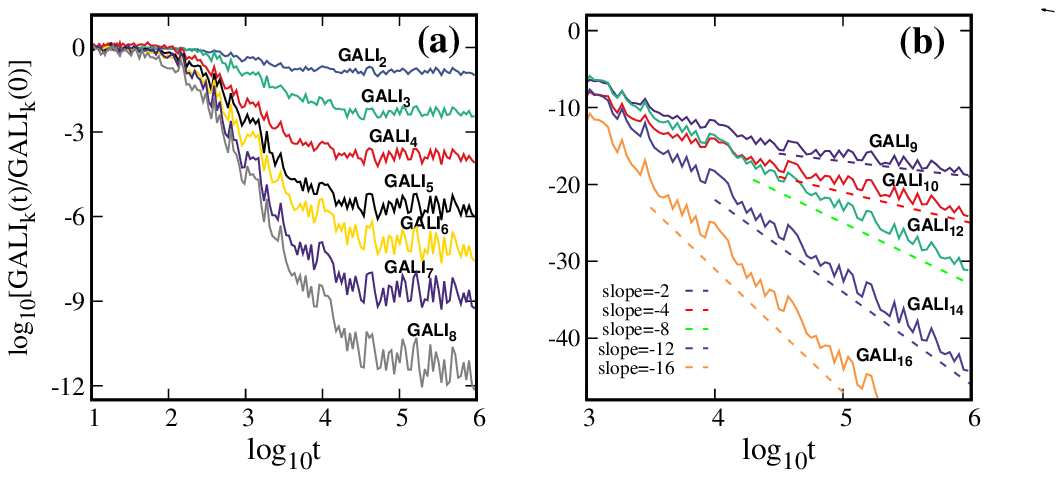}
	\caption{Time evolution of the GALI$_k$ for (a) $k=2,\ldots,8$ and (b) $k=9,\ldots,16$ of the regular orbit \textbf{$R_1$} of the $\beta-$FPUT system \eqref{eq:FPUT_B}. The plotted straight lines in (b) correspond to functions proportional to $t^{-2}$, $t^{-4}$, $t^{-8}$, $t^{-12}$, and $t^{-16}$.}
	\label{fig:FPU_GALI_R8}
\end{figure}

\FloatBarrier 
Let us now study the behavior of the GALIs for these two orbits. The regular orbit \textbf{$R_1$} lies on an $8D$ torus and therefore GALI$_2$ up to GALI$_8$ eventually oscillate around a non-zero constant value which decreases with increasing order $k$ (Fig.~\ref{fig:FPU_GALI_R8}a), while the GALIs with order greater than the dimension of the torus go to zero following the asymptotic power law decays in accordance with Eq.~\eqref{eq:GALI_reg} (Fig.~\ref{fig:FPU_GALI_R8}b). On the other hand, the GALIs of the chaotic orbit \textbf{$C_1$} follow exponential decays in accordance with  Eq.~\eqref{eq:GALI_chaos} (Fig.~\ref{fig:FPU_GALI_C8}).  
\FloatBarrier 

\FloatBarrier 
\newpage 
\section{The behavior of the GALIs for regular orbits} \label{sec:Behavior_SPOs}

For Hamiltonian ﬂows and symplectic maps, a random small perturbation of a stable periodic orbit in general leads to regular motion. Hence, finding a stable periodic orbit of the $\beta-$FPUT system \eqref{eq:FPUT_B} and performing small perturbations to it will allow us to study the behavior of GALI for several regular orbits.   

The eigenvalues and eigenvectors of the so-called monodromy matrix $\mathbf{M}(T)$ dictate the stability of the periodic orbit. $\mathbf{M}(T)$ is the fundamental solution matrix of the variational equations of the periodic orbit evaluated at a time equal to one period $T$ of the orbit (see e.g. \cite{S01b}). This matrix is symplectic, and its columns are linearly independent solutions of the equations that govern the evolution of deviation vectors from the periodic orbit, i.e.~the variational equations. Since the Hamiltonian system \eqref{eq:FPUT_B} is conservative, the mapping defined by the monodromy matrix is volume-preserving, i.e. the determinant of $\mathbf{M}(T)$ is one. In addition, the product of the matrix eigenvalues, which is equal to the determinant, is also one.  

If all the eigenvalues of $\mathbf{M}(T)$ are on the unit circle in the complex plane, then the corresponding periodic orbit is stable otherwise it is unstable. The instability can be of different types (more discussion about this can be found in \cite{S01b} and references therein) but here we are only interested in the stability of periodic orbits. So we do not pay much attention to the types of instability periodic orbits have when they become unstable.

In an $N$D autonomous Hamiltonian system, it turns out that two eigenvalues are always equal to $\lambda=1$ \cite{S01b}. In practice, the remaining $2(N-1)$ eigenvalues define the stability of the periodic orbit. In addition, we expect the eigenvalues of $\mathbf{M}(T)$ computed at any point of the stable orbit to remain the same since the stability of the orbit does not change along the orbit. Thus, we can reduce our investigation to a $2(N-1)$D subspace of the whole phase space using the PSS technique (see e.g.~\cite{LL_92}), where the corresponding monodromy matrix has $2(N-1)$ eigenvalues, none of which is by default $\lambda=1$.
  
In our analysis, we investigate the behavior of the GALIs for regular orbits in the neighborhood of two simple periodic orbits (SPOs) of the $\beta-$FPUT system \eqref{eq:FPUT_B}, which we refer to as  SPO1 and SPO2. In the remaining part of our study, we set $\beta = 1$. The existence and dynamics of these SPOs for the $\beta-$FPUT system were discussed in \cite{ABS06,AB06}.

\subsection{Regular motion in the neighborhood of SPO1}
\label{sec:SPO1}

\begin{figure}[h!]
	\centering
	\includegraphics[trim=.0cm 0.1 .0cm 0.675cm,width=8.5cm,keepaspectratio]{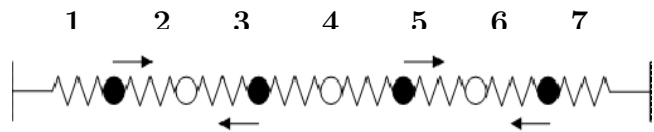}
	\caption{Example of the SPO1 with N=7}
	\label{fig:SPO1_eg_N7}
\end{figure}

The first SPO we study is called SPO1 in \cite{AB06} and it is obtained by considering the $\beta-$FPUT lattice \eqref{eq:FPUT_B} with $N$ being an odd integer so that all particles at even-numbered positions are kept stationary at all times, while the odd-numbered particles are always displaced symmetrically to each other. For example, when $N=7$ the 2nd, 4th and 6th particles are fixed, i.e.~$x_2=x_4=x_6=0$ and the odd numbered particles, i.e.~1st, 3rd, 5th and 7th are equally displaced but in opposite directions, i.e.~$x_1=-x_3=x_5=-x_7$ (see Fig.~\ref{fig:SPO1_eg_N7}).

\hfill

$ \hfill   \mathbf{1} \qquad \; \mathbf{2}	\qquad	\mathbf{3} \; \qquad \mathbf{4} \; \qquad 	\mathbf{5}\qquad 	\mathbf{6}\qquad  \mathbf{7} \> \> \qquad \hfill$
\FloatBarrier
\begin{flushleft}
In general, the SPO1 for $N$ dof is given by   
\end{flushleft}
\begin{equation}
\label{eq:SPO1} 
x_{2j}(t)=0, \qquad x_{2j-1}(t)=-x_{2j+1}(t)=x(t), 
\end{equation}
for $j=1,2, \ldots, \frac{N-1}{2}$.   
Using condition  \eqref{eq:SPO1} in the system's equations of motion we end up  with a second order nonlinear differential equation for variable $x(t)$ which describes the oscillations of all moving particles of the SPO1. In particular,  for $j = 1,3,5,\ldots ,N$ we have
\begin{equation}
\label{eq:SPO1_difeq}
\ddot{x}_j(t) = -2x(t) - 2\beta x^3(t),
\end{equation}
while $x_j(t)=0$, $j= 2 , 4 , 6 , \ldots , (N-1)$ for all times.

\begin{figure}[h!]
	\centering
	\includegraphics[width=0.7\textwidth,keepaspectratio]{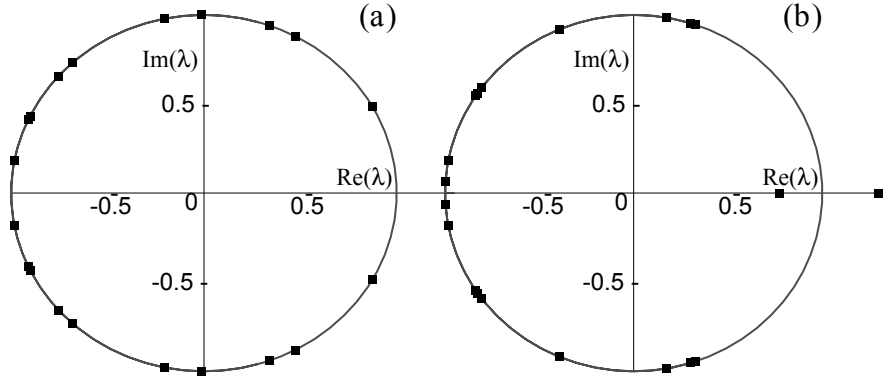}
	\caption{Representation of the eigenvalues $\lambda_i$, $i=1,2,\ldots,20$, of the monodromy matrix of the SPO1 \eqref{eq:SPO1} of the $\beta-$FPUT system \eqref{eq:FPUT_B} on the complex plane with $N=11$ for (a) the stable SPO1 with $H_N/N=0.1$, $x(0)=-0.4112$ and (b) the unstable SPO1 with $H_N/N=0.2$, $x(0)=-0.5626$.}
	\label{fig:SPO1_eigen}
\end{figure}

In Fig.~\ref{fig:SPO1_eigen} we see the arrangement of the  eigenvalues  of the monodromy matrix of the SPO1 orbit  of Hamiltonian \eqref{eq:FPUT_B} with $N=11$ when $H_N/N=0.1$ (Fig.~\ref{fig:SPO1_eigen}a) and $H_N/N=0.2$ (Fig.~\ref{fig:SPO1_eigen}b). In the first case the SPO1 is stable as all eigenvalues are on the unit circle, while in the latter the orbit is unstable as two eigenvalues are off the unit circle. 

\begin{figure}[h!]
	\centering
	\includegraphics[width=0.5\textwidth,keepaspectratio]{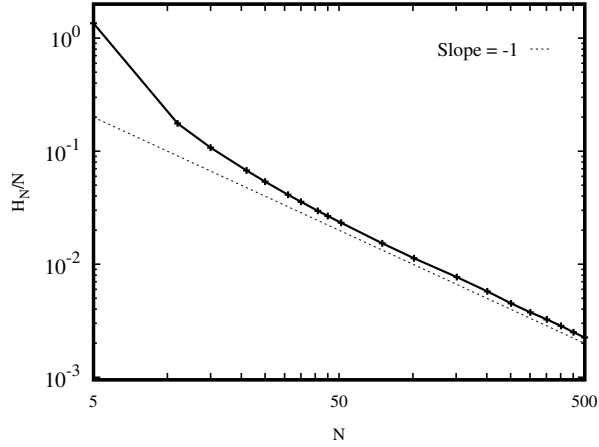}
	\caption{The energy per particle threshold of the first destabilization $H^c_N/N$ of the SPO1 of the $\beta-$FPUT system \eqref{eq:FPUT_B} for $\beta =1 $. The straight line corresponds to a function proportional to $N^{-1}$.}
	\label{fig:SPO1-Ec}
\end{figure}

\FloatBarrier 
The stability analysis of the SPO1 \cite{ABS06,AB06} showed that for small values of $H_N$ \eqref{eq:FPUT_B} the orbit is stable, but it becomes unstable when the energy increases beyond a certain threshold $H_N^c$. For instance in Fig.~\ref{fig:SPO1_eigen} the transition from stability to instability occurs at $H_N^c/N \approx 0.1755$ for $N=11$ particles. In \cite{ABS06,AB06} the authors computed the energy density threshold $H_N^c/N$ for different number of particles $N$, and they found that $H_N^c/N$ decreases as the value of $N$ increases following an asymptotic law $H_N^c/N \propto N^{-1}$. Fig.~\ref{fig:SPO1-Ec} illustrates the destabilization energy threshold for SPO1 with respect to increasing $N$.  
\FloatBarrier

The main objective of our analysis is to study the behavior of GALIs for regular orbits that are located in the neighborhood of the stable periodic orbits. In order to find these regular orbits we initially start with the stable SPO1 having ICs $x_i(0)$, $p_i(0)$, $i=1,2,\ldots, N$ and energy density $H_N/N$. Then we perturb this orbit to obtain a nearby orbit with ICs $X_i(0)$, $P_i(0)$ while keeping the total energy fixed. Practically, in order to find the perturbed orbit, we add a small random number to the positions of the stationary particles of the stable SPOs, i.e.~we have $x_j(0) + w_j$, $j= 2 , 4 , 6 , \ldots , (N-1)$ where $w_j$s are small real random numbers. Then we change the position of one of the initially moving particles to keep the same energy value $H_N$. 

The phase space distance $D$ between these two orbits,  $x_i(0)$, $p_i(0)$ and $X_i(0)$, $P_i(0)$ is given by
\begin{equation}
\label{eq:D}
D = \left\{\sum_{i=1}^{N} \left[ (x_i(0) - X_i(0))^2 + (p_i(0) - P_i(0))^2\right]
\right\}^{1/2}.
\end{equation}

As we have already shown in Sec.~\ref{sec:chaos_indicators}, in multidimensional Hamiltonian systems, the value of the GALI$_k$ with $2 \leq k \leq N$ remains practically constant for regular orbits lying on an $N$D torus and these constant values decrease as the order of the index $k$ grows \cite{SBA08}. In order to improve our statistical analysis we evaluate the average value of $ \mbox{GALI}_k(t) / \mbox{GALI}_k(0) $ over a set of $n_v=10$ different random initial deviation vectors and denote this quantity as $\langle \mbox{GALI}_k(t) / \mbox{GALI}_k(0) \rangle$. We stop our integration when the values of the GALIs more or less saturate showing some small fluctuations. Then we estimate the asymptotic GALI$_k$ value, denoted by $\overline{ \mbox{GALI}}_k$, by finding the mean value of  $\langle \mbox{GALI}_k(t) / \mbox{GALI}_k(0) \rangle$ over the last $n_t=20$ recorded values and the related error is estimated through the standard deviation of the averaging process.

\begin{figure}[h!]
	\centering
	\includegraphics[trim=.0cm 1.cm .0cm 1.5cm,width=0.8\textwidth,keepaspectratio]{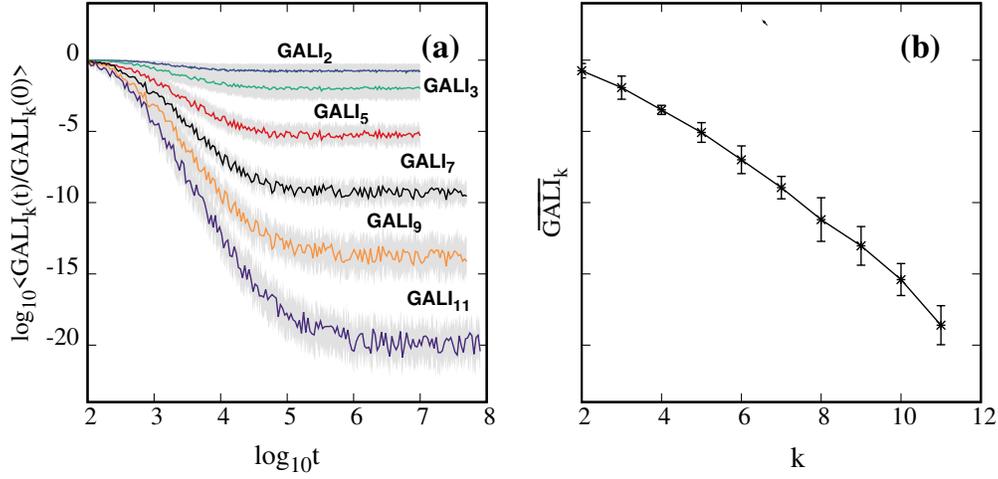}
	\caption{(a) The time evolution of the averaged GALI, $\langle \mbox{GALI}_k(t) / \mbox{GALI}_k(0) \rangle$ for $k=$2, 3, 5, 7, 9 and 11, of a regular orbit close to the stable SPO1 with $H_N/N=0.01$ and $D=0.12$. The average is computed over $n_v=10$ sets of random initial deviation vectors. (b) Estimation of the asymptotic GALI$_k$ value, $\overline{ \mbox{GALI}}_k$ with respect to the order $k$ for the results of (a). The shaded gray area around each curve in (a) and the error bars in (b) denote one standard deviation.}
	\label{fig:SPO1_11_GALIs}
\end{figure}

Fig.~\ref{fig:SPO1_11_GALIs} shows the time evolution of $\langle \mbox{GALI}_k(t) / \mbox{GALI}_k(0) \rangle$ for a regular orbit with distance $D=0.12$ from the stable SPO1 and $H_N/N=0.01$.  The gray area around these curves represents one standard deviation. We observe that GALIs of higher orders converge to lower values and need more time to settle to these values. In  Fig.~\ref{fig:SPO1_11_GALIs}b we see more clearly how these final asymptotic values, $\overline{ \mbox{GALI}}_k$, decrease with increasing $k$.

\begin{figure}[h!]
	\centering
	\includegraphics[trim=.0cm 1.cm .0cm 1.5cm,width=0.8\textwidth,keepaspectratio]{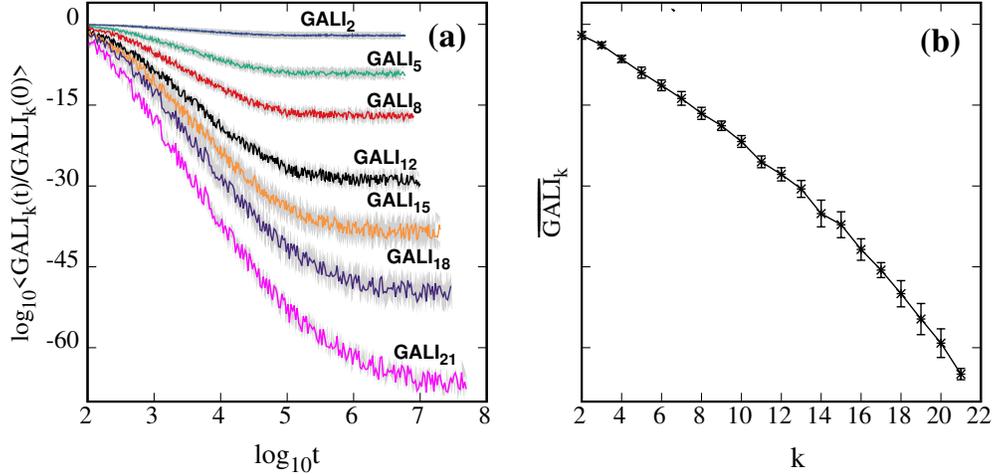}
	\caption{Similar to Fig.~\ref{fig:SPO1_11_GALIs} but for a regular orbit in the vicinity of the stable SPO1  with $H_N/N=0.04$ and $D=0.03$.}
	\label{fig:SPO1_21_GALIs}
\end{figure}

\begin{figure}[h!]
	\centering
	\includegraphics[trim=.0cm 0.65cm .0cm 1.75cm,width=0.85\textwidth,keepaspectratio]{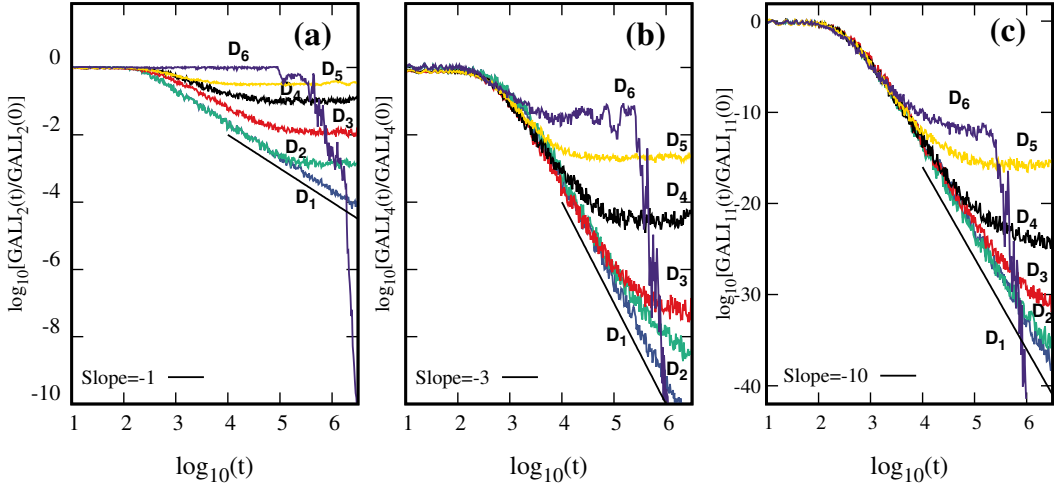}
	\caption{Time evolution of (a) $\mbox{GALI}_2(t)/\mbox{GALI}_2(0)$, (b) $\mbox{GALI}_4(t)/\mbox{GALI}_4(0)$, and (c) $\mbox{GALI}_{11}(t)/\mbox{GALI}_{11}(0)$ for orbits with $H_N/N=0.01$ and distances $D_1 = 0.008$, $D_2 = 0.01$, $D_3 = 0.06$, $D_4 = 0.1$, $D_5 = 0.22$ and  $D_6 = 0.4$ from the stable SPO1 of the $\beta-$FPUT system \eqref{eq:FPUT_B} with $N=11$. The straight lines correspond to functions proportional to (a) $t^{-1}$, (b) $t^{-3}$ and (c) $t^{-10}$.}
	\label{fig:SPO1_11_change_D_E} 
\end{figure}

\begin{figure}[h!]
	\centering
	\includegraphics[trim=.0cm 0.25cm .0cm 1.cm,width=0.8\textwidth,keepaspectratio]{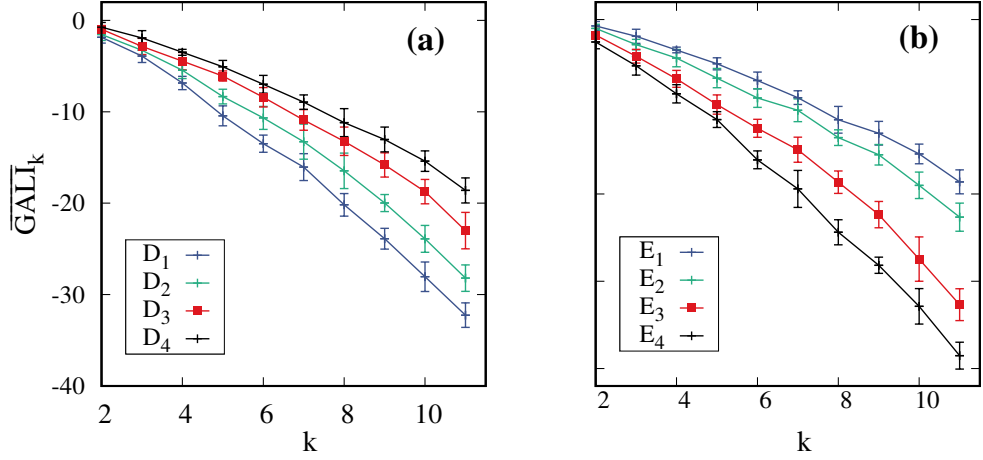}
	\caption{Estimation of the asymptotic  GALI values, $\overline{ \mbox{GALI}}_k$, as a function of their order $k$ for regular orbits in the vicinity of the stable SPO1 of the $\beta-$FPUT system \eqref{eq:FPUT_B} with $N=11$. (a) Regular orbits for $E=H_N/N=0.01$ and distances $D_1=0.01$, $D_2=0.04$, $D_3=0.08$ and $D_4=0.12$. (b) Regular orbits with $E_1=0.01$ and $D=0.12$, $E_2=0.04$ and $D=0.1$, $E_3=0.12$ and $D=0.01$, and $E_4=0.16$ and $D=0.01$. The error bars in both panels denote one standard deviation. Note that in this case the destabilization energy density of the SPO1 is $H^c_N/N \approx 0.1755$.}
	\label{fig:SPO1_11_change_D_many}
\end{figure}

\FloatBarrier
Similar results are obtained for a regular orbit in the neighborhood of SPO1 for $N=21$ (Fig.~\ref{fig:SPO1_21_GALIs}). Again we observe that the almost constant final values of the GALIs drastically decrease as the order of the index $k$ increases from GALI$_2$ to GALI$_{21}$. Also, GALI$_{21}$ starts saturating at around $t=10^{7}$ but GALI$_2$ becomes constant around $t=10$.
\FloatBarrier
Now we focus our attention on the behavior of GALIs when the regular orbits depart from a stable periodic orbit moving towards the edges of the stability island, for a constant value of $N$. We do that by perturbing the ICs of the stable periodic orbit by a small number, i.e.~by increasing $D$ \eqref{eq:D}. Fig.~\ref{fig:SPO1_11_change_D_E} shows the outcome of this process. There we see how the behavior of GALIs change with increasing distance $D$ from the stable SPO1, when we only consider one set of initial deviation vectors. In particular, we see the evolution of three different GALIs, namely GALI$_2$ (Fig.~\ref{fig:SPO1_11_change_D_E}a), GALI$_4$ (Fig.~\ref{fig:SPO1_11_change_D_E}b), and GALI$_{11}$ (Fig.~\ref{fig:SPO1_11_change_D_E}c) for several orbits in the neighborhood of the stable SPO1 with $H_N/N=0.01$ and $D_1 = 0.008$, $D_2 = 0.01$, $D_3 = 0.06$, $D_4 = 0.1$, $D_5 = 0.22$ and  $D_6 = 0.4$. All values of GALIs for the smaller distance $D=D_1$ go to zero asymptotically following the power law of Eq.~\eqref{eq:GALI_reg}, i.e.~GALI$_2$ with $t^{-1}$ [Fig.~\ref{fig:SPO1_11_change_D_E}a], GALI$_4$ with $t^{-3}$ [Fig.~\ref{fig:SPO1_11_change_D_E}b], and GALI$_{11}$ with $t^{-10}$ [Fig.~\ref{fig:SPO1_11_change_D_E}c]. The behavior of the GALIs for the perturbed orbit with the smallest distance from the SPO1, $D=D_1$, is still similar to what we expect for a stable periodic orbit. When we increase $D$ the asymptotic behavior of the GALIs start deviating from the above mentioned power decay law, and eventually GALI$_k$ becomes constant for $D=D_i$, $i=2,\ldots,5$. Finally, for very large $D$ values the perturbed orbits will be outside the stability island. This happens for $D=D_6$ which leads to chaotic motions. In this case, GALIs go to zero exponentially fast. Note that the values of the GALIs for the regular orbit grow with increasing distance $D$ from the stable SPO1. In Fig.~\ref{fig:SPO1_11_change_D_many}a we see the values of $\overline{ \mbox{GALI}}_k$ with respect to increasing $D$ when $E=H_N/N$ is kept constant to $H_N/N=0.01$, for regular orbits around the SPO1 orbit. We chose an appropriate value for $D$ in order to avoid a transition to chaotic motion. 

\FloatBarrier


There is another way to move from stability to instability, that is by increasing the energy densities $E=H_N/N$. We have seen in Fig.~\ref{fig:SPO1-Ec} that as long as we are below the destabilization energy of SPO1, the periodic orbit is stable. For instance, in the case of $N=11$ the SPO1 is stable below $H^c_N/N \approx 0.1755$. In Fig.~\ref{fig:SPO1_11_change_D_many}b we present the values of $\overline{ \mbox{GALI}}_k$ as a function of the GALI's order $k$ for regular orbits in the neighborhood of the stable SPO1 for increasing energy densities $E=H_N/N$. In particular, we consider regular orbits with $D=0.12$ for $E_1 = 0.01$, $D=0.1$ for $E_2 = 0.04$ and with $D=0.01$ for $E_3 = 0.12$ and  $E_4 = 0.16$. Therefore, the $\overline{ \mbox{GALI}}_k$ values decrease as we increase the energy density for regular orbits around the stable SPO.   

\FloatBarrier
We have investigated the behavior of the GALI$_k$ for regular orbits in two ways. We showed that the asymptotic GALI$_k$ values change when we increase the distance $D$ of the studied orbit from the stable SPO1 for constant energy $H_N$ (Fig.~\ref{fig:SPO1_11_change_D_many}a). We also saw that, the asymptotic value of GALI$_k$ decreases with increasing energy densities inside the stability island [Fig.~\ref{fig:SPO1_11_change_D_many}b]. 

\begin{figure}[h!]
	\centering
	\includegraphics[width=0.5\textwidth,keepaspectratio]{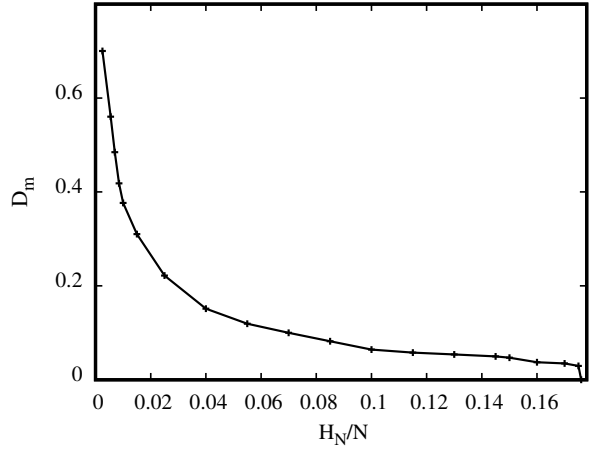}
	\caption{Estimation of the extent in phase space of the stability island around SPO1: The maximum value of $D$, denoted by $D_m$, for which regular motion occurs, as a function of the energy density $H_N/N$ of the $\beta-$FPUT Hamiltonian \eqref{eq:FPUT_B} with $N=11$. The destabilization of the SPO1 takes place at $H_N^c/N \approx 0.1755$.}
	\label{fig:SPO1_D_energy}
\end{figure}

For example, for the stable SPO1 with $N=11$ and $E_1=0.01$ regular motion occurs up to $D=0.3765$. So, the motion of the perturbed orbits above this $D$ value is chaotic. Fig.~\ref{fig:SPO1_D_energy} shows an approximation of the size of the stability island around the SPO1 orbit with different energy densities $H_N/N$ by finding the largest distance, denoted by $D_m$, in which regular motion is observed for $N=11$. 
      
\FloatBarrier

\begin{figure}[h!]
	\centering
	\includegraphics[trim=.0cm 0.25cm .0cm 1.cm,width=0.8\textwidth,keepaspectratio]{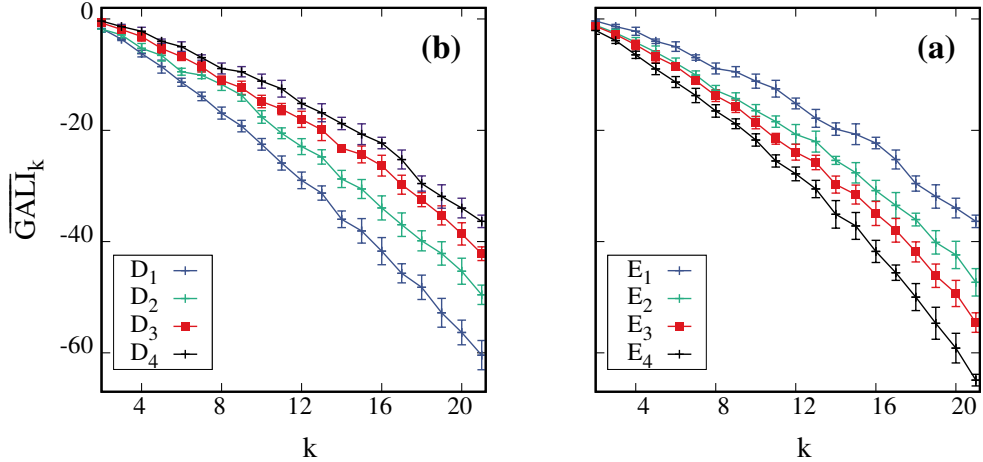}
	\caption{Similar to Fig.~\ref{fig:SPO1_11_change_D_many} but for regular orbits in the vicinity of the stable SPO1 of system \eqref{eq:FPUT_B} for $N=21$ (a) with $H_N/N=0.002$, and $D_1=0.01$, $D_2=0.04$, $D_3=0.08$, and $D_4=0.12$, and (b) $E_1=0.002$ and $D=0.12$, $E_2=0.006$ and $D=0.1$, $E_3=0.01$ and $D=0.09$, and $E_4=0.04$ and $D=0.03$. The error bars in both panels denote one standard deviation. Note that in this case the destabilization energy density of the SPO1 is $H^c_N/N \approx 0.0675$.}
	\label{fig:SPO1_21_change_D_E}
\end{figure}

Similar results have been obtained in Fig.~\ref{fig:SPO1_21_change_D_E} for $N=21$. In Fig.~\ref{fig:SPO1_21_change_D_E}a we see the asymptotic, $\overline{ \mbox{GALI}}_k$ values for regular orbits in the vicinity of the stable SPO1 of the $\beta-$FPUT system \eqref{eq:FPUT_B} for $E=H_N/N=0.002$ with respect to increasing distances $D_1=0.01$, $D_2=0.04$, $D_3=0.08$ and $D_4=0.12$, while in Fig.~\ref{fig:SPO1_11_change_D_many}b we see how $\overline{ \mbox{GALI}}_k$ changes for $E_1=0.002$ and $D=0.12$, $E_2=0.006$ and $D=0.1$, $E_3=0.01$ and $D=0.09$, and $E_4=0.04$ and $D=0.03$. 
\FloatBarrier

\subsection{Regular motion in the neighborhood of SPO2}
\label{sec:SPO2}

So far our analysis was based on results obtained in the neighborhood of one SPO type, i.e.~SPO1. We will now show that these findings are quite general as they remain similar for another SPO. In particular, to illustrate that a similar analysis is performed for regular orbits in the neighborhood of what was called SPO2 in \cite{AB06}.

\begin{figure}[!h]
	\centering
	\includegraphics[trim=.0cm 0.42cm .0cm 0.7cm,width=8.5cm,keepaspectratio]{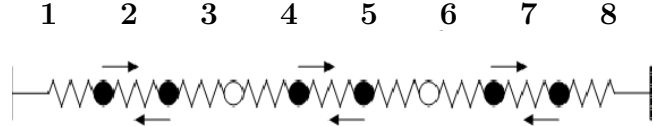}
	\caption{Example of the SPO2 with N=8}
   \label{fig:SPO2_eg_N8}
\end{figure}

The SPO2 of the $\beta-$FPUT system \eqref{eq:FPUT_B} with $N=5+3 m$, $m=0,1,2,\ldots$, particles, corresponds to an arrangement where every third particle remains always stationary and the two particles in between move in opposite directions, i.e. 
\begin{eqnarray} \label{eq:SPO2}
x_{3j}(t)= 0, \quad j=1,2,3,\ldots ,\frac{N-2}{3},\qquad\qquad\qquad \nonumber \\
x_{j}(t) = -x_{j+1}(t)=x(t), \quad j=1,4,7,\ldots,N-1. \
\end{eqnarray}
For example, for $N=8$, particles located in the 3rd and 6th position do not move, $x_3=x_6=0$, while the remaining consecutive particles are moving symmetrically and in opposite directions, i.e. $x_1=-x_2=x_4=-x_5=x_7=-x_8$ (Fig.~\ref{fig:SPO2_eg_N8}).  

\hfill

$ \hfill   \mathbf{1} \qquad \mathbf{2}	\qquad	\mathbf{3} \qquad \mathbf{4} \qquad 	\mathbf{5}\qquad 	\mathbf{6}\qquad  \mathbf{7} \qquad  \mathbf{8} \; \qquad \hfill$
\FloatBarrier
As in the case of the SPO1, we can obtain a single differential equation for the time evolution of the SPO2 when substituting \eqref{eq:SPO2} in the equations of motion of the $\beta-$FPUT system \eqref{eq:FPUT_B}. In particular we get    
\begin{equation}
\label{eq:SPO2_difeq}
\ddot{x}_j(t) = -3x(t) - 9\beta x^3(t),
\end{equation}
for the moving particles, while $x_j(t)=0$ for the stationary particles with $j=3,6,9,\ldots, N-2$. 

\begin{figure}[h!]
	\centering
	\includegraphics[trim=.0cm 2.5cm .0cm 1.7cm,width=0.7\textwidth,keepaspectratio]{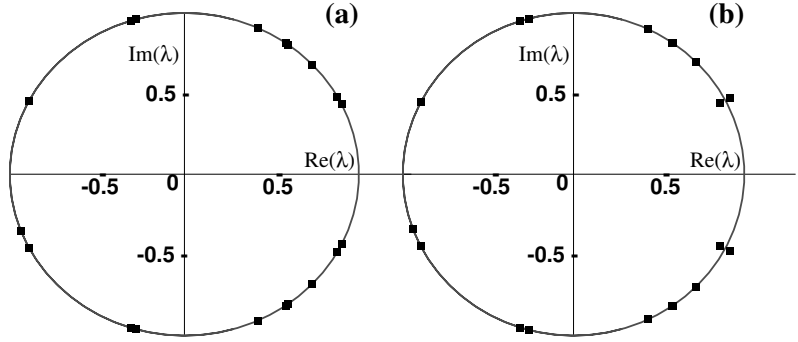}
	\caption{Similar to Fig.~\ref{fig:SPO1_eigen} but for 20 eigenvalues of the monodromy matrix of the SPO2 \eqref{eq:SPO2} of the $\beta-$FPUT Hamiltonian \eqref{eq:FPUT_B} with $N=11$ for (a) the stable SPO2 with $H_N/N=0.01$ and $x(0)=-0.0951$, and (b) the unstable SPO2 with $H_N/N=0.02$ and $x(0)=-0.1336$.}
	\label{fig:SPO2_eigen}
\end{figure}

\begin{figure}[h!]
	\centering
	\includegraphics[width=0.5\textwidth,keepaspectratio]{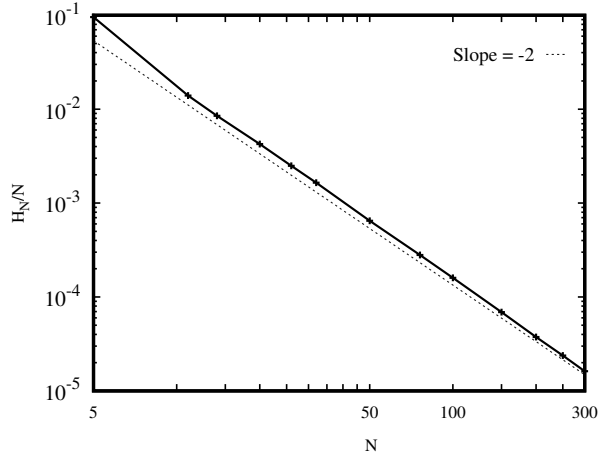}
	\caption{Similar to Fig.~\ref{fig:SPO1-Ec} but for the SPO2. The straight dashed line corresponds to a function proportional to $N^{-2}$.}
	\label{fig:SPO2-Ec}
\end{figure}

In Fig.~\ref{fig:SPO2_eigen} we present the arrangement of the eigenvalues $\lambda_i$, $i=1,2,\ldots,20$, of the monodromy matrix of the SPO2 of Hamiltonian \eqref{eq:FPUT_B} with $N=11$. In Fig.~\ref{fig:SPO2_eigen}a all eigenvalues are on the unit circle, which means that the SPO2 with $H_N/N=0.01$ and $x(0)=-0.0951$ is stable. We note that the monodromy matrix is evaluated on the PSS $x_1=-0.091$, $p_1>0$. In Fig.~\ref{fig:SPO2_eigen}b, there are four eigenvalues outside the unit circle, thus the SPO2, with $H_N/N=0.02$ and $x(0)=-0.1336$ is unstable. In this case, the monodromy matrix is evaluated on the PSS $x_1=-0.131$, $p_1>0$. The critical energy density for which the SPO2 encounters its first transition from stability to instability is much smaller that the one seen for SPO1. In particular, it is $H_N^c/N \approx 0.01395$.
\FloatBarrier 
Similarly to Fig.~\ref{fig:SPO1-Ec} for the SPO1, the stability analysis of the SPO2 \cite{ABS06} determined the relation between the energy density threshold $H^c_N/N$ and the dimension $N$ of the system (Fig.~\ref{fig:SPO2-Ec}). The first destabilization energy density of SPO2 follows a slower asymptotic decrease compared to SPO1, i.e.~$H_N^c/N \propto N^{-2}$ \cite{ABS06}.
\FloatBarrier

\begin{figure}[h!]
	\centering
	\includegraphics[width=0.5\textwidth,keepaspectratio]{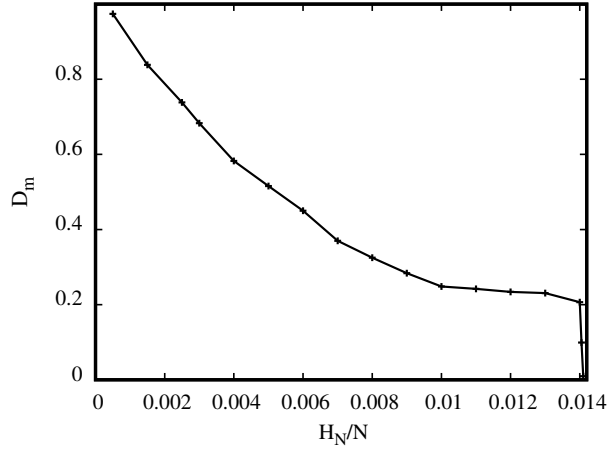}
	\caption{Similar to Fig.~\ref{fig:SPO1_D_energy} but for the stable SPO2 of the $\beta-$FPUT Hamiltonian \eqref{eq:FPUT_B} with $N=11$.}
	\label{fig:SPO2_D_energy}
\end{figure}

The estimation of the size of the stability island around the stable SPO2 with $N=11$ shows a similar behavior to what we observed for the same $N$ in the case of the SPO1 (Fig.~\ref{fig:SPO1_D_energy}). In Fig.~\ref{fig:SPO2_D_energy} we report the maximum value of $D$ (i.e.~$D_m$) for which regular motion occurs as a function of the energy density for the stable SPO2 of \eqref{eq:FPUT_B}. Note that the energy axis is smaller than the one of Fig.~\ref{fig:SPO1_D_energy} since the critical energy density for which the SPO2 becomes unstable for the first time is $H_N^c/N \approx 0.01395$.
 
\FloatBarrier

\begin{figure}[h!]
	\centering
	\includegraphics[trim=.0cm 0.25cm .0cm 1.cm,width=0.8\textwidth,keepaspectratio]{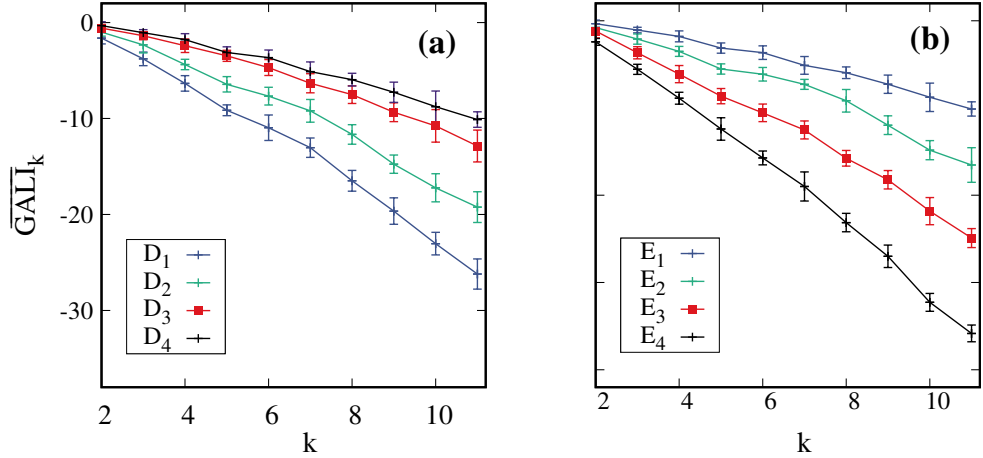}
	\caption{Similar to Fig.~\ref{fig:SPO1_11_change_D_many} but for regular orbits in the vicinity of the stable SPO2 of the $\beta-$FPUT system \eqref{eq:FPUT_B} with $N=11$. Regular orbits with ICs (a) $E=H_N/N=0.001$ and $D_1=0.01$, $D_2=0.04$, $D_3=0.08$ and $D_4=0.12$, (b) $E_1=0.001$ and $D=0.12$, $E_2=0.004$ and $D=0.12$, $E_3=0.008$ and $D=0.04$, and $E_4=0.012$ and $D=0.01$.}
	\label{fig:SPO2_11_change_D_many}
\end{figure}

Similarly to Fig.~\ref{fig:SPO1_11_change_D_many} in Fig.~\ref{fig:SPO2_11_change_D_many}a we see the dependence of $\overline{ \mbox{GALI}}_k$ on the order $k$ for increasing $D$ and constant $H_N/N$, whereas in Fig.~\ref{fig:SPO2_11_change_D_many}b we have similar results but for increasing $H_N/N$ values for regular orbits close to the stable SPO2 with $N=11$.  

\FloatBarrier

\begin{figure}[h!]
	\centering
	\includegraphics[trim=.0cm 0.25cm .0cm 1.cm,width=0.8\textwidth,keepaspectratio]{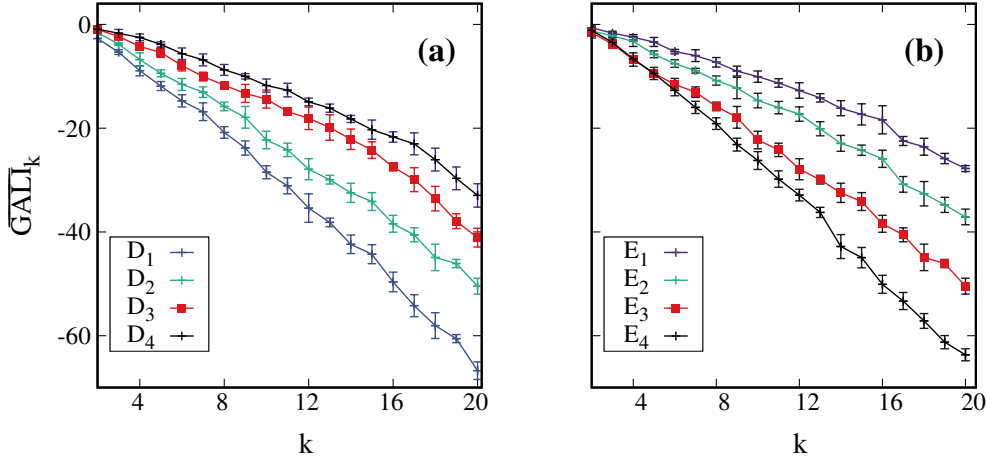}
	\caption{Similar to Fig.~\ref{fig:SPO2_11_change_D_many} but for $N=20$. Regular orbits for (a) $E=H_N/N=0.002$ and  $D_1=0.01$, $D_2=0.04$, $D_3=0.08$ and $D_4=0.12$, (b) $E_1=0.0004$ and $D=0.1$, $E_2=0.0008$ and $D=0.04$, $E_3=0.002$ and $D=0.04$, and $E_4=0.004$ and $D=0.002$. }
	\label{fig:SPO2_20_change}
\end{figure}

In Fig.~\ref{fig:SPO2_20_change} we see the $\overline{ \mbox{GALI}}_k$ values for regular orbits around the stable SPO2 for $N=20$. Note that the destabilization threshold $H_N^c/N \approx 0.00425$ of the SPO2 for $N=20$ is much smaller than in the case with $N=11$. 
\FloatBarrier

The results of Figs.~\ref{fig:SPO2_11_change_D_many} and \ref{fig:SPO2_20_change} clearly shows that the behavior of the asymptotic GALI values for regular orbits in the neighborhood of the stable SPO1 we have observed in Sec.~\ref{sec:SPO1}, remains valid for the stable SPO2. The $\overline{ \mbox{GALI}}_k$ values increase when the regular orbit approaches the edge of the stability island (Figs.~\ref{fig:SPO2_11_change_D_many}a and \ref{fig:SPO2_20_change}a) while it decreases when this orbit moves towards the destabilization energy ( Figs.~\ref{fig:SPO2_11_change_D_many}b and \ref{fig:SPO2_20_change}b).     

\section{Statistical analysis of deviation vectors}\label{sec:devvec}

Let us now turn our attention to the properties of the deviation vectors needed for the computation of the GALIs. Regular motions take place on a torus and the time evolution of all the initially linearly independent deviation vectors brings them on the tangent space of this torus, having in general different directions. On the other hand, in the case of chaotic orbits, the deviation vectors gradually align to the direction defined by the mLE following the exponential law given in Eq.~\eqref{eq:GALI_chaos}.

\begin{figure}[h!]
	\centering
	\includegraphics[trim=.0cm 0.65cm .0cm 1.75cm,width=0.85\textwidth,keepaspectratio]{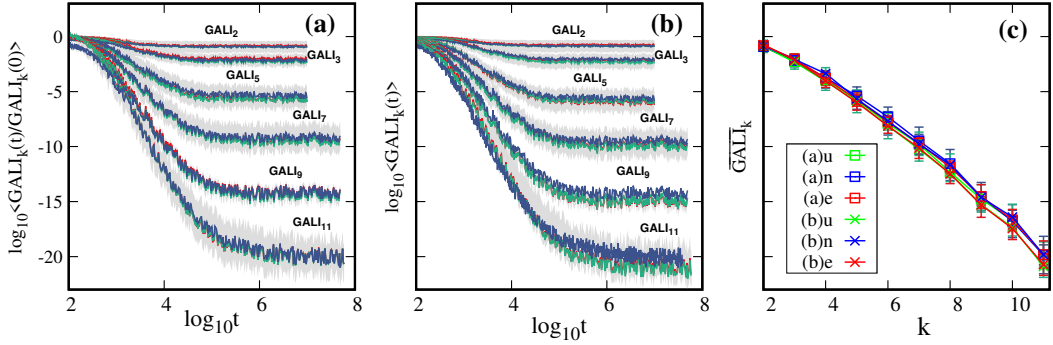}
	\caption{The average over different initial sets of deviation vectors time evolution of (a) $\langle \mbox{GALI}_k(t) / \mbox{GALI}_k(0) \rangle$ and (b) $\langle \mbox{GALI}_k(t) \rangle$ for $k=2$, 3, 5, 7, 9, 11, for a regular orbit in the neighborhood of the stable SPO1 ($D=0.12$, $H_N/N=0.01$) of Hamiltonian \eqref{eq:FPUT_B} with $N=11$. The coordinates of the initially (a) unit  [(b) orthonormalized] deviation vectors were chosen from a uniform [u] (green curves), normal [n] (blue curves) and exponential [e] (red curves) distributions. In the legend of panel (c), `(a)' and `(b)' denotes the first and second panel, and `u', `n', and `e' denote uniform, normal and exponential distributions respectively.}
	\label{fig:dif_dev_vec}
\end{figure}

We start our study by noting that the values of the GALIs are practically independent of the choice of the initial deviation vectors. To illustrate this property, we consider different sets of initial deviation vectors whose coordinates are drawn from three different types of probability distributions and compute the corresponding GALIs. In particular, we consider a uniform distribution in the interval $[-0.5, 0.5]$ (green curves in Fig.~\ref{fig:dif_dev_vec}), a normal distribution (blue curves in Fig.~\ref{fig:dif_dev_vec}) with mean 0 and standard deviation 1 in the interval $[-0.5, 0.5]$, and an exponential distribution with mean 1 (red curves in Fig.~\ref{fig:dif_dev_vec}). Fig.~\ref{fig:dif_dev_vec}a displays the time evolution of $\langle \mbox{GALI}_k(t) / \mbox{GALI}_k(0) \rangle$ for a regular orbit in the neighborhood of the stable SPO1 of Hamiltonian \eqref{eq:FPUT_B} for $N=11$ and $H_N/N=0.01$. The average is done over $n_v=10$ sets of normalized unit (denoted by [u]) initial deviation vectors. Fig.~\ref{fig:dif_dev_vec}b shows a similar computation but over $n_v=10$ sets of orthonormalized (denoted by [o]) deviation vectors, i.e.~GALI$_k(0)=1$. In Fig.~\ref{fig:dif_dev_vec}c we compare asymptotic values, $\overline{ \mbox{GALI}}_k$, obtained from the results of Figs.~\ref{fig:dif_dev_vec}a and \ref{fig:dif_dev_vec}b. The curves are more or less the same. This indicates that the choice of the initial deviation vectors does not affect the values of the GALIs. The three curves corresponding to the different initial distribution of the vectors' coordinates practically overlap both in Fig.~\ref{fig:dif_dev_vec}a and \ref{fig:dif_dev_vec}b which is a clear indication that the time evolution of the GALI$_k$ is similar, irrespective of the used distribution.  
\FloatBarrier

\begin{figure}[h!]
	\centering
	\includegraphics[trim=.0cm 0.25cm .0cm 1.cm,width=0.8\textwidth,keepaspectratio]{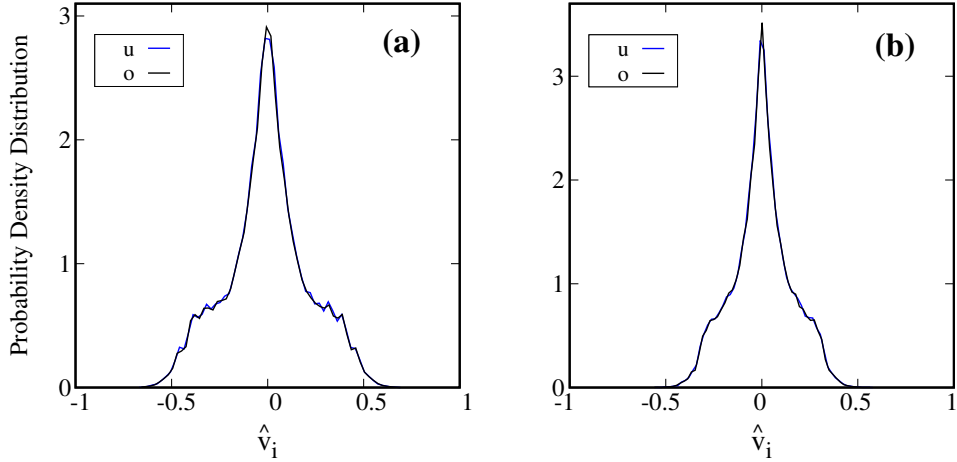}
	\caption{Probability density distributions of the coordinates $\hat{v}_i$ of (a) 11, and (b) 21 initially unit (blue curves - `u') and orthonormalized (black curves - `o') deviation vectors for a regular orbit of Hamiltonian \eqref{eq:FPUT_B} close to the stable SPO1, with (a) $H_N/N=0.01$, $D=0.12$ for $N=11$, and (b) $H_N/N=0.002$, $D=0.04$ for $N=21$.} 
	\label{fig:distributions_1}
\end{figure}

Furthermore, we analyze the distribution of the coordinates of the vectors $\hat{v}_j$ and the corresponding angles between these vectors $\theta_{ij}$ (seen in Eq.~\eqref{eqn:matrix_cos}). In particular, we consider the two cases $N=11$ and $N=21$. In Fig.~\ref{fig:distributions_1}a we show vector coordinates distributions obtained from the evolution of GALI$_k$ for a regular orbit of Hamiltonian \eqref{eq:FPUT_B} with $N=11$ which is close to the stable SPO1, with $H_N/N=0.01$ and $D=0.12$. When we follow the time evolution of GALI$_k$, $k=2,\ldots,11$ using unit or orthonormal initial deviation vectors, the coordinates of the $k$ deviation vectors and the corresponding angles on the torus have the same distribution. We note that for GALI$_k$ with high order $k$ we obtain better statistics as we have in our disposal a large number of vectors and corresponding angles. Thus, considering, for example, GALI$_{k}$ with the largest possible order, i.e.~GALI$_{11}$, we obtain a fine picture of the distributions. More specifically, in order to follow the evolution of GALI$_{11}$ we use $11$ initial deviation vectors with $22\times11=242$ coordinates and $11!/(9!2!)=55$ angles. In Fig.~\ref{fig:distributions_1}a we plot the probability density distributions of the coordinates of the unit deviation vectors needed for the evaluation of GALI$_{11}$ for a regular orbit close to the SPO1 of the Hamiltonian \eqref{eq:FPUT_B} with $N=11$. The distributions are generated from the coordinates of $11$ sets of deviation vectors over $10$ snapshots when the GALI$_{11}$ has reached its asymptotic value. A similar behavior is observed when increasing the number of particles. The time evolution of GALI$_{21}$ for regular orbits of \eqref{eq:FPUT_B} with $N=21$ close to the stable SPO1 $H_N/N=0.002$, $D=0.04$ using $21$ initial unit [u] and orthonormal [o] deviation vectors led to the creation of Fig.~\ref{fig:distributions_1}b where we plot the distributions obtained from the coordinates of $10$ sets of vectors as in Fig.~\ref{fig:distributions_1}a. The two curves, in both Figs.~\ref{fig:distributions_1}a and \ref{fig:distributions_1}b practically overlap.
\FloatBarrier

\begin{figure}[h!]
	\centering
	\includegraphics[trim=.0cm 0.25cm .0cm 1.cm,width=0.8\textwidth,keepaspectratio]{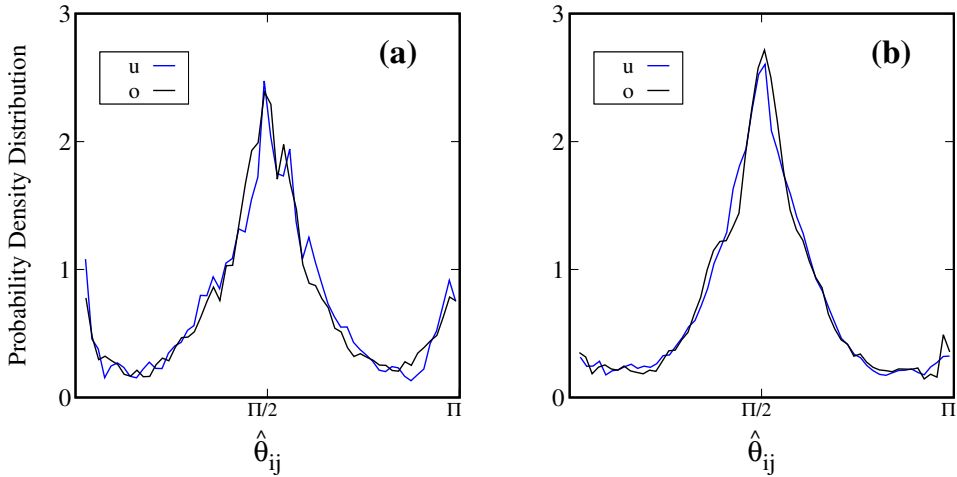}
	\caption{Probability density distributions of the angles $\theta_{ij}$ corresponding to the deviation vectors in Fig.~\ref{fig:distributions_1} for (a) $N=11$ and (b) $N=21$.}
	\label{fig:FPU_Angles_1}
\end{figure}

We further investigated how the distribution of the angles between each deviation vectors $\theta_{ij}$ behaves and a behavior similar to that seen in Fig.~\ref{fig:distributions_1} is obtained. Fig.~\ref{fig:FPU_Angles_1}a [Fig.~\ref{fig:FPU_Angles_1}b] displays the probability density distributions of the angles corresponding to the deviations in Fig.~\ref{fig:distributions_1}a [Fig.~\ref{fig:distributions_1}b]. The curves are almost the same, with distributions of the angles corresponding to vectors whose coordinates are drawn from the initial unit [u] and orthonormal [o].   
\FloatBarrier

Based on the results from Fig.~\ref{fig:dif_dev_vec} and Fig.~\ref{fig:distributions_1} we can argue that the distribution of the coordinates of the initial deviation vectors does not affect the asymptotic behavior of the GALI. Therefore, we can, for example, use initial unit deviation vectors whose coordinates are drawn from a uniform distribution for our analysis. 

\begin{figure}[h!]
	\centering
	\includegraphics[trim=1.5cm 0.25cm 2.5cm 0.cm,width=0.7\textwidth,keepaspectratio]{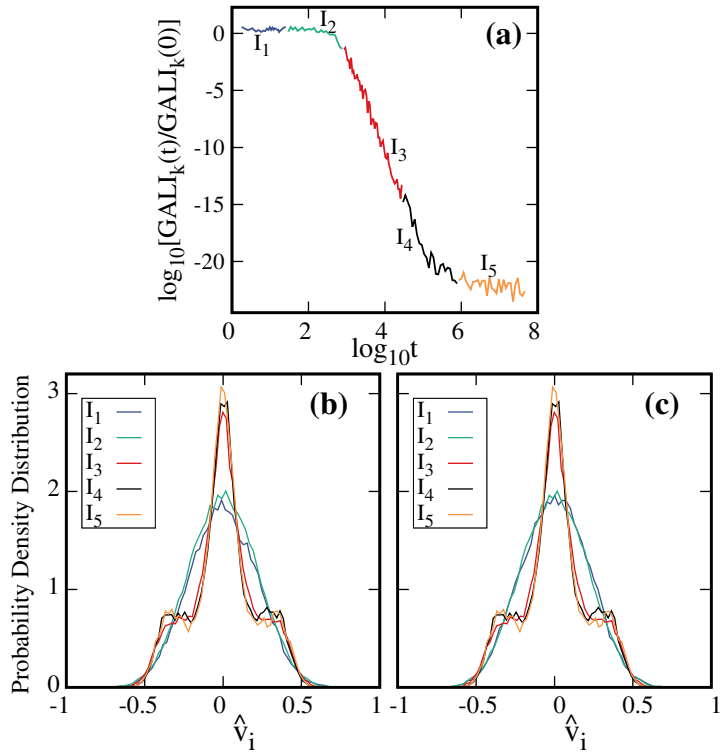}
	\caption{(a) The time evolution of the GALI$_{11}$ for a regular orbit in the neighborhood of the stable SPO1 orbit of Hamiltonian \eqref{eq:FPUT_B} for $N=11$ with $H_N/N=0.004$ and $D=0.04$. (b) The coordinate distributions of the 11 unit deviation vectors for the intervals I$_1$, I$_2$, I$_3$, I$_4$ and I$_5$ of (a). (c) Similar to (b) but for a different random set of initial deviation vectors.}
	\label{fig:distribution_evolve}
\end{figure}

So far we have investigated the behavior of the distribution of the coordinates of deviation vectors when the vectors have fallen on the tangent space of the torus, but it is interesting to also investigate the dynamics of the deviation vectors for the whole evolution of the regular orbit. In order to do so, we follow the evolution of GALI$_{11}$ (Fig.~\ref{fig:distribution_evolve}a) for the regular orbit close to the stable SPO1 of \eqref{eq:FPUT_B} for $N=11$, with $H_N/N=0.004$ and $D=0.04$, and analyze in Figs.~\ref{fig:distribution_evolve}b and \ref{fig:distribution_evolve}c the distributions of the coordinates of the deviation vectors by dividing the time evolution into five evenly spaced intervals, I$_1$: $0 \leq \log_{10}t < 1.5$ (blue curves),  I$_2$: $1.5\leq \log_{10}t < 3$ (green curves), I$_3$: $3 \leq \log_{10}t < 4.5$ (red curves), I$_4$: $4.5 \leq \log_{10}t < 6$ (black curves) and I$_5$: $6 \leq \log_{10}t \leq 7.5$ (yellow curves). Fig.~\ref{fig:distribution_evolve}b shows the distribution of the coordinates of the deviation vectors for the five intervals. In Fig.~\ref{fig:distribution_evolve}c we see a similar distribution as in Fig.~\ref{fig:distribution_evolve}b but with different random sets of initial deviation vectors. In both cases, we can say that the time evolution of the GALIs does not depend on the sets of initial deviation vectors. Moreover, as the GALIs approach their constant asymptotic value (interval I$_5$) the distribution of the coordinates of the deviation vectors has a sharply peaked shape with a high concentration in the middle.              

\begin{figure}
	\centering
	\includegraphics[trim=.0cm 0.25cm .0cm 1.cm,width=0.8\textwidth,keepaspectratio]{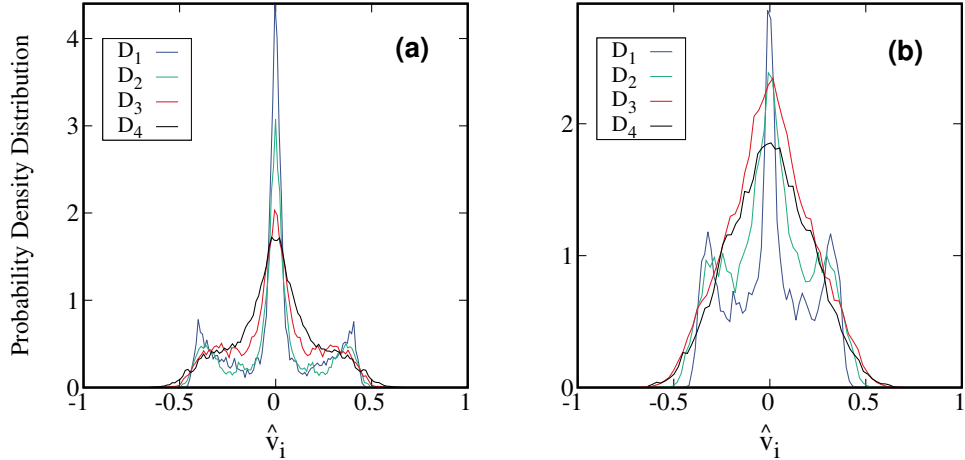}
	\caption{The probability density distributions of the coordinates of 11 unit deviation vectors used for the time evaluation of GALI$_{11}$ for regular orbits close to the stable (a) SPO1 and (b) SPO2 of the $\beta-$FPUT Hamiltonian \eqref{eq:FPUT_B} with $N=11$. Regular orbits have (a) $E=H_N/N=0.01$ and distances $D$ \eqref{eq:D} from the stable SPO1 $D_1=0.01$, $D_2=0.06$, $D_3=0.12$ and $D_4=0.22$ and  (b) $E=H_N/N=0.001$ and distances $D$ from the stable SPO2 $D_1=0.01$, $D_2=0.04$, $D_3=0.08$ and $D_4=0.12$.}
	\label{fig:distributions_ED_SPO1}
\end{figure}

\begin{figure}
	\centering
	\includegraphics[trim=.0cm 0.25cm .0cm 1.cm,width=0.8\textwidth,keepaspectratio]{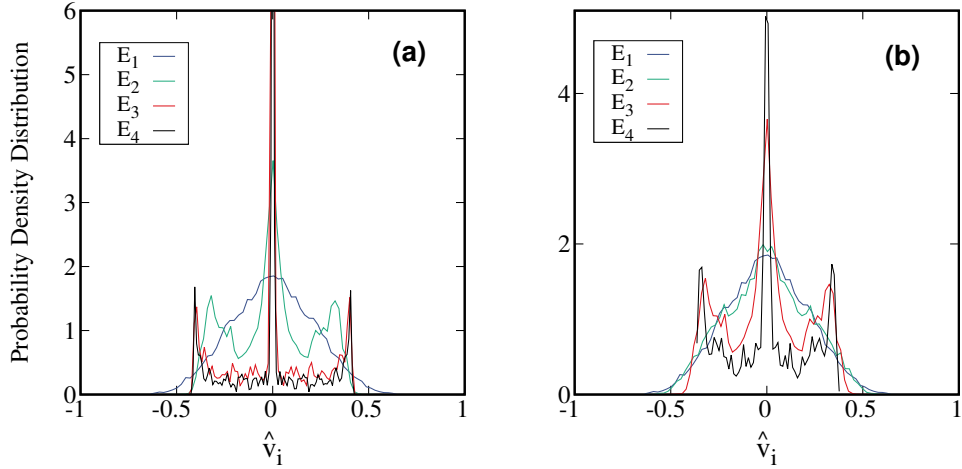}
	\caption{Similar to Fig.~\ref{fig:distributions_ED_SPO1} but for regular orbits with (a) $E_1=0.001$ and $D= 0.015$, $E_2=0.01$ and $D= 0.015$, $E_3=0.1$ and $D= 0.015$, and $E_4=0.175$ and $D=0.001$. (b) $E_1=0.001$ and $D= 0.12$, $E_2=0.004$ and $D= 0.12$, $E_3=0.008$ and $D= 0.04$, and $E_4=0.012$ and $D=0.01$.}
	\label{fig:distributions_ED_SPO2}
\end{figure}

\FloatBarrier
Finally we evaluated the asymptotic coordinate distributions of the deviation vectors used to compute the GALIs for regular orbits in the neighborhood of the stable SPO1 Fig.~\ref{fig:distributions_ED_SPO1}a and Fig.~\ref{fig:distributions_ED_SPO2}a and SPO2 Fig.~\ref{fig:distributions_ED_SPO1}b and Fig.~\ref{fig:distributions_ED_SPO2}b orbits. In our computations we use 10 sets of 11 initially linearly independent unit deviation vector. In Fig.~\ref{fig:distributions_ED_SPO1} we see the dependence of the final coordinate distribution of these vectors when the distance $D$ from the SPO is increased (for fixed $H/N$ values) for the stable SPO1 (Fig.~\ref{fig:distributions_ED_SPO1}a) and SPO2 (Fig.~\ref{fig:distributions_ED_SPO1}b) orbits. In Fig.~\ref{fig:distributions_ED_SPO2} we see similar results as in Fig.~\ref{fig:distributions_ED_SPO1} but when the orbits' energy density increases for cases close to the stable SPO1 (Fig.~\ref{fig:distributions_ED_SPO2}a) and SPO2 (Fig.~\ref{fig:distributions_ED_SPO2}b) periodic orbits. In all considered cases for small values of $D$ and $E$, the distributions have a peaked shape with a concentration in their middle. As the regular orbits get closer to the boundary of the stability island (increase D in Fig.~\ref{fig:distributions_ED_SPO1} and increase E in Fig.~\ref{fig:distributions_ED_SPO2}) the distributions become more sharply peaked with a very high concentration in their centers.     

\FloatBarrier
\FloatBarrier
\section{A multidimensional area-preserving mapping}
So far we have discussed the behavior of the GALIs for the $\beta-$FPUT Hamiltonian model \eqref{eq:FPUT_B} which has a continuous time $t$. In particular, we have studied the asymptotic behavior of GALIs for regular orbits in the neighborhood of the stable SPOs when their distance $D$ from them increases. Here, we will implement the same methodology for regular orbits of a multidimensional area-preserving mapping in order to investigate the generality of our findings.   

As we already discussed in Sec.~\ref{sec:Simple_DS}, the standard mapping describes a universal, generic of area-preserving mapping with divided phase space in which the integrable islands of stability are surrounded by chaotic regions. A $2N$D system of coupled mappings is: 

\begin{eqnarray}
x'_j &=& x_j + y'_j, \nonumber \\ 
y'_j &=& y_j + \dfrac{K_j}{2\pi} sin{\big(2\pi x_j\big)} - \dfrac{\mu}{2\pi} \Bigg \{ sin\Big[2\pi \big(x_{j+1}-x_j\big)\Big] +  sin\Big[2\pi \big(x_{j-1}-x_j\big)\Big] \Bigg\},
\label{eq:NDSM}
\end{eqnarray}
where $('{})$  indicates the new values of variables after one mapping iteration and $K_j$ is a dimensionless parameter that influences the degree of chaos, while $\mu_j$ is the coupling parameter between neighboring mappings where $j=1,2,\ldots, N$. All values have (mod $1$), i.e.~$0 \le x_j < 1$, $0 \le y_j < 1$ and also periodic boundary conditions are imposed: $x_0 = x_N$  and $x_{N+1} = x_1$. 

\begin{figure}[h!]
	\centering
	\includegraphics[width=0.45\textwidth,keepaspectratio]{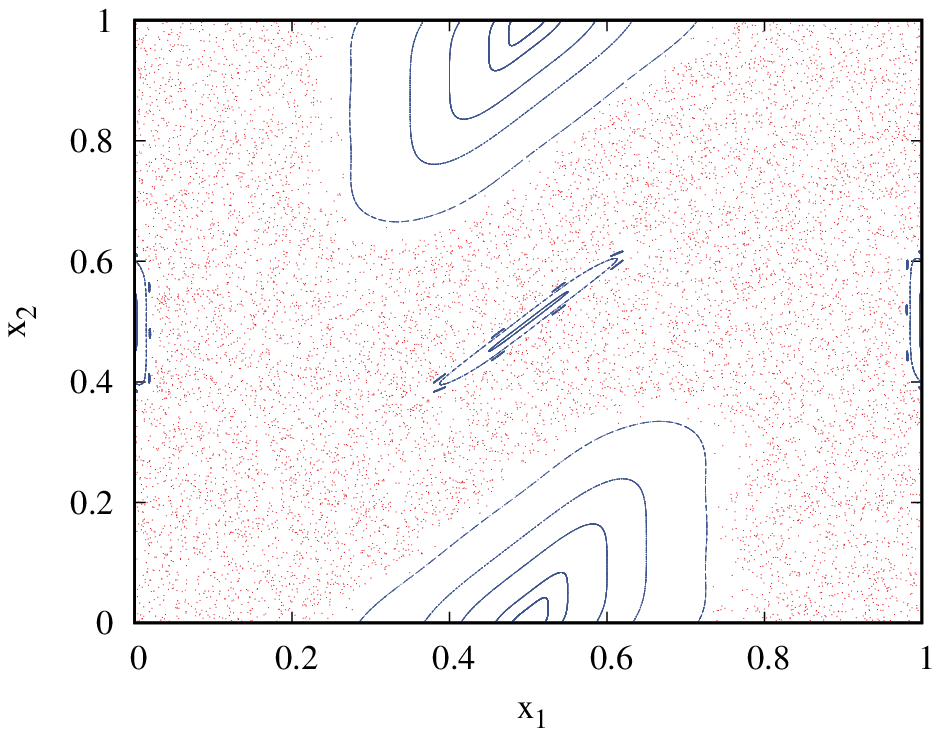}
	\caption{Phase space portraits of the standard mapping \eqref{eq:2DSM} with $K_j=2$ and $\mu=10^{-5}$.}
	\label{fig:STM_2D_PSS}
\end{figure}

For this coupled standard mapping model, the behavior of the GALIs for a given IC depends on the parameters $K_j$, as $K_j$ strongly influences the regular or chaotic nature of the mapping's orbits. Here, the objective is to show that what we have observed so far for the behavior of the GALIs in the case of the $\beta-$FPUT model (Sec.~\ref{sec:Behavior_SPOs}) and in particular the results of Figs.~\ref{fig:SPO1_11_GALIs} and \ref{fig:SPO1_11_change_D_E}, hold also for the discrete system \eqref{eq:NDSM}. In other words, we investigate the behavior of the GALIs when the transition of periodic orbits from regular to chaotic motion happens for the mapping \eqref{eq:NDSM} using the same analysis. In order to do this, we start from the center of the big island $(x_j,y_j)=(0.5,0)$  \cite{BMC2009}, $j=1,2,\ldots,N$ for a small value of $K_j$ and coupling $\mu$ and we locate a stable periodic orbit. The location of this island is shown in Fig.~\ref{fig:STM_2D_PSS} for the simple case of the 2D mapping \eqref{eq:2DSM}
\FloatBarrier

\begin{figure}[h!]
	\centering
	\includegraphics[width=0.45\textwidth,keepaspectratio]{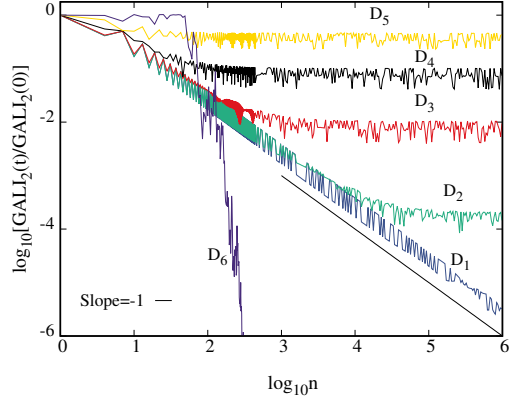}
	\caption{The time evolution of $\mbox{GALI}_2(t) / \mbox{GALI}_2(0)$ for orbits of the mapping \eqref{eq:NDSM} with $K_j=2$ and $\mu=10^{-5}$, and distances $D_1 = 0$, $D_2 = 0.005$, $D_3 = 0.15$, $D_4 = 0.1$, $D_5 = 0.2$ from the center $(x_j,y_j)=(0.5,0)$, $j=1,2,\ldots,20$ of the stability island for 1 set of initial deviation vectors. The straight line  correspond to a function proportional to $n^{-1}$.}
	\label{fig:STM_5_change_D}
\end{figure}

\begin{figure}[h!]
	\centering
	\includegraphics[trim=.0cm 0.25cm .0cm 1.75cm,width=0.8\textwidth,keepaspectratio]{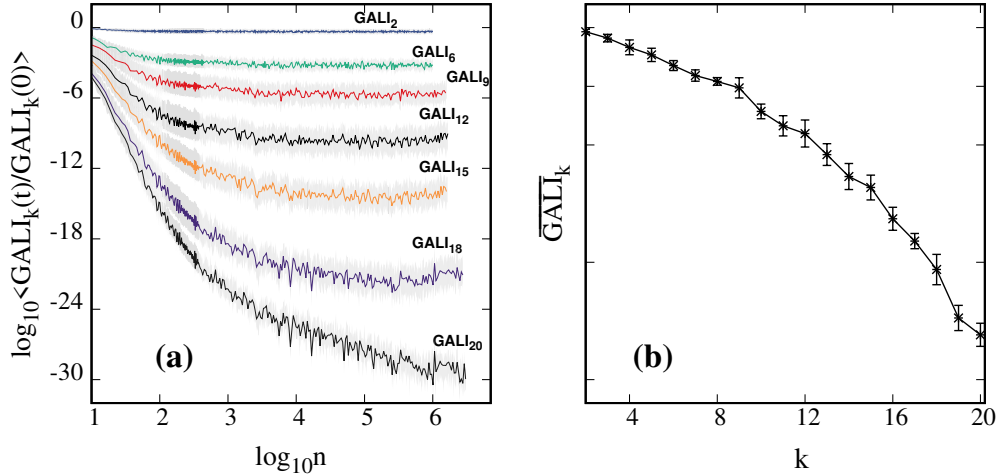}
	\caption{(a) The time evolution of the average, over $n_v=10$ sets of random initial deviation vectors, $\langle \mbox{GALI}_k(t) / \mbox{GALI}_k(0) \rangle$ for $k=$2, 6, 9, 12, 15, 18 and 20, for a regular orbit of the mapping \eqref{eq:NDSM} with $N=20$. The orbit's ICs are given in \cite{20DSM-IC} (b) Estimation of the asymptotic GALI values of panel (a), $\overline{ \mbox{GALI}}_k$, as a function of the order $k$. The shaded gray area around each curve in panel (a) and the error bars in panel (b) denote one standard deviation.}
	\label{fig:STM_20_GALIs}
\end{figure}

As we have seen in Fig.~\ref{fig:SPO1_11_change_D_E} when the stable orbit is very close to the SPO, the GALIs tend to zero following the power law \eqref{eq:GALI_st} (curve $D_1$ in Fig.~\ref{fig:SPO1_11_change_D_E}). Then as the orbit's IC is moved further and further away from the SPO the GALIs asymptotically tend to a positive value (curves $D_2$ to $D_5$ in Fig.~\ref{fig:SPO1_11_change_D_E}) while when the orbit's IC is outside the stability island the motion becomes chaotic and the GALIs will go to zero exponentially fast as it happens for the curve $D_6$ in Fig.~\ref{fig:SPO1_11_change_D_E}. All these behaviors can also be observed for the mapping, Fig.~\ref{fig:STM_5_change_D} clearly illustrate these results. There we see the behavior of the GALIs for increasing distances from the center of the island of stability. In particular, the evolution of the GALI$_2$ for different orbits starting in the neighborhood of the stable periodic orbit of \eqref{eq:NDSM} with parameters $K_j=2$ and $\mu=10^{-5}$, and distances $D_1 = 0$, $D_2 = 0.005$, $D_3 = 0.15$, $D_4 = 0.1$, $D_5 = 0.2$ and  $D_6 = 0.3$ are shown in Fig.~\ref{fig:STM_5_change_D}. We note that GALI$_2$ for the periodic orbit ($D_1=0$) goes to zero asymptotically following the power law $t^{-1}$ \eqref{eq:GALI_reg}. Then, as the value of $D$ is increased the GALI$_2$ gradually starts to saturate to a constant value ($D_2$ to $D_5$). Finally, the perturbed orbit becomes unstable and the GALI$_2$ goes to zero exponentially fast ($D_6=0.3$).
\FloatBarrier
In Fig.~\ref{fig:STM_20_GALIs} we follow the evolution of the GALI$_k$ in a similar way to Figs.~\ref{fig:SPO1_11_GALIs} and \ref{fig:SPO1_21_GALIs}. From the results of Fig.~\ref{fig:STM_20_GALIs}a we see that the GALI$_k$ with higher order $k$ needs more time to become constant. For example, GALI$_{20}$ asymptotically saturates around $t=10^{5.5}$ while GALI$_2$ saturates faster around $t=10$ time units. From Fig.~\ref{fig:STM_20_GALIs}b we see that the final asymptotic value $\overline{ \mbox{GALI}}_k$ drastically decrease as the order of the index $k$ increases. 
\FloatBarrier

\chapter{Summary and discussion} \label{Chapter-Four}

In this work, we performed several numerical investigations of multidimensional dynamical systems by implementing several chaos detection techniques. In the first chapters of the thesis, we gave a brief overview of the basic notions of Hamiltonian dynamics along with a presentation of some basic numerical techniques for investigating chaos. The creation of the Poincar{\'e} surface of section is an important tool to understand the behavior of dynamical systems by representing trajectories of the full $2N$D phase space by an object in a $2(N-1)$D spaces, which is more suited for lower-dimensional dynamical systems. On the other hand, chaos indicators like LEs, SALI and GALI, can also be used for the same purpose, having the advantage of efficiently discriminating between regular and chaotic motion in high-dimensional systems. Initially, we discussed the LEs which measure the average rate of growth or shrinking of small perturbations to the orbits of a dynamical system. The mLE is a powerful tool to determine the chaotic and regular nature of an orbit, while the whole spectrum of LEs provides additional information on the dynamics of the system. Then we consider the SALI which is related to the area defined by two deviation vectors. It is an efficient and simple method to determine the ordered or chaotic behavior of orbits in dynamical systems. Generalizing the idea of the SALI leads to a computationally more efficient technique called the GALI method. The GALI of order $k$ (GALI$_k$) represents the volume of a parallelepiped formed by $k$ initially linearly independent deviation vectors of unit length. 

For the computation of these chaos indicators, a symplectic integration scheme was used to follow the time evolution of the equations of motion and variational equations of the Hamiltonian. In our presentation, we applied these indicators on the 2D H{\'e}non-Heiles Hamiltonian, 2D and 4D symplectic mappings as well as on simple dissipative systems.

The main results of our work was presented in Chapter \ref{Chapter-Three}, where we thoroughly investigated the behavior of the GALI chaos indicator for regular motions in multidimensional Hamiltonian systems. In the case of stable periodic orbits, GALI$_k$ goes to zero following the power law described in \cite{MSA12}, while it tends to zero exponentially fast following a rate that depends on the values of different LEs for unstable orbits. In our study, more attention was given to regular motions that lie on $N$D tori, i.e.~GALI$_k$, $k =2,3,\dots, N$ asymptotically attain a non zero constant value. To analyze these asymptotic values of GALIs we considered several regular orbits in the vicinity of two basic simple periodic orbits of the $\beta-$FPUT model \eqref{eq:FPUT_B}, namely, SPO1 and SPO2 orbits \cite{AB06}, for various numbers of the system's degrees of freedom.

The main results of our study are the following: the asymptotic GALI values, $\overline{ \mbox{GALI}}_k$, $k=2,3,\ldots,N$ for regular orbits on $N$D tori of multidimensional Hamiltonian models depend on:  

\begin{itemize}
	\item The order $k$ of the index. The constant values of the GALIs decrease as $k$ increases (Figs.~\ref{fig:SPO1_11_GALIs}, \ref{fig:SPO1_21_GALIs}, and \ref{fig:STM_20_GALIs}). 
	\item The phase space distance $D$ of the regular orbit from the nearby stable periodic orbit. The asymptotic GALI values increase when we increase the distance $D$ for a fixed energy $H_N$ (Figs.~\ref{fig:SPO1_11_change_D_many}a and \ref{fig:SPO2_20_change}a). 
	\item The orbit's energy $H_N$. The $\overline{ \mbox{GALI}}_k$ values decrease when we approach the destabilization energy (Figs.~\ref{fig:SPO1_11_change_D_many}b and \ref{fig:SPO2_20_change}b). 
\end{itemize}

In addition, we computed the evolution of GALIs using initially linearly independent deviation vectors generated from different random distributions (uniform, normal and exponential). We choose the deviation vectors in two ways: firstly, by normalizing each vector and secondly, by setting the angles between vectors to be right angles (i.e. orthonormal vectors). In all cases, we observed that the volume of the parallelepiped formed by the deviation vectors does not depend on the distribution from which the vectors were created and the way we choose our vectors (Fig.~\ref{fig:dif_dev_vec}). This means that the values of GALIs do not depend on the type of initial deviation vectors but rather on the initial conditions. Furthermore, we showed that the shape of the coordinate distributions of the deviation vectors and the angles between them when the GALIs have reached their asymptotic values is also independent of the way the initial deviation vectors were produced (Fig.~\ref{fig:distributions_1} and Fig.~\ref{fig:FPU_Angles_1}).

Finally, in order to corroborate the generality of our results, we evaluated the asymptotic constant values of GALIs for regular orbits of the $2N$D multidimensional area-preserving mapping \eqref{eq:NDSM} and showed that similar results to the ones seen for the $\beta-$FPUT model were also obtained.

\newpage
\hfill 
\section*{Dissemination of the results of this work}
Parts of this work were included in:
\begin{enumerate}
	\item Paper in a peer review journal 
		\begin{itemize}
		\item ``On the behavior of the Generalized Alignment Index (GALI) method for regular motion in multidimensional Hamiltonian systems", \textbf{Moges H.T.}, Manos T. and Skokos Ch.: 2019, Nonlin. Phenom. Complex Syst. (in press), preprint version: \url{https://arxiv.org/abs/2001.00803}
	 	\end{itemize}
	\item Conference presentations
	\begin{itemize}
		\item ``Investigation of Chaos by the Generalized Alignment Index (GALI) Method", Poster presentation in the International Conference on “Mathematical Modeling of Complex Systems”, Pescara (Italy), July 3-11, 2019.
		\item ``Investigation of Chaos by the Generalized Alignment Index (GALI) Method", Poster presentation in the 12th Annual University of Cape Town/Stellenbosh Faculty of Science Postgraduate Symposium, Stellenbosch (South Africa), September 11, 2019.
		\item ``Investigating the properties of regular motion in multidimensional nonlinear lattices by the Generalized Alignment Index (GALI) method", Oral presentation in the 62nd Annual Congress of the South African Mathematical Society, Cape Town (South Africa), December 2-4, 2019.
	\end{itemize}  
\end{enumerate}
	

\printindex


\end{document}